\documentclass[a4paper,fleqn]{cas-dc}

\usepackage[numbers]{natbib}

\usepackage{graphicx}
\usepackage{amsmath}
\usepackage{amssymb}
\usepackage{tikz}
\usepackage{xspace}
\usepackage{booktabs}
\usepackage{hyperref}
\usepackage{mathtools}
\usepackage{dsfont}

\newcommand{\msmarcovo}{\textsc{Ms Marco-v1}\xspace}

\newcommand{\sift}{\textsc{Sift}\xspace}
\newcommand{\siftom}{\textsc{Sift1M}\xspace}
\newcommand{\siftob}{\textsc{Sift1B}\xspace}

\newcommand{\seismic}{\textsc{Seismic}\xspace}
\newcommand{\seismicw}{\textsc{SeismicWave}\xspace}
\newcommand{\kannolo}{\textsc{kANNolo}\xspace}
\newcommand{\vectorium}{\textsc{Vectorium}\xspace}
\newcommand{\faiss}{\textsc{Faiss}\xspace}
\newcommand{\warp}{\textsc{Warp}\xspace}
\newcommand{\emvb}{\textsc{Emvb}\xspace}
\newcommand{\igp}{\textsc{Igp}\xspace}

\newcommand{\hnsw}{\textsc{Hnsw}\xspace}
\newcommand{\rerankindex}{\textsc{RerankIndex}\xspace}
\newcommand{\vamana}{\textsc{Vamana}\xspace}

\newcommand{\kannolohnsw}{\textsc{kANNolo}-\textsc{Hnsw}\xspace}
\newcommand{\kannolorerank}{\textsc{kANNolo}-\textsc{Rerank}\xspace}

\newcommand{\splade}{\textsc{Splade}\xspace}
\newcommand{\esplade}{\textsc{E-Splade}\xspace}
\newcommand{\spladeT}{\textsc{Splade}-v3\xspace}
\newcommand{\colbert}{\textsc{ColBERT}\xspace}
\newcommand{\colbertvt}{\textsc{ColBERTv2}\xspace}

\newcommand{\svb}{\textsc{Stream VByte}\xspace}

\newcommand{\cocond}{\textsc{CoCondenser}\xspace}
\newcommand{\cocondo}{\textsc{CoCondenser}}
\newcommand{\snowf}{\textsc{SnowFlake}\xspace}

\newcommand{\tachiom}{\textsc{Tachiom}\xspace}

\newcommand{\dragon}{\textsc{Dragon}\xspace}

\newcommand{\hnswlib}{\texttt{hnswlib}\xspace}
\newcommand{\parlayann}{\textsc{ParlayANN}\xspace}
\newcommand{\annlib}{\textsc{ANNLib}\xspace}

\newcommand{\rabitq}{\textsc{RaBitQ}\xspace}
\newcommand{\rabitqext}{\textsc{RaBitQ}-ext\xspace}
\newcommand{\dotpacking}{\textsc{DotPacking8}\xspace}
\newcommand{\jumpdot}{\textsc{JumpDot}\xspace}
\newcommand{\pq}{\textsc{PQ}\xspace}

\newcommand{\vibe}{\textsc{Vibe}\xspace}
\newcommand{\glass}{\textsc{Glass}\xspace}
\newcommand{\symphonyqg}{\textsc{SymphonyQG}\xspace}
\newcommand{\lorann}{\textsc{LoRANN}\xspace}
\newcommand{\svsleanvec}{\textsc{Svs}-\textsc{LeanVec}\xspace}
\newcommand{\rabitqhnsw}{\textsc{Hnsw}-\textsc{RaBitQ}\xspace}

\newcommand{\agnews}{\textsc{AgNews}\xspace}
\newcommand{\msmarcoqwen}{\textsc{MsMarco}-\textsc{Qwen}\xspace}
\newcommand{\dpr}{\textsc{Dpr}\xspace}
\newcommand{\laion}{\textsc{Laion}\xspace}

\newcommand{\lsp}{\textsc{Lsp}\xspace}

\newcommand{\sindi}{\textsc{Sindi}\xspace}
\newcommand{\pyann}{\textsc{PyANN}\xspace}
\newcommand{\msmarcosplade}{\textsc{MsMarco}-\textsc{SPLADEv3}\xspace}
\newcommand{\msmarcoesplade}{\textsc{MsMarco}-\textsc{E-Splade}\xspace}
\newcommand{\nq}{\textsc{Nq}\xspace}

\begin{document}
\let\WriteBookmarks\relax
\def\floatpagepagefraction{1}
\def\textpagefraction{.001}

\shorttitle{Modular and Efficient Approximate Nearest Neighbors Search with Vectorium and kANNolo}
\shortauthors{S. Martinico \emph{et al.}}

\title[mode = title]{Modular and Efficient Approximate Nearest Neighbors Search with Vectorium and kANNolo}

\author[1,2]{Silvio Martinico}[orcid=0009-0005-7280-6147]
\cormark[1]
\ead{silvio.martinico@isti.cnr.it}
\credit{Conceptualization, Methodology, Software, Validation, Formal analysis, Investigation, Data curation, Visualization, Writing -- original draft, Writing -- review \& editing}

\author[2]{Franco Maria Nardini}[orcid=0000-0003-3183-334X]
\ead{francomaria.nardini@isti.cnr.it}
\credit{Conceptualization, Methodology, Software, Validation, Formal analysis, Investigation, Data curation, Visualization, Writing -- review \& editing}

\author[2]{Cosimo Rulli}[orcid=0000-0003-0194-361X]
\fnmark[1]
\ead{cosimo@linkup.so}
\credit{Conceptualization, Methodology, Software, Validation, Formal analysis, Investigation, Data curation, Visualization, Writing -- review \& editing}

\author[1]{Rossano Venturini}[orcid=0000-0002-9830-3936]
\ead{rossano.venturini@unipi.it}
\credit{Conceptualization, Methodology, Software, Validation, Formal analysis, Investigation, Data curation, Visualization, Writing -- review \& editing}

\affiliation[1]{organization={Department of Computer Science, University of Pisa},
                addressline={Largo Bruno Pontecorvo 3},
                postcode={56127},
                city={Pisa},
                country={Italy}}

\affiliation[2]{organization={Institute of Information Science and Technologies ``Alessandro Faedo'' (ISTI), National Research Council of Italy (CNR)},
                addressline={Via Giuseppe Moruzzi 1},
                postcode={56127},
                city={Pisa},
                country={Italy}}

\cortext[cor1]{Corresponding author}
\fntext[1]{Present address: Linkup Technologies, 28 avenue des P\'epini\`eres, 94260 Fresnes, France.}

\begin{abstract}
Approximate Nearest Neighbors (ANN) search over neural embeddings powers modern retrieval systems such as Retrieval-Augmented Generation. Embeddings come in three representations (dense, sparse and multi-vector), and efficient search combines an index structure, a similarity function and a quantization scheme. Existing libraries fix one combination, hard-code each one separately, or factor only one side of the problem, and none covers all three representations. We present \vectorium{} and \kannolo, two open-source Rust libraries that split this design space into a data-management layer and an indexing layer. \vectorium{} abstracts the vector, the distance function, the quantizer and the dataset independently, and provides $11$ encoders, from $32$-bit floats to sub-bit codes. On top of it, \kannolo{} implements Hierarchical Navigable Small-World (\hnsw) graph and Inverted Vector File (IVF) indexes that are generic over representation, encoder and metric, and a two-stage index that composes any first-stage index with any rescoring dataset. Since every combination is specialized at compile time, this modularity adds no run-time overhead, and the resulting indexes reach state-of-the-art performance: our Product Quantization matches \faiss{} in accuracy at competitive speed, our \rabitq{} and \hnsw{} match or outperform their reference implementations, and our IVF outperforms \faiss. Against six dense and five sparse systems on seven public collections, our indexes give the best memory--throughput trade-off on dense data in $9$ of $12$ collection--accuracy pairs and perform well on out-of-distribution queries. On sparse data, the systems built on \vectorium{} lead on all three collections, with \seismic{} and \seismicw{} the fastest and \kannolo's indexes the most compact.
\end{abstract}


\maketitle

\section{Introduction}
\label{sec:intro}

Vector search is a core component of many Artificial Intelligence (AI) applications. Retrieval-Augmented Generation (RAG), semantic search, recommendation and multimodal retrieval all rely on the same operation: embed a query, then find the items of a collection whose embeddings are closest to it under a similarity function. These systems run on millions or billions of high-dimensional vectors, and at this scale an exhaustive scan is impractical. Approximate Nearest Neighbors (ANN) search therefore trades a small loss in accuracy for orders of magnitude of throughput.

Embeddings come in different representations. A dense dual encoder produces a compact real vector per document. A learned sparse encoder produces a high-dimensional vector with only a small number of nonzero coordinates. A multi-vector encoder produces one dense vector per token and scores a document with the MaxSim operator. The three representations offer different trade-offs between memory, effectiveness and efficiency. Which of them is preferable depends on the task or on the data modality: for example, sparse representations are often well suited to text, while dense ones usually work better for images. The representations can also be combined to improve the quality of the results~\cite{martinico2026multivector}. Researchers and practitioners therefore need to move freely between the three. They need to explore these trade-offs, design new retrieval techniques, build their RAG or agentic pipelines, and study how the representations interact.

Serving any of the three efficiently requires two main ingredients: an index data structure for efficient search, and a quantization scheme that reduces the memory occupied by the vectors. Each ingredient involves further choices: the index structure (e.g. a proximity graph or an inverted index), the similarity function (e.g. Euclidean distance or inner product), the quantization scheme, and the optimizations applied on top of them (e.g. compressing the edges of a graph index). Representation, index, metric and compression are largely independent of one another. Together they span a large cross-product of configurations.

\paragraph{The gap.} Libraries differ in how much of that cross-product a user can reach. Earlier libraries either implement one index over one compression scheme~\cite{hnswlib,diskann}, which offers no flexibility, or cover more combinations by hard-coding each of them. \faiss~\cite{ivf_gpu,douze2024faiss}, to date the most widely adopted ANN library for dense data, exposes one class per combination of index structure and quantizer. Each class has its own construction and search code path (Section~\ref{sec:related_libraries}). The code therefore grows with every combination, which hurts extensibility and maintainability. Recent works have developed more modular frameworks, but each of them factors only one side of the problem. \parlayann~\cite{parlayann} implements several graph algorithms on shared parallel primitives, and \annlib~\cite{shen2026annlibdevelopmentframeworkefficient} separates the graph algorithm from the data structure that stores the graph. Both are generic over the point type and the distance, but provide few concrete implementations of either: \annlib{} leaves both to the user, and \parlayann{} ships three dense point types with no quantizers. \glass~\cite{pyglass} factors the other side. It selects the quantized representations of dense vectors used to build, search and refine independently of one another, but its graph layout is fixed inside each index. Two limits therefore remain. The first is how far the factoring goes, which determines how easy a library is to understand, maintain and extend. No library abstracts each level of the data side (the vector, the distance function, the quantizer and the dataset) independently of the others and builds its indexes on top of those abstractions. With such a design, a new component would reach every combination without code specific to each of them. In addition, speed is often obtained with distance kernels hand-written for each instruction set, as in \hnswlib, \faiss{} and \glass. This code is duplicated across instruction sets and grows with every combination it serves. The second limit is the representation of the data. ANN libraries target one representation, in most cases the dense one. Support for two is rare and reduces performance on each~\cite{nmslib}. No ANN library covers all three under a unified interface, so an optimization written for one representation does not reach the others.

\paragraph{Our contribution.} We present \vectorium{} and \kannolo, two Rust libraries that split this design space into two layers. \vectorium{} (Section~\ref{sec:data_management}) is the data-management layer. It defines how a vector is represented in memory, how two vectors are scored under a metric, how a collection is compressed, and how it is stored and accessed. \kannolo{} (Section~\ref{sec:indexing}) is the indexing layer. It implements standard ANN indexes, \hnsw{} and IVF, each generic over the representations, encoders, metrics and datasets below it. It also composes them into new indexes through a two-stage index.

Every component is generic, and the factoring adds no run-time overhead. The compiler specializes each index for the encoding it runs on, so the search loop contains no indirection. A new encoder is therefore available to every index as soon as it is written, and a new index works with every encoding already in the library. The resulting implementations reach state-of-the-art performance: where they implement the same algorithm as a hand-tuned library, they are competitive with it or faster (Sections~\ref{sec:data_management} and~\ref{sec:indexing}). In exchange, compilation takes longer and binaries are larger. The distance computations follow the same principle. Each is a few lines of generic code that the compiler vectorizes for the target hardware, with explicit SIMD only for the packed encodings. They remain competitive with hand-optimized implementations (Section~\ref{sec:product_quantization}). Each component can thus be read, maintained and extended on its own. Finally, \vectorium{} can be used independently of \kannolo. It also underlies the learned sparse retrieval engine \seismic~\cite{seismic,seismic_journal} and the multi-vector retrieval engine \tachiom~\cite{tachiom}, and it can support the development of any new index technique by providing its data-management layer.

The paper is organized as follows. Section~\ref{sec:related} covers the background and the related work on the three representations and on the index and compression techniques this work builds on. Section~\ref{sec:setup} fixes the notation, the evaluation measures, the collections and the measurement conventions. Sections~\ref{sec:data_management} and~\ref{sec:indexing} present \vectorium{} and \kannolo{} respectively, each with the experiments that support its design decisions. Section~\ref{sec:exp} evaluates the two libraries end to end, and Section~\ref{sec:conclusions} concludes.

\subsection{Contributions}

The contributions of this work are the following:

\begin{itemize}
    \item We introduce \vectorium, a data-management library for dense, sparse and multi-vector data. It offers $11$ encoders behind one interface, from lossless \texttt{f32} down to sub-bit codes. They include well-known quantizers that we reimplement, and our own encoders for learned sparse vectors, \texttt{BlockedSparseEncoder} and \texttt{DotPacking8Encoder}~\cite{nardini2026compactefficientindexeslearned}. The reimplementations match or beat their reference implementations: our Product Quantizer matches \faiss{} in accuracy at competitive speed, and our \rabitq{} scan is $1.4$--$2.4\times$ faster than the original authors' implementation (Section~\ref{sec:data_management}).

    \item We present \kannolo, extended from its first version~\cite{kannolo} (Section~\ref{sec:related_libraries} lists the differences). It provides \hnsw{} and IVF indexes generic over \vectorium's representations and encodings, a graph layer shared by all graph indexes, and an experimental study of the optimizations a modern \hnsw{} needs. Its \hnsw{} and IVF match or outperform the reference implementations, \hnswlib{} and \faiss{} (Section~\ref{sec:indexing}).

    \item We present the \rerankindex, \kannolo's two-stage index, which composes any first-stage index with any rescoring dataset. It was proposed for a sparse-to-multi-vector pipeline~\cite{martinico2026multivector}. We generalize it into a framework that combines any representations, encodings and indexes. We show that it expresses the two-stage designs behind many recent graph indexes, and that it improves on its first-stage graph index used alone.

    \item An evaluation against six dense and five sparse state-of-the-art systems on seven publicly available collections (Section~\ref{sec:exp}). Two of the sparse systems, \seismic{} and \seismicw, are built on \vectorium. On dense data, our indexes offer the best memory--throughput trade-off in $9$ of $12$ collection--accuracy target pairs. At $99\%$ accuracy our two-stage \hnsw{} reaches $2.5$--$3.3\times$ the throughput of \hnswlib{} with an index half the size. On the out-of-distribution queries, it is one of only two systems, with \lorann, to reach $99\%$ accuracy. On sparse data, the systems built on \vectorium{} lead. On all three collections, \seismic{} and \seismicw{} are the fastest systems at every target, and our indexes are the most compact. Our two-stage index is the fastest graph index: it is $5.2$--$48.4\times$ faster than \pyann{} with a third of its memory.

    \item We release both libraries open source under the MIT license, \kannolo\footnote{\url{https://github.com/TusKANNy/kannolo}} and \vectorium,\footnote{\url{https://github.com/TusKANNy/vectorium}} with Python bindings that give access to dense, sparse and multi-vector retrieval through a single interface.
\end{itemize}

\section{Background and Related Work}
\label{sec:related}

This section introduces the vector representations used in retrieval and the quantization schemes that compress them. It then states the approximate search problem and surveys the index families and adjacency-list compression methods that Sections~\ref{sec:data_management} and~\ref{sec:indexing} build on. It closes with the libraries and systems that cover this design space, and with the relation of \vectorium{} and \kannolo{} to the first version of \kannolo{} and to our earlier publications.

\subsection{Vector Representations and Retrieval Models}
\label{sec:related_representations}

To scale, retrieval systems rely on cheap data representations. These representations are built offline and evaluated efficiently at query time to gather a set of relevant results, which slower but more accurate methods such as cross-encoders can optionally re-rank. Vectors are the most widespread such representation, since they express a set of features compactly.

Before representations were learned, they were engineered, from term weighting such as BM25~\cite{bm25} in text retrieval to local descriptors such as SIFT~\cite{sift} in computer vision. Neural representations, or embeddings, have superseded them. They fall into three families: dense, sparse and multi-vector. The last is also called late-interaction, after the MaxSim scoring it relies on.

\paragraph{Dense embeddings.} A dual encoder built on a pretrained transformer~\cite{attention,bert} maps a text into a single real vector of fixed dimension. The vector typically has a few hundred to a few thousand components, all of them populated, and texts are ranked by inner product or cosine similarity. DPR~\cite{dpr} established the architecture for passage retrieval. \dragon~\cite{dragon} reaches state-of-the-art zero-shot effectiveness through diverse data augmentation, and \snowf~\cite{snowflake} is a recent open family of general-purpose text encoders. Most ANN work targets this representation.

\paragraph{Learned sparse embeddings.} A learned sparse encoder maps a text into a vector over the vocabulary in which only a few dozen to a few hundred coordinates are nonzero. The representation is therefore semantically informed and can still be indexed by the inverted-index machinery of lexical retrieval~\cite{sparse1,sparse2,sparse3}. uniCOIL~\cite{unicoil1,unicoil2} learns a weight per term over a document expanded by doc2query~\cite{doc2query}. \splade~\cite{splade,formal2021splade} adds sparsity-inducing regularization and log-saturated term weights, and was later refined for effectiveness and for efficiency~\cite{lassance2024splade,formal2024towards}. Learned sparse retrieval (LSR) has a second advantage besides that machinery: its representations are rooted in a vocabulary, so they are interpretable.

\paragraph{Multi-vector (late-interaction) embeddings.} Pioneered by \colbert~\cite{colbert}, multi-vector representations keep one dense vector per token. A document is scored by the MaxSim operator, the sum over query tokens of the best inner product against any document token,
\[
  \operatorname{MaxSim}(q, x) \;=\; \sum_{i=1}^{|q|} \; \max_{1 \le j \le |x|} \; \langle q_i,\, x_j \rangle .
\]
Every query token interacts with every document token. This gives strong effectiveness but makes exhaustive scoring costly: a document costs $|x|$ vectors of storage and $|q|\cdot|x|$ inner products of work. \colbertvt~\cite{colbertv2} compresses the token vectors as centroid-plus-residual codes.

\subsection{Vector Quantization}
\label{sec:related_quantization}

Vector search usually relies on in-memory indexes. These indexes store the vectors themselves, so the footprint of the vectors is one of the main bottlenecks of the search. \emph{Quantization} techniques reduce this footprint by mapping the continuous values of a vector onto a smaller discrete set.

\emph{Scalar quantization} casts every component into a narrower numeric type. For instance, it reduces $32$-bit floating point numbers to the $16$-bit half-precision \texttt{f16} and \texttt{bf16} formats, or to a fixed-point integer under a data-dependent range. The simplest uniform quantizers scale every component on a single global range. Research has since moved to more sophisticated techniques that keep as much information as possible in the few bits available. Locally-adaptive Vector Quantization (LVQ)~\cite{lvq} takes the range from the minimum and maximum component of each individual vector, so every vector is quantized against its own bounds. LVQ also adds a residual second level to the resulting codes, used for reranking.

\rabitq~\cite{rabitq} makes binary quantization, a single bit per component, viable. Each vector is centered by the collection's mean and randomly rotated. This gives a residual $r = P(x-\mu)$ whose information is spread evenly over the coordinates. The code stores the sign of each component of $r$, together with two scalars of metadata. From these, the distance is estimated \emph{without bias} and with a theoretical error bound. The codes can therefore be used for ranking on their own, and not only as a filter. Extended \rabitq~\cite{rabitq_ext} generalizes the scheme to $b$ bits per component: a sign bit and $b-1$ magnitude bits.

In scalar quantization the distance computation is usually an \emph{asymmetric distance computation} (ADC), meaning that the query is left unquantized. Quantizing the query as well can reduce the computation to bit operations, as in \rabitq.

Scalar quantization has its limits at low bit budgets. It spends at least one bit on each component, and the points it can represent lie on a regular grid, which the distribution of the data may not follow. \emph{Vector quantization} techniques learn a set of codes, the \emph{codebook}, and encode each vector as a reference to its closest code. Product Quantization (PQ)~\cite{pq} makes this viable at scale. It splits the $d$ dimensions into $m$ subspaces of $\text{dsub} = d/m$ dimensions and learns a codebook of $2^{\text{nbits}}$ centroids per subspace by $k$-means. A vector is encoded as the $m$ indices of its nearest centroids, which take $m \cdot \text{nbits}$ bits regardless of $d$. Scoring is \emph{asymmetric}: the query is not quantized. Every partial distance the scan can need is precomputed into a per-query table of $m \times 2^{\text{nbits}}$ entries, so a candidate costs $m$ lookups and $m-1$ sums. Refinements rotate the space before splitting it (OPQ~\cite{opq}), quantize residuals iteratively~\cite{rq,huijben2024residual}, drop the independence between subspaces~\cite{babenko2014additive}, or weight the error by its effect on the inner product~\cite{scann1}. The fast-scan kernels of \faiss{} shrink a subspace to $16$ centroids so that a sub-table fits in a vector register~\cite{douze2024faiss}.

Quantization compresses the values of a vector. A sparse vector, however, is made of two streams: the weights and the integer identifiers of the components they belong to. The identifiers cannot be quantized in a lossy way, since a perturbed identifier names a different component and not a nearby one. Once the weights are quantized down to a byte, the identifiers account for the larger share of the footprint. They are therefore compressed losslessly, with the integer codes of inverted indexes. \dotpacking~\cite{nardini2026compactefficientindexeslearned} was introduced to cut this identifier overhead. It bit-packs the gaps between consecutive identifiers and computes the inner product \emph{directly on the compressed form}, so a compressed scan runs close to the speed of an uncompressed one.

\subsection{Approximate Nearest Neighbor Search}
\label{sec:related_ann}

Vector search is usually framed as $k$-Nearest-Neighbor ($k$-NN) search.
Given a collection $\mathcal{X} = \{x_1,\dots,x_n\}\subset \mathbb{R}^d$ of $n$ vectors in dimension $d$, and given a query $q$ and an integer $k$, $k$-NN search asks for the $k$ elements of $\mathcal{X}$ that a similarity function ranks highest~\cite{cover1967nearest}. Under \emph{squared-Euclidean distance} the answer minimizes $\lVert q - x\rVert^2$. Under \emph{inner product} the answer maximizes $\langle q, x\rangle$, and the task is also called \emph{Maximum Inner Product Search} (MIPS). However, $k$-NN commonly serves as an umbrella term for any task that asks for the best-matching vectors under a given similarity function.

Answering such a query exactly costs a scan of the collection, and the classical space-partitioning structures of Section~\ref{sec:related_ann_indexes} do not avoid that cost in high dimension. Approximate Nearest Neighbor (ANN) search therefore trades a limited loss of quality for a large gain in throughput~\cite{indyk1998approximate,datar2004lsh,ferhatosmanoglu2001approximate}. An index restricts the scan to a subset of $\mathcal{X}$, and indexes are compared by the quality/throughput trade-off they offer~\cite{ann_bench}. Many formulations of ANN exist, distance-bounded and error-bounded among them. The most widely adopted family of ANN techniques, which this paper studies, offers no guarantee at all. Its results are evaluated empirically, by the fraction of the exact top-$k$ they recover at a given cost (Section~\ref{sec:measures}).

\subsubsection{Indexes for Dense Retrieval}
\label{sec:related_ann_indexes}
\label{sec:related_ivf}

The index families below were developed for dense vectors. The classical ones divide the search space and avoid exhaustive distance computations by pruning irrelevant regions: hierarchical partitioning trees~\cite{bentley1975multidimensional,vp_tree,ball_tree,m_tree,flann} and Locality-Sensitive Hashing (LSH) families~\cite{indyk1998approximate,min_hash,angular,datar2004lsh}. Both struggle with the curse of dimensionality. They degrade towards a linear scan, or they need large memory overheads to hold accuracy.

\paragraph{Inverted Vector File.} One partition-based structure nonetheless remains in wide use, at scale and on accelerators~\cite{ivf_gpu}. An Inverted Vector File (IVF) index partitions the collection into $\text{nlist}$ clusters, the Voronoi cells~\cite{delaunay} of a set of k-means centroids, and stores each cluster contiguously. A query is compared against the centroids, and only the $\text{nprobe}$ most promising clusters are scanned. This is the geometric analogue of scanning the posting lists of a query's terms in an inverted index~\cite{inv_index}. IVF is attractive for operational as well as algorithmic reasons: construction is one k-means, and a cluster scan is a dense computation over contiguous memory. Two refinements are common. First, the centroid set can itself be indexed, by a graph or by a second level of clustering, once routing to it becomes expensive. Second, the vectors of a cluster can be stored as centroid-relative \emph{residuals} $r = v - c$, which a lossy quantizer reconstructs more accurately. This is the IVFADC scheme introduced with Product Quantization~\cite{pq}. ScaNN~\cite{scann1} trains the quantizer on the error of the score it estimates, where other quantizers use the reconstruction error. SOAR~\cite{scann2} assigns a vector to several clusters.

\paragraph{Proximity graphs.} Elsewhere the field has shifted towards proximity graph indexes, sometimes using a tree to initialize the traversal~\cite{sptag,elpis,tree_graph_survey}. Empirical benchmarks place them at the top of the accuracy/throughput trade-off for in-memory search~\cite{wang2021comprehensive,survey_graph_2025,ann_bench}. These indexes represent vectors as nodes and navigate towards a query by best-first traversal. They combine short-range edges for local resolution with long-range edges that let a traversal cross the space. Hierarchical structures such as \hnsw~\cite{nsw,hnsw} and optimized single-layer graphs such as NSG~\cite{fu2017fast,rng} and \vamana~\cite{vamana} follow this design. Exact $k$-NN graph construction requires a prohibitive quadratic number of distance computations, so these indexes build their graphs approximately, from cheap initializations such as NN-Descent~\cite{nndescent}. Modern indexes prune neighbors to reduce redundancy and promote long-range edges~\cite{pbknng,nssg,hnsw,vamana}. They thus favor diversity of direction over proximity alone, and they bound node degree to balance traversal speed against the memory footprint of the adjacency lists.

One of the most widely adopted members of the family is the Hierarchical Navigable Small World (\hnsw) graph~\cite{hnsw}. It organizes the collection into a hierarchy of proximity graphs. The lowest level, the \emph{ground level}, holds every vector. Each level above it holds a decreasing random sample of the level below and provides long-range links that shorten navigation. A query descends greedily through the upper levels to locate its neighborhood cheaply. Beam search over the ground level then refines the result.

Three parameters govern the trade-off. The degree bound $M$ limits how many outgoing edges a node keeps per level ($2M$ on the ground level). The beam width used to find a new node's neighbors at insertion is $ef_\text{construction}$, and the beam width at query time is $ef_\text{search}$. Neighbor selection uses a robust pruning heuristic that approximates a relative-neighborhood graph. An edge is kept only if no already-selected neighbor is closer to its endpoint than the endpoint is to the node. This diversifies a node's neighborhood by direction and not by distance alone.

\hnsw's construction is incremental: nodes are inserted one at a time, each running the ordinary search to find its neighbors in the graph.

The standard beam search of \hnsw halts once the best remaining frontier candidate is worse than the worst element of the current results. Two families of replacement criteria exist. \emph{Relaxation-based} criteria widen that comparison so the search continues a little past it. Adaptive Beam Search is one of them, with a provable bound on navigable graphs~\cite{ee_distance_adaptive}. \emph{Saturation-based} criteria such as Patience-in-Proximity~\cite{pip} halt once the result set stops changing. A third line of work learns the decision from the state of the search~\cite{ee_learned}.

Today, graph indexes still represent the state of the art in dense retrieval. However, they rely on scattered memory accesses and do little computation, so they are usually memory-bound. To alleviate this problem, a large part of the literature has moved towards two-stage pipelines~\cite{lvq,leanvec,pyglass,rabitq}. A set of candidates is first generated over quantized vectors and then refined against uncompressed or higher-precision representations. The graph is traversed quickly with the cheap representations, and the larger, more expensive ones are read only for a small subset of candidates. Another line of research goes further and stores heavily compressed representations of the neighbors directly inside the adjacency lists. This replication gives locality in the graph and trades memory for search speed. The two main examples of this line of research are NGT-QG~\cite{ngt1,ngt2} by Yahoo Japan and \symphonyqg~\cite{symphonyqg}.

\subsubsection{Indexes for Learned Sparse Retrieval}
\label{sec:related_sparse}

There are two main families of indexes for learned sparse embeddings: one inherited from lexical retrieval, and one borrowed from the dense techniques above.

The first is the \emph{inverted index}. A learned sparse vector is a vector over the vocabulary, so it can be indexed by the machinery that BM25 and its predecessors built for term-weighted documents (Section~\ref{sec:related_representations}). The index holds one inverted list per component, and a query touches only the lists of its own nonzero components. This structure allows pruning at query time without any geometric index over the collection. Block-Max Pruning~\cite{bmp} and Dynamic Superblock Pruning~\cite{dsbp} bound the contribution of a block of postings and skip the block when the bound cannot enter the result set. Lightweight Superblock Pruning (\lsp)~\cite{carlson2026lsp} is a lighter-weight variant of the latter. \seismic~\cite{seismic,seismic_journal} partitions each inverted list into geometrically cohesive blocks with summary vectors, and evaluates only the blocks whose summary is promising. \sindi~\cite{li2025sindi} is a recent alternative in the same setting.

The other family is taken from the dense world. A proximity graph needs only a similarity function that it can evaluate, so the indexes of Section~\ref{sec:related_ann_indexes} apply to learned sparse embeddings without changes. \pyann~\cite{pyann}, an approach based on \hnsw, won the Sparse Track of the NeurIPS~2023 Big ANN challenge~\cite{annneurips23}.

\seismicw combines the two worlds: it augments \seismic with a $\kappa$-NN graph over documents~\cite{seismic_knn} to expand the set of results and refine the search.

\subsection{Adjacency List Compression}
\label{sec:related_int_compression}

In graph indexes the vectors account for only part of the footprint. Adjacency lists carry a significant memory overhead of their own, especially when the vectors are low-dimensional or heavily quantized~\cite{ottaviano_ann_compression}. Since the order of a list does not matter, it can be sorted and stored as the \emph{gaps} between consecutive identifiers~\cite{gap_encoding}, which a variable-length integer code stores in few bytes. Combined with graph reordering~\cite{laxman2016compression,egb}, \svb~\cite{streamvbyte} compresses the lists by up to $2.3\times$ with a query time equal to or lower than that of the uncompressed graph~\cite{benocci2026lossless,ann_graph_permutation}.

\subsection{Libraries and Systems}
\label{sec:related_libraries}

The families above are composable: any index structure can be paired with any quantizer, over any of the three representations of Section~\ref{sec:related_representations}. Libraries differ in how much of this cross-product they cover.

\faiss~\cite{ivf_gpu,douze2024faiss} is the most widely used and most extensively tested of them for dense data. It factors the space through a family of loosely related classes, each with its own construction and search code path: \texttt{IndexIVFFlat} and \texttt{IndexIVFPQ}, the scalar-quantizer and \texttt{ByResidual} variants of both, and their counterparts with a graph quantizer. Other libraries narrow the problem: each covers one algorithm thoroughly and leaves the cross-product aside. \hnswlib~\cite{hnswlib}, maintained by \hnsw's authors under the \texttt{nmslib} organization~\cite{nmslib}, implements one algorithm over uncompressed dense vectors. DiskANN~\cite{diskann} implements \vamana~\cite{vamana} with a focus on out-of-core search. \parlayann~\cite{parlayann} provides deterministic parallel builds for several graph algorithms. \annlib~\cite{shen2026annlibdevelopmentframeworkefficient} builds on the same primitives and separates the graph algorithm from the data structure that stores the graph. It composes filtered search, deletions and snapshots as modules over them, for dense vectors under a point type and distance function supplied by the user. Finally, the entries that populate public benchmarks~\cite{n2,pyann} are typically specialized to one algorithm and one representation.

Vector databases and search engines form a different layer. Milvus~\cite{milvus}, Qdrant~\cite{qdrant}, Vespa~\cite{vespa} and Elasticsearch~\cite{elasticsearch} serve dense and learned sparse embeddings behind one interface, and Weaviate~\cite{weaviate} serves dense embeddings. Qdrant, Vespa, Elasticsearch and Weaviate also score late-interaction token vectors with MaxSim, mostly as a reranking stage. They are services built for deployment, with persistence, filtering and distribution, and each representation is served by a separate index type. The libraries above, and ours, form the layer where indexes are designed, composed and compared.

\kannolo{} was introduced at ECIR~2025~\cite{kannolo} as a modular Rust library offering \hnsw{} over both dense and sparse representations. This paper presents the system after that first release, with large structural differences. The data-management layer is now a separate library, \vectorium{} (Section~\ref{sec:data_management}). It owns representations, distance kernels, encoders, storage, clustering and the index contract. \kannolo{} (Section~\ref{sec:indexing}) implements \hnsw, IVF and the two-stage \rerankindex{} generically over the representations and encodings that \vectorium{} supplies. Beyond the ECIR version, this paper therefore adds the layer separation and its cost model, the encoder family and its quantization schemes, the multi-vector representation, the IVF family, and the graph backend abstraction with its optimizations and compressed adjacency layouts. It also adds the measurements that motivate each choice. Three components have been published separately and are summarized here: the \rerankindex{} with its two pruning rules~\cite{martinico2026multivector}, generalized here to any pair of indexes and datasets; the \dotpacking{} encoding and the \jumpdot{} kernel~\cite{nardini2026compactefficientindexeslearned}, which \vectorium{} integrates; and the graph compression scheme of~\cite{benocci2026lossless}, developed in \kannolo.

\section{Preliminaries and Experimental Setup}
\label{sec:setup}

This section fixes the notation, the measures, the collections and the measurement conventions used by Sections~\ref{sec:data_management} and~\ref{sec:indexing}. These conventions apply throughout those two sections unless a study states otherwise.

\subsection{Notation}
\label{sec:problem}

We keep the notation of Section~\ref{sec:related_ann}: a collection $\mathcal{X}$ holds $n$ vectors of dimension $d$, each identified by its position $i$, and $q$ is a query. For a proximity graph, $N$ is its number of nodes ($N=n$ at the ground level), $M$ its maximum degree and $ef_\text{search}$ the beam width of its search. $m$ is the number of subspaces of a Product Quantizer, $\lambda$ the early-termination relaxation factor, and $\beta$ the early-exit patience of the reranking stage.

\subsection{Evaluation Measures}
\label{sec:measures}

\paragraph{Accuracy@$k$.} The percentage of the exact top-$k$ that a search recovers over a set of queries $\mathcal{Q}$,
\[
  \text{Accuracy@}k \;\coloneq\; 100 \cdot \frac{1}{|\mathcal{Q}|} \sum_{q \in \mathcal{Q}} \frac{|\,\mathcal{R}_k^q \cap \mathcal{E}_k^q\,|}{k},
\]
where $\mathcal{R}_k^q$ is the set the index returns for query $q$ and $\mathcal{E}_k^q$ is the exact top-$k$ set.

\paragraph{Throughput.} Queries per second (QPS), measured in wall-clock time over the whole query set. Unless a study says otherwise, queries are issued one at a time, and the reported figure is the reciprocal of the mean per-query latency.

\paragraph{Footprint.} Bytes per vector, as computed by the \texttt{SpaceUsage} accounting of \vectorium. This accounting traverses a structure and sums the bytes it owns. When owned bytes cannot be compared with those of an external system, we report peak resident memory instead and state this in the text.

\paragraph{Build time.} Wall-clock seconds for the complete construction, excluding serialization on both sides of any comparison.

\subsection{Development Collections}
\label{sec:collections}

Table~\ref{tab:collections} lists the collections used for the design studies of Sections~\ref{sec:data_management} and~\ref{sec:indexing}. They span two representations, three dense dimensionalities and two metrics, so each design decision is tested in the regimes where it can behave differently.

\siftom~\cite{sift} contains hand-crafted, integer-valued features compared under squared-Euclidean distance. Its subsampled billion-scale counterpart \siftob{} provides the $10$k--$100$M range of sizes used in the scaling study of Section~\ref{sec:visited_set}. The three remaining collections are all built on the passage corpus of \msmarcovo~\cite{msmarco} and differ only in the encoder used to embed it: \dragon~\cite{dragon} and \snowf~\cite{snowflake} are dense encoders, and \splade\,\cocond~\cite{splade,cocondenser} is a sparse encoder. After its first mention, each collection is referred to by the name of its encoder alone.

\begin{table}[t]
\centering
\small
\setlength{\tabcolsep}{4pt}
\caption{Development collections; $\ell_2$/IP denotes squared-Euclidean / inner product. $^{\star}$ the average number of non-zero components is $\approx 120$.}
\label{tab:collections}
\begin{tabular}{@{}lcrcrr@{}}
\toprule
\textbf{Collection} & \textbf{Type} & $d$ & \textbf{Metric} & $n$ & \textbf{Queries} \\
\midrule
\siftom      & dense  & $128$      & $\ell_2$ & $1$M   & $10\,000$ \\
\siftob      & dense  & $128$      & $\ell_2$ & $1$B & --- \\
\dragon      & dense  & $768$      & IP       & $8.8$M & $6\,980$ \\
\snowf       & dense  & $1\,024$     & IP       & $8.8$M & $6\,980$ \\
\cocond      & sparse & $^{\star}30\,522$ & IP & $8.8$M & $6\,980$ \\
\bottomrule
\end{tabular}
\end{table}

\subsection{Hardware and Software}
\label{sec:hardware}

All measurements run on a dual-socket Intel Xeon Silver 4314 (Ice Lake-SP, $2.40$\,GHz) with $16$ cores and $32$ hardware threads per socket, for $64$ threads in total. Each socket has $24$\,MiB of shared L3 cache. The machine has two NUMA nodes and $503$\,GiB of memory, and runs Ubuntu $24.04$ LTS. The processor supports the \texttt{AVX-512} instruction-set extensions, including \texttt{AVX-512VPOPCNTDQ}. The portability measurements of Appendix~\ref{app:kernels} also run on this machine with a build restricted to \texttt{AVX2}. They also run on a Raspberry Pi~5 (Cortex-A76, Armv8.2 with $128$-bit NEON, $16$\,GiB), built from the same source with the same nightly toolchain.

\vectorium{} and \kannolo{} are built with the Rust nightly toolchain of 2026-06-19 (\texttt{rustc}~1.98.0-nightly), which is required for the portable-SIMD and algebraic floating-point intrinsics of Section~\ref{sec:distance_computation}. They are compiled in release mode with \texttt{target-cpu=native}. \faiss{} appears as a baseline in two versions: $1.15.0$ for the Product-Quantization comparison of Section~\ref{sec:product_quantization}, and $1.14.3$ for the IVF comparison of Section~\ref{sec:ivf_impl}. \hnswlib{} appears at release $0.9.0$ in the validation of Section~\ref{sec:hnsw_validation}. The end-to-end comparison of Section~\ref{sec:exp} uses release $0.8.0$, the one pinned by \vibe.

\subsection{Measurement Conventions}
\label{sec:methodology}

\paragraph{Parallel scans.} The exhaustive scans of Section~\ref{sec:data_management} run on all $64$ threads with \texttt{numactl -{}-interleave=all}, on a machine verified to be idle. When an external system is compared, both sides run under the same settings.

\paragraph{Micro-benchmarks.} Micro-benchmarks run single-threaded, pinned to one core. We report the minimum over five repetitions. Hardware performance counters (retired instructions, cycles, last-level-cache misses) are restricted to the measured loop, so instruction counts and IPC measure the kernel alone and exclude the harness.

\paragraph{Search timing.} We time one query at a time. The reference implementations we compare against parallelize across queries without shared state between them, and so do ours. A batched figure would therefore measure thread scaling and not the index itself. Every index class of \kannolo's Python interface exposes this as a batch search with a configurable number of threads.

\paragraph{Ground truth.} The exact top-$k$ under the collection's own metric, computed in \texttt{f32}. It is either shipped with the collection or computed once and reused across every configuration.

\section{\vectorium}
\label{sec:data_management}

\vectorium is the data management layer beneath every index in the system. It owns how a vector is represented in memory, how two vectors are scored under a given metric, how a collection is compressed, and how that collection is stored and accessed. One set of storage and scoring primitives is thus shared across systems and index families, and no index family reimplements it. \vectorium{} also defines the \texttt{Index} contract that the indexes above it implement, and ships the exact flat index; Section~\ref{sec:system_overview} describes both together with \kannolo's indexes.

\subsection{Overview and Design Principles}
\label{sec:data_overview}

\begin{figure*}[t]
    \centering
    \includegraphics[width=\textwidth]{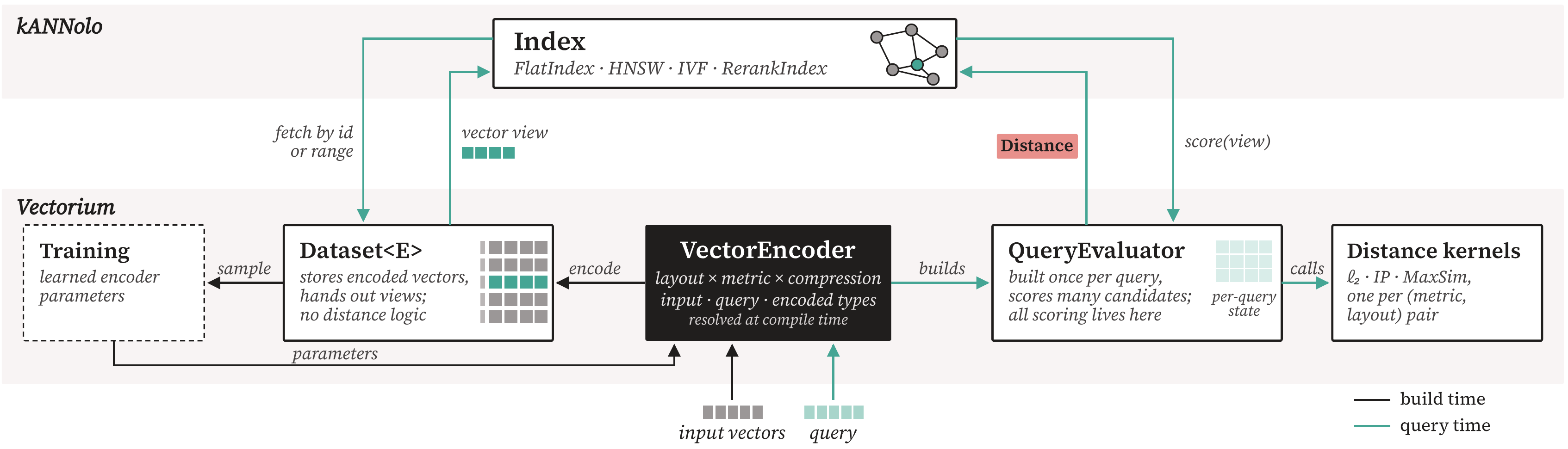}
    \caption{The layered architecture of \vectorium{} and its interface to the indexes of \kannolo.}
    \label{fig:vectorium_architecture}
\end{figure*}

\vectorium is built from four core abstractions, each layered on top of the previous one. A \emph{vector view} is data in flight. It is a lightweight, zero-copy view over a vector's storage and describes how a given vector type is laid out in memory (Section~\ref{sec:vector_representations}). A \texttt{Distance} is a score together with its ordering. It keeps all generic frontier and heap code independent of the metric (Section~\ref{sec:distance_computation}). A \texttt{VectorEncoder}, paired with its \texttt{QueryEvaluator}, unifies \emph{compression} and \emph{scoring}. It maps input vectors to a stored encoding and defines how a query is scored against it, so all distance logic is kept in one place (Section~\ref{sec:encoders_quantization}). A \texttt{Dataset} is storage, parameterized by an encoder. It holds encoded vectors and hands out views of them (Section~\ref{sec:datasets}). Figure~\ref{fig:vectorium_architecture} shows how the four are stacked and how the three orthogonal axes of layout, metric and compression cut across them.

Three principles govern their design. First, encoders and datasets are \emph{type parameters}, so every combination of layout, metric and compression is a distinct monomorphized type. Distance kernels inline into the caller, no dynamic dispatch appears on the hot path, and a single generic implementation, such as the \hnsw{} index of Section~\ref{sec:hnsw_impl}, serves every encoding. Second, storage layout (dense, sparse, packed, multi-vector), metric (squared-Euclidean or inner product) and compression scheme vary independently, and any meaningful combination is obtained by naming its type, with \emph{no new code}. Third, storage is decoupled from scoring. A \texttt{Dataset} holds no distance logic, and \emph{all} scoring is done by the encoder's evaluator, so one dataset and one index search loop serve every compression scheme.

\paragraph{The cost of genericity.} Monomorphization moves the cost of the factoring from run time to build time. The Python bindings of \kannolo~0.15.1 instantiate every combination they expose. A clean release build of these bindings takes $113$ s on our machine, $83$ s of it for \kannolo{} itself, and produces a $5.3$\,MiB library. The beam search of Section~\ref{sec:graph_search_opts} is compiled $164$ times, once per combination of graph layout, encoder and metric. These copies make up $3.1\%$ of the generated code. Each copy inlines its distance kernel into the search loop, so scoring makes no indirect call. Calling the kernel through a trait object would make each distance $1.07$--$1.52\times$ slower on cache-resident vectors and up to $1.07\times$ slower under random access (harness of Table~\ref{tab:batch6}). A Rust program compiles only the combinations it uses. The Python bindings expose a fixed set of index classes, so their number of copies grows with the product of the graph layouts, encoders and metrics they cover. A combination outside that set needs a new binding class and a rebuild.

Table~\ref{tab:encoder_capabilities} summarizes the concrete encoders that \vectorium\ ships and the position of each on the three axes. Its \emph{bits/dim} column also maps the compression spectrum. The range goes from the lossless \texttt{f32} cast at $32$ bits/dim down to one bit per dimension for the binary code of \rabitq, and below one bit for a sufficiently short PQ code.

\begin{table*}[t]
\centering
\small
\caption{Capability matrix of \vectorium's concrete encoders. $d$ is the vector dimension, $m$ the number of PQ subspaces (one \texttt{u8} code each), $b=\texttt{total\_bits}\in\{2,4,8\}$ the \rabitqext{} code width.}
\label{tab:encoder_capabilities}
\begin{tabular}{@{}llllr@{}}
\toprule
\textbf{Encoder} & \textbf{Layout} & \textbf{Stored type} & \textbf{Metric} & \textbf{Bits/dim} \\
\midrule
\texttt{PlainDenseQuantizer}          & dense        & \texttt{f32}          & $\ell_2$/IP & 32 \\
\texttt{ScalarDenseQuantizer}         & dense        & \texttt{f16}, \texttt{bf16}, $16$/$8$-bit fixed point & $\ell_2$/IP & 16/8 \\
\texttt{ProductQuantizer}    & dense        & \texttt{u8} codes     & $\ell_2$/IP & $8m/d$ \\
\texttt{RabitqQuantizer}     & dense        & \texttt{u64} sign code & $\ell_2$/IP & $1+64/d$ \\
\texttt{RabitqExtQuantizer}  & dense        & \texttt{u64} $b$-bit fields & $\ell_2$/IP & $b+64/d$ \\
\texttt{ScalarSparseQuantizer}        & sparse       & \texttt{f32}, \texttt{f16}, \texttt{bf16}, $16$/$8$-bit fixed point & $\ell_2$/IP & 32/16/8 \\
\texttt{BlockedSparseEncoder}       & packed sparse & \texttt{u16}/\texttt{f16} blocks & IP & 16 \\
\texttt{DotPacking8Encoder}     & packed sparse & packed gaps; fixed-point, \texttt{f16} or scalar values & IP          & var. \\
\texttt{ScalarMultiVecQuantizer}      & multi-vector & \texttt{f32}, \texttt{f16}, \texttt{bf16}, $16$/$8$-bit fixed point & MaxSim  & 32/16/8 \\
\texttt{MultiVecProductQuantizer}          & multi-vector & \texttt{u8} codes     & MaxSim      & $8m/d$ \\
\texttt{MultiVecTwoLevelProductQuantizer}  & multi-vector & centroid + \texttt{u8} & MaxSim     & $(8m{+}32)/d$ \\
\bottomrule
\end{tabular}
\end{table*}

\subsection{Vector Representations}
\label{sec:vector_representations}

The lowest layer of \vectorium\ fixes how a vector is held in memory while it is encoded, stored or scored, independently of any encoder or metric. Every representation is a zero-copy view over its backing storage, cheap enough to pass by value in the innermost distance loops. There are four. A \emph{dense} view covers one contiguous, fixed-dimension run of values. A \emph{sparse} view covers two parallel runs: the component indices and their nonzero values. A \emph{packed} view covers an opaque run of words that only the encoder that produced it can interpret, such as the bit-packed blob of the \dotpacking{} encoder (Section~\ref{sec:dotpacking}). A \emph{multi-vector} view covers a flat buffer of token vectors that share one dimension. Each view is generic over its value type, so one encoder body covers \texttt{f32}, \texttt{f16}, \texttt{bf16} and fixed-point values alike.

An encoder (Section~\ref{sec:encoder_abstraction}) names three representations separately: the \emph{input} pushed in at build time, typically plain \texttt{f32}; the \emph{query}; and the \emph{encoded} vector stored after compression. Because the three are distinct, a query can be scored against a compressed stored vector without either side matching the other's representation. An asymmetric scoring scheme is thus expressed in the type system and does not depend on a convention.

\subsection{Distance Computation}
\label{sec:distance_computation}

The next layer turns two vector representations into a score. A single \texttt{Distance} trait fixes what a score \emph{is} and how scores are ordered. A flat family of kernels computes the scores, each operating directly on the slices behind a view (Section~\ref{sec:vector_representations}). The encoders' evaluators invoke these kernels (Section~\ref{sec:encoder_abstraction}).

A distance is a thin wrapper around a single \texttt{f32}. The \texttt{Distance} contract only requires it to be totally ordered, and \texttt{SquaredEuclideanDistance} and \texttt{DotProduct} implement it. \texttt{DotProduct} \emph{reverses} its ordering, because a larger inner product denotes a better match, while for squared-Euclidean distance a smaller value does. ``Smaller is better'' therefore holds under either metric. Every generic consumer above this layer, including the candidate heaps, is written once and never branches on the metric in use.

\paragraph{A unified relaxation predicate.} Early termination and pruning ask whether a candidate is within a \emph{relaxed} bound, whose direction depends on the metric. The \texttt{Distance} trait hides the direction behind a single predicate. A minimize metric accepts a candidate when $d \le \text{threshold} + |\text{threshold}|\cdot\lambda$, a maximize metric when $d \ge \text{threshold} - |\text{threshold}|\cdot\lambda$, and $\lambda=0$ recovers the exact comparison. The threshold enters in \emph{absolute value} because an estimated distance can be negative, as the unbiased estimator of \rabitq{} returns for near-duplicate pairs. On such a bound, the multiplicative form $\text{threshold}\cdot(1+\lambda)$ would tighten the window when it should widen it.

\paragraph{The kernel family.} The scoring work is done by a small set of free functions, one per (metric, layout) pair that occurs. Each comes in a checked form that validates its invariants and an unchecked form that assumes them. Sparse data is scored against a \emph{densified} query: the evaluator expands the sparse query once into a dense buffer, into which each candidate's stored components index directly. Sparse-versus-sparse comparisons, needed only during construction, use a sorted merge. The sparse layouts compute squared-Euclidean distance as $\lVert q\rVert^2 + \lVert v\rVert^2 - 2\langle q,v\rangle$, so they visit only the \emph{intersection} of the active components of the two operands and not their union. The same layer hosts the \texttt{maxsim} kernel of the multi-vector encoders (Section~\ref{sec:multivec_encoders}), which decodes each document token to \texttt{f32} once and reuses it across all query tokens.

\paragraph{Portable auto-vectorization.} Explicit SIMD in \vectorium{} is confined to the packed encodings, described at the end of this paragraph. The kernel family, which scores every other encoding, contains none. Hand-written vector code would have to be duplicated for each instruction set and again for every (metric, layout) pair, and each copy would fix the register width and the reduction schedule in the source. Each kernel is a few lines of plain Rust. The compiler, with native CPU optimizations (Section~\ref{sec:hardware}), lowers it to the vector ISA of the target. Two choices make this possible. First, the reductions use Rust's \emph{algebraic} (fast-math) floating-point operators, which let the compiler reassociate and contract the multiply--add chains. A strictly ordered IEEE reduction fixes the sum to one scalar schedule, which can be neither widened into SIMD lanes nor fused. This relaxation applies only to scoring: the ordering and relaxation predicates above compare exactly. Second, the unchecked kernels present loops of known, equal trip count, with no bounds checks that would prevent widening. The packed encodings are scored differently. \rabitq{} (Section~\ref{sec:rabitq}) scores by popcounts and shift-and-mask unpacking, and \dotpacking{} (Section~\ref{sec:dotpacking}) decodes bit-packed gaps. These kernels use Rust's \emph{portable} SIMD, which fixes the lane width but is still independent of the ISA. Only the packed sparse encodings (Sections~\ref{sec:blocked_sparse} and~\ref{sec:dotpacking}) call per-ISA intrinsics (AVX2, and NEON on Armv8).

Without fast-math the kernels do not vectorize at all. The strict variant lowers to a scalar reduction over a single accumulator, and the algebraic one to packed fused multiply--adds over several independent accumulators. The algebraic \texttt{f32} dot-product and squared-Euclidean kernels are $13\times$ faster at $d{=}128$ and $18$--$20\times$ faster at higher dimensions, and exhaustive top-$10$ results stay identical on \sift{} and \dragon. The same source vectorizes unchanged when rebuilt for \texttt{AVX2} alone and on Armv8 with NEON (Appendix~\ref{app:kernels}). Expert-tuned intrinsics may still outperform compiled code; Section~\ref{sec:product_quantization} measures this difference against \faiss.

\paragraph{Batched evaluation.} Every kernel that can exploit batching also comes in a \emph{batch-of-six} form, exposed on the \texttt{QueryEvaluator} contract (Section~\ref{sec:encoder_abstraction}). It scores one query against six vectors in a single fused pass. The gain comes from \emph{query reuse}: each query element is loaded once and used six times. On cache-resident \texttt{f32} vectors the batched kernel is $1.32$--$1.73\times$ faster than six single-candidate calls. Under random access, as among graph neighbors, it is $1.24$--$1.59\times$ faster for \texttt{f32} and \texttt{f16} (Appendix~\ref{app:kernels}). The contract provides the batched call as a \emph{default} fan-out, and an encoder overrides it only where a single-pass kernel is faster.

\subsection{Encoders and Quantization}
\label{sec:encoders_quantization}

A \texttt{VectorEncoder} binds a compression scheme to a layout and a metric. It defines how an input vector is mapped to its stored encoding and how a query is scored against that encoding through the kernels of the previous layer. It is the single unit in which storage, compression and scoring meet. The subsubsections that follow state the contract and then present the concrete encoders along the compression axis, from lossless to increasingly lossy.

\subsubsection{The Encoder and Evaluator Abstraction}
\label{sec:encoder_abstraction}

The \texttt{VectorEncoder} is the central contract of \vectorium. An encoder supplies everything that a dataset or index needs about an encoding, so the dataset or index never needs to know which encoding it is. This comprises its distance (Section~\ref{sec:distance_computation}), its three representations (Section~\ref{sec:vector_representations}), the evaluator that computes scores, and the query-side parameters used to build that evaluator. All of them are resolved at compile time.

\paragraph{The evaluator.} All scoring goes through a \texttt{QueryEvaluator}. It is created once from a query and then scores many candidates, so the once-per-query work (e.g. building PQ's distance tables) is amortized over all the distances it computes. Evaluators are constructed in two ways: from a user query at search time, and from a stored dataset vector at build time. In the second case the encoded vector is first decoded back to \texttt{f32}. One scoring machinery thus serves both search and index construction.

\paragraph{Query-side parameters.} An evaluator may take a parameter that belongs to the \emph{query} and not to the stored data. For example, for the $1$-bit \rabitq{} encoder (Section~\ref{sec:rabitq}) it sets how finely the query is quantized. Because this parameter is not part of the stored encoding, one stored dataset can serve every width concurrently from parallel threads.

\paragraph{Encoding, decoding and stored-vector pairs.} Encoding appends directly into the dataset's own backing storage, with no intermediate allocation. Decoding reconstructs an \texttt{f32} vector for the build-time workflows that need the original space. Graph construction (Section~\ref{sec:indexing}) also scores pairs of stored vectors against each other many times. The contract therefore includes a pairwise method, which encoders override to score their stored form directly, without decoding either operand.

\subsubsection{Scalar Quantization}
\label{sec:scalar_quantization}

Scalar quantization (Section~\ref{sec:related_quantization}) populates the compression axis by changing the numeric type in which values are stored. \texttt{ScalarDenseQuantizer} is a component-wise cast from an input value type to a stored value type, generic over the metric. \texttt{ScalarSparseQuantizer} is the same cast over the sparse layout. Neither has a codebook or a learned range. When the two value types coincide, the cast is the identity and the quantizer is lossless; the \texttt{PlainDenseQuantizer} and \texttt{PlainSparseQuantizer} aliases name this case. When the stored type is narrower, the same code path yields a reduced-precision encoding.

The stored type may be any value type that can be converted from \texttt{f32}: \texttt{f32} itself, the half-precision \texttt{f16} and \texttt{bf16}, or the fixed-point \texttt{FixedU8Q} and \texttt{FixedU16Q}. Whatever the stored precision, the query is kept in \texttt{f32}.

\texttt{f16} is the default storage precision throughout the system, on the assumption that it costs no measurable accuracy. We measure accuracy, footprint and exhaustive search throughput as a function of stored precision. To do so, we run a full top-$10$ scan against the exact ground truth on three collections with three value distributions: \siftom (integers), \dragon (real numbers) and \cocond (real, positive numbers).

Table~\ref{tab:bsq1} supports this choice. \texttt{f16} ranks like \texttt{f32}: Accuracy@$10$ is unchanged on \siftom{} and drops by $0.06$ points on \dragon{} and $0.08$ on the sparse set. These drops are far below any difference that an approximate index later introduces. \texttt{f16} also \emph{halves} the dense footprint and, because the scan is memory-bandwidth-bound, roughly \emph{doubles} throughput. \texttt{bf16} loses an order of magnitude more accuracy than \texttt{f16} on \dragon{} and \cocond. Its $7$ mantissa bits, against $10$ for \texttt{f16}, resolve values more coarsely. On \siftom{} the two tie, because the small-integer \sift{} values are representable on the coarser grid of \texttt{bf16}. Finally, the sparse row shows the asymmetry of a compression that touches only the values. It removes only about a third of the per-nonzero payload, because each entry still carries an uncompressed \texttt{u16} component index.

\begin{table}[t]
\centering
\small
\caption{Precision against accuracy and footprint. Exhaustive \texttt{FlatIndex} top-$10$ search against the exact \texttt{f32} ground truth; throughput parallel across queries on $64$ threads (Section~\ref{sec:methodology}).}
\label{tab:bsq1}
\setlength{\tabcolsep}{4pt}
\begin{tabular}{@{}llrrr@{}}
\toprule
\textbf{Dataset} & \textbf{Prec.} & \textbf{B/vec} & \textbf{Accuracy@10 (\%)} & \textbf{QPS} \\
\midrule
\siftom              & \texttt{f32}  &  512 &  100.00 &  819 \\
($128$-d, $\ell_2$)  & \texttt{f16}  &  256 &  100.00 & 1\,401 \\
                     & \texttt{bf16} &  256 &  100.00 & 1\,109 \\
\midrule
\dragon               & \texttt{f32}  & 3\,072 & 100.00 &   10 \\
($768$-d, IP)        & \texttt{f16}  & 1\,536 &  99.94 &   22 \\
                     & \texttt{bf16} & 1\,536 &  99.59 &   20 \\
\midrule
\cocond     & \texttt{f32}  &  728 & 100.00 &   24 \\
(sparse, IP)         & \texttt{f16}  &  488 &  99.92 &   26 \\
                     & \texttt{bf16} &  488 &  99.23 &   25 \\
\bottomrule
\end{tabular}
\end{table}

\subsubsection{Product Quantization}
\label{sec:product_quantization}

Product Quantization (Section~\ref{sec:related_quantization}) is the first encoder in this section with a \emph{learned codebook}. \texttt{ProductQuantizer} implements it with $\text{KSUB}=256$ centroids per subspace, so a code is one \texttt{u8} and a sub-table takes $1$\,KiB. Narrower codes would have to be bit-unpacked on every lookup. Wider ones would inflate both the codebook and the per-query table with no clear gain. The encoded form is therefore $m$ bytes, and the footprint is $8m/d$ bits per dimension. The subspace count is a compile-time parameter, so the code length is part of the encoder's type. The per-vector stride of a PQ dataset is therefore a compile-time constant. The divisibility invariants that the kernels rely on ($m \bmod 4 = 0$ and $d \bmod m = 0$) are asserted once at construction and assumed in all later code.

The $m$ codebooks are $m$ independent clustering problems, each a k-means over $\text{dsub}$-dimensional projections. Training therefore parallelizes across subspaces with no synchronization, as one task per subspace over the k-means primitive of Section~\ref{sec:data_infrastructure}.

The evaluator builds the $m\times256$ lookup table once per query. A small \texttt{ProductQuantizerDistance} trait, implemented once per metric, selects the table builder at compile time, so both metrics share one scoring path.
Building the table costs $m \cdot 256 \cdot \text{dsub} = 256d$ multiply--adds, roughly the cost of $256$ uncompressed full-dimension distances. Each subsequent candidate costs $m$ loads and $m-1$ adds. A PQ evaluator therefore recovers the cost of its table after a few hundred candidates, which is the regime in which index search uses it.

\paragraph{Centroid layout.} K-means naturally produces centroids as an \emph{array of structures} (AoS), with the $\text{dsub}$ values of centroid $i$ contiguous. With that layout, the table build needs one \emph{horizontal reduction per table entry}. At a typical $\text{dsub}$ of $8$ or $16$, this final horizontal sum costs more than the multiply--adds before it, and it is repeated $256$ times per subspace. \vectorium\ stores the codebooks transposed, as a flat \emph{structure of arrays} (SoA) of shape $[m \times \text{dsub} \times \text{KSUB}]$. At dimension $k$, the values of consecutive centroids are then contiguous and load in one vector instruction. The loop nest is inverted, with dimensions in the outer loop and centroids in the inner loop. The accumulation becomes \emph{vertical}: partial distances accumulate in parallel lanes, with no horizontal reduction anywhere. The kernel tiles the centroids into groups of $T{=}32$ whose accumulators stay in registers across all dimension passes, so the table is written once, at the end. This is the same query reuse as in the batched kernels: here a query element is reused across $T$ centroid lanes, there across six candidates. As with the kernels of Section~\ref{sec:distance_computation}, the loop body is plain Rust and the compiler performs the vectorization. Without tiling, each of the $\text{dsub}$ passes would read and write the running partial of every centroid, ${\approx}2\cdot\text{dsub}\cdot\text{KSUB}$ table accesses per subspace, against $\text{KSUB}$ writes with tiling.

Timed alone on the three deployed shapes (Appendix~\ref{app:pq_table}), the shipped kernel builds the table $4.5$--$5.2\times$ faster than the natural AoS build and $3.9$--$4.4\times$ faster than an AoS build with fast-math. Almost all of the gain comes from the layout.

\paragraph{Scoring.} A candidate costs $m$ table lookups, one per code byte, which are summed. Their addresses depend on the data, so they cannot be vectorized. Scoring consumes the codes four at a time into four independent accumulators. PQ keeps the default batched call of Section~\ref{sec:distance_computation}. The query has already been consumed into the table, so there is nothing to reuse, and six codes index a $1$\,KiB sub-table spanning sixteen cache lines at six unrelated positions.

\paragraph{Validating against a reference implementation.} We check our PQ encoder end to end against \faiss's reference implementation, \texttt{IndexPQ}. Both sides train on the \emph{same} one-million-vector sample with the same parameters: \texttt{KSUB}$=256$, $25$ k-means iterations, one restart, non-spherical, exhaustive assignment. We disable the subsampling of \faiss's \texttt{max\_points\_per\_centroid}, which costs it $0.6$ accuracy points. We link \faiss's dynamic-dispatch build, which its documentation recommends for deployment, and both sides search one query at a time (Section~\ref{sec:methodology}). Code lengths are $\{0.5, 1, 2\}$ bits per component, i.e. $m \in \{8,16,32\}$ on \siftom\ and $m \in \{48,96,192\}$ on \dragon.

Table~\ref{tab:pq_faiss} reports the result. \emph{Accuracy matches}. No configuration differs by more than $0.43$ points, which is about the spread of PQ's own retraining (up to $0.40$ points across three trainings, Table~\ref{tab:rabitq_pq}). The sign of the difference changes across the six configurations, and footprint is identical to three decimals. \emph{Search} is faster for us when the codes fit in cache, and faster for this \faiss{} build when they do not. Our advantage is $1.62$, $1.30$ and $1.20\times$ as the \siftom{} code array grows from $8$ to $32$\,MB. On \dragon{} the two are within a few percent at $425$ and $850$\,MB, and at $1.7$\,GB \faiss{} is $1.17\times$ faster. Our advantage holds while the code array fits in the $48$\,MiB of L3 that the two sockets of the machine provide together (Section~\ref{sec:hardware}), since the parallel scan runs on both. Once the scan is memory-bandwidth-bound, the measurement no longer reflects the lookup kernel. \emph{Training} is faster in \faiss, by $1.13$--$1.57\times$, and the gap is widest where $\text{dsub}$ is largest. Its assignment step scores all $256$ centroids in one call, which fills a vector register exactly at $\text{dsub}{=}16$; ours iterates over the centroids one at a time. \emph{Encoding} changes sharply with $\text{dsub}$. \faiss\ is $2.8$--$4.1\times$ faster than us at $\text{dsub} \in \{4,8\}$, but $16$--$17\times$ \emph{slower} at $\text{dsub}{=}16$. At that width its encoder switches to an algorithm that writes out the full $256m$-entry distance table of every vector before reducing any of them.

Both results follow from how \faiss{} implements PQ. Its $8$-bit scan is hand-written once per instruction set, in $1\,146$ lines across eleven files. Its table builds dispatch to functions specialized at $\text{dsub} \in \{1,2,4,8\}$. This code also serves $\texttt{nbits}=16$ and arbitrary widths, while we fix the width at $8$. Its $736$ lines of $8$-bit SIMD code are directly comparable to ours, which consist of one $16$-line function with no intrinsics plus $45$ lines for the two table builds. The specialized code makes \faiss{} up to $17\%$ faster on the largest \dragon{} scan. However, each hard-coded threshold can select the wrong kernel, as the $16\times$ slower encoder at $\text{dsub}{=}16$ shows. On this machine the per-instruction-set dispatch also selects a slower kernel. The \texttt{AVX-512} scan loads its table through a microcoded gather instruction. It runs $1.2$--$2.0\times$ slower than \faiss's own scalar loop on \siftom, though up to $1.18\times$ faster on the largest \dragon{} codes. Forced to its scalar loop, \faiss{} scans the two smallest \siftom{} code arrays $1.23$ and $1.25\times$ faster than ours. From $32$\,MB to $850$\,MB the two are within $3.4\%$, and at $1.7$\,GB the \texttt{AVX-512} kernel stays $1.13\times$ ahead.

\begin{table*}[t]
\centering
\small
\caption{Validation of \texttt{ProductQuantizer} against \faiss\ \texttt{IndexPQ} (dynamic-dispatch build, plain ADC path) at $\texttt{nbits}=8$ on both sides. Exhaustive top-$10$ scan over all queries, conventions of Section~\ref{sec:methodology} on both sides, minimum of three repetitions. Footprint is identical throughout and omitted; bold marks a difference above $5\%$.}
\label{tab:pq_faiss}
\begin{tabular}{@{}ll rr @{\hspace{2em}} rr @{\hspace{2em}} rr @{\hspace{2em}} rr@{}}
\toprule
& & \multicolumn{2}{c}{\textbf{Accuracy@10 (\%)}} & \multicolumn{2}{c}{\textbf{Train (s)}} & \multicolumn{2}{c}{\textbf{Encode (s)}} & \multicolumn{2}{c}{\textbf{QPS}} \\
\cmidrule(lr){3-4}\cmidrule(lr){5-6}\cmidrule(lr){7-8}\cmidrule(lr){9-10}
\textbf{Collection} & $m$ & ours & \faiss & ours & \faiss & ours & \faiss & ours & \faiss \\
\midrule
\siftom             & $8$   & 32.1 & 31.7 & 9.4  & \textbf{6.0}  & \textbf{0.20} & 3.31 & \textbf{7\,587} & 4\,676 \\
($d{=}128$, $\ell_2$) & $16$  & 55.3 & 55.6 & 13.6 & \textbf{11.0} & 0.31 & \textbf{0.11} & \textbf{4\,125} & 3\,172 \\
                    & $32$  & 71.6 & 71.5 & 24.3 & \textbf{21.4} & 0.55 & \textbf{0.13} & \textbf{2\,155} & 1\,802 \\
\midrule
\dragon & $48$  & 53.8 & 53.8 & 55.3 & \textbf{37.2}  & \textbf{10.4} & 172.3 & 140.8 & 136.0 \\
($d{=}768$, IP)     & $96$  & 68.6 & 69.0 & 81.1 & \textbf{65.0}  & 16.6 & \textbf{5.9} & 64.7 & 65.9 \\
                    & $192$ & 83.2 & 83.3 & 144.0 & \textbf{126.9} & 29.3 & \textbf{7.2} & 26.7 & \textbf{31.2} \\
\bottomrule
\end{tabular}
\end{table*}

\subsubsection{\rabitq: Binary and Multi-Bit Codes}
\label{sec:rabitq}

\rabitq{} quantizes every component individually, so its minimum code width is one bit per component: $d/8 + 8$ bytes once the metadata word is counted. \vectorium\ ships it as two encoders that share their preprocessing, metadata layout and metric handling. \texttt{RabitqQuantizer} is the $1$-bit case, and \texttt{RabitqExtQuantizer}~\cite{rabitq_ext} widens the code to $\texttt{total\_bits} \in \{2,4,8\}$ bits per component.

\paragraph{Rotation and stored metadata.} \vectorium\ uses an $O(d\log d)$ rotation of four rounds. Each round applies a random $\pm1$ sign flip, a fast Hadamard transform over the largest power-of-two prefix, and a Kac's-walk butterfly~\cite{Kac1947RandomWA} that mixes the remaining tail. It is seeded by the caller for reproducibility. A $1$-bit code stores the sign of each component of $r = P(x-\mu)$, sixty-four to a \texttt{u64} word. A $b$-bit code adds $b-1$ magnitude bits per component. The estimator needs two per-document constants: $\lVert r\rVert^2$, and $\lVert r\rVert$ already divided by the rescale factor that maps magnitudes onto code levels. In both encoders they are bit-cast into the two halves of one extra \texttt{u64} word, which keeps a division off the per-candidate path. The rescale factor is estimated once at training time by averaging over random unit vectors. The encoding does not depend on the metric, which is a query-time property. Under squared-Euclidean distance the query is centered by the same mean. Under inner product it is only rotated, and the term $\langle \mu,q\rangle$ is added back in closed form, with one dot product per query.

\paragraph{Code layout.} All $b$ bits of a component are stored in one aligned bit-field of a \emph{single} component-major slab of $b\cdot d/8 + 8$ bytes. A scan therefore reads each query element only once. It spends one widening conversion and one fused multiply--add per $16$ components, the \texttt{f32} lane count of a $512$-bit vector register. That count is the same at every width; only the number of shift-and-mask extractions changes. For this reason the scan cost grows much more slowly with the code width than in a table-lookup scheme (Table~\ref{tab:rabitq_pq}). Only the aligned widths $\{2,4,8\}$ are supported, which together with $1$ bit cover the useful range densely.

\paragraph{Query quantization.} The $1$-bit encoder scalar-quantizes the query to \texttt{query\_bits} $\in \{1,2,4,8\}$ bits per component and scores symmetrically, as popcounts over bit planes. \texttt{query\_bits} is passed at search time, so one stored index serves every setting concurrently. We sweep it over one stored index on each of \siftom, \dragon{} and \snowf. Accuracy saturates at $4$ bits. Going from $1$ to $2$ bits gains up to $+9.3$ points of Accuracy@$10$ (from $53.9$ to $63.2$ on \dragon). Going from $4$ to $8$ gains at most $+0.36$ points and costs up to $20\%$ of the throughput ($148$ against $118$ QPS on \dragon). \vectorium{} therefore defaults to $4$, and the comparisons below use this setting. The extended encoder does not quantize the query at all. This gave better accuracy and better throughput, because a quantized query costs scan time in proportion to its width.

\paragraph{Validating the implementation against the reference.} We test our \rabitq{} implementation against the authors' own. The reference library ships IVF and \hnsw{} indexes with an encoder tailored to them. To isolate quantizer, rotator and estimator, we add a flat scan to it. Both sides then run the library's production scan path, and only the encoder differs. The comparison runs on \siftom\ and \dragon, with all queries and top-$10$, under the parallel-scan conventions of Section~\ref{sec:methodology} on both sides. At $\texttt{total\_bits}=1$ both sides quantize the query to $4$ bits, the width that the reference fixes at compile time. At wider codes neither side quantizes the query. Every row therefore compares the same estimator on the same query representation. The reference seeds its rotation from the system entropy source, so its figures are the mean of three repetitions. Their spread, up to $1.1$ points on \siftom\ and $0.32$ on \dragon, limits how precisely the differences below can be interpreted.

Table~\ref{tab:rabitq_reference} reports the outcome. \emph{Accuracy matches}. The largest difference at any width is $0.80$ points on \siftom\ and $0.29$ on \dragon, with no systematic sign, and neither exceeds the run-to-run spread of the reference on that collection. Our \emph{footprint is smaller at every width}, by $8$ bytes/vector at $1$ bit and $16$ at wider codes. The reference stores more per-vector metadata beside each code. This includes an error bound that lets its graph index prune on a cheap lower bound before computing a full distance; we do not have this first pass. \emph{Our throughput is higher}, by $1.44$--$1.74\times$ on \siftom\ and $1.44$--$2.37\times$ on \dragon, and it comes from the layout, width and accumulator choices described above. Build time is not compared, because centering on one global mean, as our encoders do, had to be added to the reference in our harness.

\begin{table*}[t]
\centering
\small
\caption{Validation against the authors' reference implementation of \rabitq. Exhaustive top-$10$ scan, all queries, conventions of Section~\ref{sec:methodology} on both sides. Best of three repetitions, the reference's accuracy the mean of three since its rotation is seeded non-deterministically. Grey figures are our factor of advantage per column.}
\label{tab:rabitq_reference}
\begin{tabular}{@{} l @{\hspace{1em}} *{4}{c} @{\hspace{2em}} r@{\hspace{2.5em}}r@{\hspace{2.5em}}r@{\hspace{2.5em}}r @{\hspace{3em}} *{4}{c} @{}}
\toprule
& \multicolumn{4}{c}{\textbf{Accuracy@10 (\%)}} & \multicolumn{4}{c}{\textbf{QPS}} & \multicolumn{4}{c}{\textbf{Bytes/vec}} \\
\cmidrule(lr){2-5} \cmidrule(lr){6-9} \cmidrule(lr){10-13}
& 1 bit & 2 bit & 4 bit & 8 bit & 1 bit & 2 bit & 4 bit & 8 bit & 1 bit & 2 bit & 4 bit & 8 bit \\
\midrule
\multicolumn{13}{@{}l}{\siftom ($d{=}128$, $\ell_2$)} \\
ours & 25.7 & 55.5 & 84.5 & 97.5 & \textbf{1\,988}\rlap{\hspace{2pt}\scriptsize\color{gray}1.64$\times$} & \textbf{1\,664}\rlap{\hspace{2pt}\scriptsize\color{gray}1.44$\times$} & \textbf{1\,608}\rlap{\hspace{2pt}\scriptsize\color{gray}1.74$\times$} & \textbf{1\,443}\rlap{\hspace{2pt}\scriptsize\color{gray}1.74$\times$} & \textbf{24} & \textbf{40} & \textbf{72} & \textbf{136} \\
ref. & 24.9 & 55.5 & 83.9 & 97.0 & 1\,209 & 1\,157 & 922 & 831 & 32 & 56 & 88 & 152 \\
\midrule
\multicolumn{13}{@{}l}{\dragon ($d{=}768$, IP)} \\
ours & 66.6 & 82.3 & 93.9 & 99.3 & \textbf{151}\rlap{\hspace{2pt}\scriptsize\color{gray}1.44$\times$} & \textbf{60}\rlap{\hspace{2pt}\scriptsize\color{gray}1.85$\times$} & \textbf{52}\rlap{\hspace{2pt}\scriptsize\color{gray}2.37$\times$} & \textbf{30}\rlap{\hspace{2pt}\scriptsize\color{gray}1.86$\times$} & \textbf{104} & \textbf{200} & \textbf{392} & \textbf{776} \\
ref. & 66.9 & 82.5 & 94.1 & 99.3 & 105 & 32 & 22 & 16 & 112 & 216 & 408 & 792 \\
\bottomrule
\end{tabular}
\end{table*}

\paragraph{Comparing \rabitq{} and Product Quantization.} Both encoders are available from one bit per component upwards, so a user must choose between them in this range. The \rabitq{} papers compare it with PQ and its variants at equal code length, and find its distance estimates more accurate and faster to compute~\cite{rabitq,rabitq_ext}. We compare the two as the encoder of an exhaustive scan on three dense collections of increasing dimensionality (Table~\ref{tab:collections}), with codes of the same size. We report accuracy, throughput and build time from harnesses that differ only in the encoder. Footprints are matched by setting $m = \texttt{total\_bits}\cdot d/8$. This leaves each \rabitq{} configuration $8$ bytes/vector \emph{larger}, because of its metadata word. PQ trains on a one-million-vector sample, and the $1$-bit \rabitq{} rows use a $4$-bit query. We follow the parallel-scan conventions of Section~\ref{sec:methodology} with one repetition, except for PQ's accuracy, which is the mean of three trainings.

Table~\ref{tab:rabitq_pq} reports the outcome. The better encoder depends on both the data distribution and $d$, and the balance moves monotonically towards \rabitq{} as $d$ grows. At the lowest bit rate, the accuracy difference $\Delta = \text{PQ} - \text{\rabitq}$ is $+29.5$ points at $d=128$, $+1.9$ at $d=768$ and $-2.3$ at $d=1\,024$. Within each collection it roughly shrinks as the code widens. All differences but one, \dragon{} at $4$ bits, are larger than the largest spread of PQ's three trainings. On \siftom{}, whose hand-crafted features are strongly clustered, PQ is more accurate at every width. On \dragon{} it leads only up to $4$ bits, and on \snowf{} \rabitq{} leads in \emph{both} accuracy and throughput at every footprint. Throughput shows the opposite trend to accuracy, and the gap widens with the code width. Across the \siftom{} matrix the throughput of \rabitq{} spans $1.4\times$, while that of PQ spans $9.8\times$. At the widest code \rabitq{} is $3.6$, $8.9$ and $10.8\times$ faster on the three collections. PQ is the faster encoder only on \siftom{} below ${\approx}40$ bytes/vector. At matched footprint, $m$ scales with $d$. The per-document work of PQ is $m$ dependent lookups into a table of $m\cdot256$ floats. This table takes $16$\,KiB and fits in L1 at \siftom's $m{=}16$, but takes $1$\,MiB at \snowf's $m{=}1\,024$. The per-document work of \rabitq{} is a fixed number of vectorized multiply--adds over contiguous bytes.

Build times differ by orders of magnitude. \rabitq{} has no training phase, while PQ must fit $m$ codebooks before it can encode anything: training takes $493$ of the $634$ seconds on \snowf{} at $m{=}1\,024$. The ratio runs from $62\times$ to $1\,062\times$ and generally widens with $m$. Fewer k-means iterations or a smaller training sample reduce PQ's build time by at most one order of magnitude before accuracy degrades appreciably, while the gap spans two to three orders.

\begin{table*}[t]
\centering
\small
\caption{\rabitq against Product Quantization at matched footprint. Exhaustive \texttt{FlatIndex} top-$10$ search, all queries, Accuracy@$10$ (\%) against the exact ground truth, footprint from \texttt{SpaceUsage}, conventions of Section~\ref{sec:methodology}. \emph{Build} is the complete construction cost. PQ's training is not deterministic: its accuracy is the mean of three trainings, which differ by at most $0.40$ points. QPS is from one repetition, so it can be lower than the best-of-three figures of Table~\ref{tab:rabitq_reference}.}
\label{tab:rabitq_pq}
\begin{tabular*}{\textwidth}{@{} l l @{\extracolsep{\fill}} r r r r r r r r @{}}
\toprule
& & \multicolumn{2}{c}{\textbf{Bytes/vec}} & \multicolumn{2}{c}{\textbf{Accuracy@10 (\%)}} & \multicolumn{2}{c}{\textbf{QPS}} & \multicolumn{2}{c}{\textbf{Build (s)}} \\
\cmidrule(lr){3-4}\cmidrule(lr){5-6}\cmidrule(lr){7-8}\cmidrule(lr){9-10}
\textbf{Dataset} & \textbf{Width} & \rabitq & PQ & \rabitq & PQ & \rabitq & PQ & \rabitq & PQ \\
\midrule
\siftom                & 1 bit / $m{=}16$   &   24 &   16 & 25.7 & \textbf{55.2} & 1\,988.1 & \textbf{3\,966.3} & \textbf{0.031} &   14.3 \\
$d{=}128$, $\ell_2$    & 2 bit / $m{=}32$   &   40 &   32 & 55.5 & \textbf{71.5} & 1\,661.6 & \textbf{2\,060.2} & \textbf{0.063} &   25.2 \\
                       & 4 bit / $m{=}64$   &   72 &   64 & 84.5 & \textbf{87.8} & \textbf{1\,606.1} & 889.1 & \textbf{0.058} &   40.0 \\
                       & 8 bit / $m{=}128$  &  136 &  128 & 97.5 & \textbf{99.6} & \textbf{1\,442.3} & 402.8 & \textbf{0.060} &   63.7 \\
\midrule
\dragon                 & 1 bit / $m{=}96$   &  104 &   96 & 66.6 & \textbf{68.5} & \textbf{148.3} & 60.8 & \textbf{1.262} &   98.1 \\
$d{=}768$, IP          & 2 bit / $m{=}192$  &  200 &  192 & 82.3 & \textbf{83.2} & \textbf{56.8} & 26.3 & \textbf{2.676} &  175.6 \\
                       & 4 bit / $m{=}384$  &  392 &  384 & 93.9 & \textbf{94.2} & \textbf{46.6} & 12.1 & \textbf{2.601} &  289.5 \\
                       & 8 bit / $m{=}768$  &  776 &  768 & \textbf{99.3} & 98.4 & \textbf{30.2} & 3.4 & \textbf{2.615} &  475.0 \\
\midrule
\snowf                 & 1 bit / $m{=}128$  &  136 &  128 & \textbf{82.1} & 79.8 & \textbf{143.1} & 46.4 & \textbf{1.842} &  132.3 \\
$d{=}1\,024$, IP         & 2 bit / $m{=}256$  &  264 &  256 & \textbf{90.4} & 88.0 & \textbf{44.6} & 19.7 & \textbf{3.756} &  233.8 \\
                       & 4 bit / $m{=}512$  &  520 &  512 & \textbf{96.6} & 95.4 & \textbf{29.3} & 9.0 & \textbf{3.637} &  386.2 \\
                       & 8 bit / $m{=}1\,024$ & 1\,032 & 1\,024 & \textbf{99.7} & 98.6 & \textbf{17.3} & 1.6 & \textbf{3.682} &  634.0 \\
\bottomrule
\end{tabular*}
\end{table*}

\subsubsection{Blocked Sparse Encoding}
\label{sec:blocked_sparse}

\texttt{BlockedSparseEncoder} stores the same \texttt{u16} components and \texttt{f16} values as the scalar quantizer, in a layout designed for scoring. Each vector is split into blocks of $16$ (component, value) pairs: $32$ bytes of components followed by $32$ bytes of values, which fill one $64$-byte cache line. A query with at least $33$ nonzeros is densified, as with the scalar quantizer. A shorter query is kept as a sorted list and scored with \jumpdot~\cite{nardini2026compactefficientindexeslearned}. The kernel walks the blocks of the candidate in order and skips every block whose last component precedes the current query component. It compares the query component against the $16$ components of a block at once. Unlike the kernel family, this kernel uses explicit SIMD (AVX2, and NEON on Armv8). 
\subsubsection{Compressed Sparse Encoding: \dotpacking}
\label{sec:dotpacking}

The scalar quantizer of Section~\ref{sec:scalar_quantization} compresses only the \emph{values} of a sparse vector. Each nonzero still carries a raw \texttt{u16} component index, and on learned-sparse embeddings~\cite{splade,formal2021splade,cocondenser} this index is the largest remaining cost once the values are compressed. \texttt{DotPacking8Encoder} is the only encoder in the library that compresses it. The component stream is a sorted list of integer identifiers, stored as the gaps between consecutive ones. It is compressed with \dotpacking~\cite{nardini2026compactefficientindexeslearned}, the bit-packing code for inner products of Section~\ref{sec:related_quantization}. Blocks of eight gaps are packed at the width of their largest gap. A byte shuffle decodes a block into a SIMD register, and a prefix sum over the gaps gives the component identifiers, which gather the query weights. A compressed vector is thus scored without being materialized. \vectorium{} integrates the authors' implementation. It stores the values in $8$-bit fixed point, \texttt{f16}, or per-component $4$-, $8$- or $16$-bit scalar codes, so both halves of the vector can be compressed.

The gaps between consecutive component identifiers of a vector, and hence its compressed size, depend on how the coordinate axes are numbered. Any permutation $\pi$ of the axes may be applied to the stored components, provided the same $\pi$ is applied to every query. The encoder computes such a $\pi$ at training time, numbering components by decreasing document frequency, and stores it with its inverse. It relabels components through $\pi$ before packing and undoes the relabeling on decode, so the permutation is invisible to the rest of the system.

\subsubsection{Multi-Vector (Late-Interaction) Encoders}
\label{sec:multivec_encoders}

The encoders above compress one vector per document. Late-interaction retrieval~\cite{colbert,colbertv2} (Section~\ref{sec:related_representations}) represents a document as a \emph{set} of token vectors and scores it by MaxSim. \vectorium{} added this representation last, after the abstractions of the preceding subsections were fixed. It therefore tests directly whether these abstractions accept a new representation at the cost the design claims. Supporting it required four additions, one per layer, all behind an optional compile-time feature: the multi-vector view of Section~\ref{sec:vector_representations}, the \texttt{maxsim} kernel of Section~\ref{sec:distance_computation}, a multi-vector encoder contract, and the multi-vector dataset of Section~\ref{sec:datasets}. The \texttt{Dataset} contract itself, its access and prefetch primitives, persistence, and the \texttt{Index} and \texttt{RerankIndex} contracts of Section~\ref{sec:indexing} were reused unmodified. A multi-vector collection is a \texttt{Dataset} whose encoded vector is a token buffer, and a MaxSim score is treated like a \texttt{DotProduct}.

\paragraph{Encoders.} The same compression ladder is reproduced for this representation. \texttt{ScalarMultiVecQuantizer} applies the cast of Section~\ref{sec:scalar_quantization} to every token value. It iterates over document tokens in the outer loop, so a stored \texttt{f16} token is widened once per document token and not once per query token. \texttt{MultiVecProductQuantizer} encodes each token with the Product Quantizer of Section~\ref{sec:product_quantization}, at $m$ bytes per token. For $Q$ query tokens it builds a single table of shape $[m \times \text{KSUB} \times Q]$ with the query-token index \emph{innermost}. A code byte thus selects the contribution of that subspace to every query token at once, and a document token is scored by $m$ vector accumulations in place of $m\cdot Q$ scalar lookups. This applies the reasoning of the SoA centroid layout one level up. \texttt{MultiVecTwoLevelProductQuantizer} assigns each token to one of a set of k-means centroids and applies Product Quantization only to the residual. A token therefore costs $4$ bytes of centroid identifier plus $m$ bytes of code. The coarse contribution, every document token's centroid against every query token, is a small dense matrix product delegated to a GEMM. The residual contribution uses the accumulation above. Of all the encoders, this one tests the storage/scoring separation of Section~\ref{sec:data_overview} most. Its stored token is a byte payload whose ``dimension'', $4+m$, only the encoder can read. The payload is laid out in \emph{blocks}, all coarse identifiers and then all codes, so each scoring phase walks its own contiguous region. An optional variant normalizes each residual and appends the discarded norm as an \texttt{f16} third block. This raises the width to $4+m+2$ and requires no change to the storage layer. The MaxSim kernel is monomorphized over the query-token count $Q$ for the powers of two from $8$ to $512$. This turns its two $Q$-element buffers into stack arrays and its inner loops into loops with a fixed trip count. A runtime-$Q$ variant covers other lengths; typical late-interaction queries, such as the $32$ tokens of \colbertvt, do not need it.

\kannolo's and \vectorium's indexes could be used to search multi-vector representations. However, MaxSim is expensive enough to compute that neither is well suited to this representation. We therefore leave its indexing to engines specialized for it~\cite{plaid,emvb}, such as \tachiom~\cite{tachiom}. This engine builds its late-interaction index on the multi-vector views and dataset, the two-level encoder and its MaxSim kernel, and routes queries to its token centroids through \kannolo's \hnsw. Late interaction is also used as the reranking step of a two-stage pipeline. This use is supported through a multi-vector \texttt{Dataset} inside \rerankindex{} (Section~\ref{sec:indexing}) and is evaluated across all three encoders in~\cite{martinico2026multivector}. There, a learned sparse first stage, searched with \kannolo's \hnsw{} or with \seismic, is followed by \colbertvt{} MaxSim rescoring. This pipeline matches or exceeds the effectiveness of the token-level systems \emvb~\cite{emvb}, \warp~\cite{warp} and \igp~\cite{igp}, and is up to $24\times$ faster than the fastest of them at comparable memory.

\subsection{Datasets}
\label{sec:datasets}

A \texttt{Dataset} holds a whole collection of encoded vectors. It is the pure storage of Section~\ref{sec:data_overview}, parameterized by a \texttt{VectorEncoder}. It reports its size, returns a vector by identifier or by storage range, iterates over its vectors, and issues prefetch hints. It knows nothing else about the encoding.

Four dataset types cover the four layouts of Section~\ref{sec:vector_representations}. Dense data is held in one contiguous buffer. Sparse data uses the compressed-sparse-row form: an offsets array plus parallel component and value arrays. Packed encodings are concatenated into one blob of words with the same offsets table; this is the storage form of \dotpacking{} and of the blocked encoder. Multi-vector documents are stored as an offsets table over a flat token buffer.

Datasets convert into one another in two ways. A \emph{zero-copy} conversion keeps the stored bytes and only reinterprets the encoder. An example is switching a dense or sparse dataset between squared-Euclidean distance and inner product, which share a layout. A \emph{re-encoding} conversion walks the data once to change its stored form, for example casting \texttt{f32} to \texttt{f16} or encoding a plain dataset into Product-Quantization codes. This conversion is also where an encoder receives the parameters it cannot learn from the data (Section~\ref{sec:rabitq}). Every dataset persists and reloads without a rebuild, in a fixed-width little-endian format that is stable across hosts.
\paragraph{Identifier/range duality.} Each stored vector can be fetched by its identifier or by its span in the backing storage. For the sparse layout the difference matters. Fetching by identifier first reads the vector's boundaries from the offsets table. This is a \emph{dependent} load, which must complete before the vector itself can be requested. An index that walks a candidate list can resolve each range once, prefetch the vector body, and fetch it by range only when it scores it. The memory latency then overlaps with other work. For dense data the two methods coincide. The prefetch hint touches at most the first $256$ bytes of the span. This covers a short or compressed vector and leaves the tail of a long one to the hardware prefetcher.

Under random access on \cocond, range addressing is $1.46\times$ faster than identifier addressing, and $2.07\times$ with a prefetch issued ahead of use; in storage order the three strategies tie (Appendix~\ref{app:range_access}).

\subsection{Clustering}
\label{sec:data_infrastructure}

The layers above also depend on a primitive that is neither a representation, a metric nor an encoder: k-means. It learns the per-subspace codebooks of the Product Quantizer (Section~\ref{sec:product_quantization}), the coarse level of the two-level multi-vector quantizer (Section~\ref{sec:multivec_encoders}), and the partition from which an IVF index is built (Section~\ref{sec:indexing}). It comes in a plain and a spherical variant. The spherical variant $\ell_2$-normalizes the centroids after each update, so an inner-product collection is clustered by \emph{angle}.

\texttt{KMeans} is generic over the representation. It clusters any dataset that can average a set of its vectors and subtract a centroid from one of them; the dense and sparse datasets currently support both operations. It returns the centroids as a \texttt{Dataset} whose encoder the caller chooses. The metric of that encoder is the clustering metric, which is independent of the metric used later to search the collection. K-means therefore has no distance loop of its own. Assignment is a search over the centroid dataset with \vectorium's distance kernels, either an exhaustive scan or a top-$1$ query against any \texttt{Index} built over the centroids. The training corpus and the centroids may use different storage types. A collection already stored as \texttt{f16} is therefore clustered directly, with no widening pass, while means accumulate in \texttt{f32}. The update step groups assignments with a stable counting sort and sums each cluster in parallel, which fixes the summation order. With a fixed seed and exhaustive assignment, the centroids are bit-identical on a given target at any thread count.

On sparse data the mean of a cluster is much denser than its members. After every update, each centroid can therefore be pruned to its largest-magnitude components, keeping a given factor times the average number of nonzeros in its cluster.
\section{\kannolo}
\label{sec:indexing}

\kannolo{} is the indexing layer above \vectorium. It implements \vectorium's index contract (Section~\ref{sec:system_overview}) with two standard index families: a graph index (\hnsw~\cite{hnsw}) and an Inverted Vector File (IVF) index~\cite{ivf_gpu}. It also provides the search kernels, construction algorithms and adjacency layouts these indexes are built from, and the two-stage \rerankindex{} that composes them. Both families are generic over the representations, encoders, metrics and datasets, and name no encoding of their own, so one implementation of each serves dense and sparse data. Multi-vector data enters \kannolo{} as the rescoring dataset of \rerankindex{} (Section~\ref{sec:multivec_encoders}).

\subsection{System Overview}
\label{sec:system_overview}

\vectorium{} defines the \texttt{Index} contract, and \kannolo{} is one library that implements it. Any other library built on \vectorium{} can implement it as well. The contract covers search only: given a query and per-family search parameters, it returns the top $k$ results as \texttt{(distance, id)} pairs. Construction differs across families, so each family exposes its own. IVF, for example, also chooses the encoding its clusters are stored with (Section~\ref{sec:ivf_impl}). \vectorium{} also ships the default implementation, \texttt{FlatIndex}, which makes a dataset searchable by exhaustive scan in $\Theta(n \log k)$ time. Because this index lives in the data layer, \vectorium{} can work on its own with no dependency on the indexing layer, for example to assign points to centroids during k-means (Section~\ref{sec:data_infrastructure}).

\paragraph{RerankIndex.}
\texttt{RerankIndex} is the two-stage composition of Section~\ref{sec:related_ann_indexes} written as a single type. It pairs a first-stage index with a companion reranking dataset. The two may use different vector encodings and distance metrics. For example, stage one can be an \hnsw{} index over sparse representations that retrieves candidates cheaply, while stage two holds multi-vector embeddings and rescores the candidates with that representation, as in~\cite{martinico2026multivector}. Alternatively, stage one can search a PQ-compressed copy of the same collection that stage two holds at full precision. The only requirement is that the two representations are aligned on the same vector identifiers, and the caller is responsible for it.

The pipeline proceeds in two stages. First, the first-stage index retrieves $k_\text{c}$ candidates using its own search parameters. Then the reranking dataset scores every candidate against the query using the full representation, and the top $k_\text{final}$ are returned. Two optional optimizations reduce the cost of the second stage~\cite{martinico2026multivector}:

\begin{itemize}
    \item \textbf{Candidate Pruning (CP, $\alpha$).} The first-stage scores are read here as similarities, so a higher value is better. Before reranking, candidates whose first-stage score falls below $(1 - \alpha)$ times the score of the $k_\text{final}$-th candidate are discarded. This exploits the correlation between first-stage and second-stage scores to prune clearly suboptimal entries before the expensive reranking step.

    \item \textbf{Early-Exit Reranking (EE, $\beta$).} The first-stage results are already sorted by first-stage score. During reranking, if $\beta$ consecutive candidates fail to enter the current top-$k_\text{final}$ heap, the reranking loop terminates. The assumption is that the remaining candidates are unlikely to improve the ranking.
\end{itemize}

A third option covers the case where the two stages score complementary parts of one representation, i.e. where the vector is (or is approximated by) the sum of the corresponding first- and second-stage vectors. The reranking dataset then holds a \emph{residual}, and the final score is the sum of the two scores. This decomposition holds only under an additive metric, and the caller must ensure that it is meaningful.

\texttt{Dataset}, \texttt{Index} and \texttt{RerankIndex} compose. Any index implementing the contract, flat or one of \kannolo's, can serve as a \texttt{RerankIndex}'s first stage, and any \vectorium{} dataset as its second. This yields many high-performance indexes, as Section~\ref{sec:exp} shows. Table~\ref{tab:compositions} lists the meaningful compositions for each representation, with the published method each one corresponds to. The types allow many more. A first stage over \texttt{f32} rescored with \pq{} gains nothing in a deployed index, but it can still be built. Such combinations can serve experiments and analysis, for example to isolate the effect of one stage.

Figure~\ref{fig:running_example} follows one composition from encoder to search. It is the dense two-stage index of Section~\ref{sec:exp}: a $1$-bit \rabitq{} first stage over a permuted \hnsw{} graph, rescored on an \texttt{f16} copy of the collection. The graph is built over the \texttt{f16} vectors, and its dataset is then replaced by the codes, so the codes never take part in construction. The result is itself an \texttt{Index}, searched through the same call as its first stage. Each stage is named by a type alias. Replacing \texttt{Codes} with a \pq{} or \rabitqext{} dataset, and its configuration accordingly, yields another dense row of Table~\ref{tab:compositions} with no other change.

\begin{figure}[t]
\centering
\footnotesize
\begin{verbatim}
type Rescoring =
    DenseDataset<PlainDenseQuantizer<f16, DotProduct>>;
type Codes = DenseDataset<RabitqQuantizer<DotProduct>>;

let exact: Rescoring = DenseDataset::from_raw(
    values, n, PlainDenseQuantizer::new(dim));
let cfg = HNSWBuildConfiguration::default()
    .with_num_neighbors(16)
    .with_ef_construction(200);
let graph: HNSW<Rescoring, Graph> =
    HNSW::build_index(exact.clone(), &cfg);
let first: HNSW<Codes, Graph> = graph
    .permute_and_encode::<PlainNeighbors>()
    .convert_dataset_into_ref(RabitqConfig::default());
let index = RerankIndex::new(first, exact);

let top10 = Index::search(&index, (q, q), 10,
    &RerankSearchParams::new(100, params, ()));
\end{verbatim}
\caption{Building and searching the dense two-stage index of Section~\ref{sec:exp}. \hnsw{} is built over \texttt{f16} vectors and permuted. Its vectors are replaced by $1$-bit \rabitq{} codes, and candidates are rescored on the \texttt{f16} copy, with $k_\text{c}=100$ and $k_\text{final}=10$.}
\label{fig:running_example}
\end{figure}

\begin{table*}[t]
\centering
\footnotesize
\caption{Index compositions expressible with \vectorium{} and \kannolo. \emph{Known as} names the published method with the same decomposition, where one exists; \emph{Evidence} points to where the composition is measured. \emph{Scalar} stands for \texttt{f32}, \texttt{f16}, \texttt{bf16} and $16$/$8$-bit fixed point. \rabitq{} stands for both the $1$-bit \rabitq{} and \rabitqext. IVF clusters store either the vectors or, on dense data, their residuals from the centroid. A rescoring encoder is more precise than the first-stage one. $^{\dagger}$The \hnsw{} and IVF indexes of the \rabitq{} authors' library. Both prune candidates with a cheap estimate and its error bound before computing the full one; \kannolo{} does not (Sections~\ref{sec:rabitq} and~\ref{sec:ivf_rabitq_validation}).}
\label{tab:compositions}
\begin{tabular*}{\textwidth}{@{\extracolsep{\fill}}l>{\raggedright\arraybackslash}p{3.3cm}lll@{}}
\toprule
\textbf{First-stage index} & \textbf{First-stage encoder} & \textbf{Rescoring} & \textbf{Known as} & \textbf{Evidence} \\
\midrule
\multicolumn{5}{@{}l}{\emph{Dense}} \\
\quad Flat & scalar & --- & --- & Table~\ref{tab:bsq1} \\
\quad Flat & \pq & --- & PQ~\cite{pq} & Tables~\ref{tab:pq_faiss},~\ref{tab:rabitq_pq} \\
\quad Flat & \rabitq & --- & --- & Tables~\ref{tab:rabitq_reference},~\ref{tab:rabitq_pq} \\
\quad Flat & \pq, \rabitq & scalar, \pq, \rabitq & --- & --- \\
\quad \hnsw & scalar & --- & \hnsw~\cite{hnsw} & Sections~\ref{sec:hnsw_validation},~\ref{sec:exp} \\
\quad \hnsw & \pq & --- & HNSWPQ & Figure~\ref{fig:batch_prefetch_comparison} \\
\quad \hnsw & \rabitq & --- & --- & --- \\
\quad \hnsw & \pq & scalar, \pq, \rabitq & HNSWPQ + refine & Section~\ref{sec:exp_encoder_transfer} \\
\quad \hnsw & \rabitq & scalar, \pq, \rabitq & \rabitqhnsw$^{\dagger}$ & Sections~\ref{sec:exp},~\ref{sec:exp_rerank_precision} \\
\quad IVF, flat centroids & scalar, \pq & --- & IVFFlat, IVFPQ~\cite{pq} & Figure~\ref{fig:ivf_recall_qps_grid} \\
\quad IVF, \hnsw{} centroids & scalar, \pq & --- & IVF-\hnsw & Figure~\ref{fig:ivf_recall_qps_grid} \\
\quad IVF, flat or \hnsw{} centroids & \rabitq & --- & IVF-\rabitq$^{\dagger}$~\cite{rabitq,rabitq_ext} & Table~\ref{tab:ivf_rabitq} \\
\quad IVF, IVF centroids & scalar, \pq & --- & --- & --- \\
\quad IVF, any centroids & scalar, \pq, \rabitq & scalar, \pq, \rabitq & --- & --- \\
\midrule
\multicolumn{5}{@{}l}{\emph{Sparse}} \\
\quad Flat & scalar, blocked, \dotpacking & --- & --- & Table~\ref{tab:bsq1} \\
\quad \hnsw & scalar, blocked, \dotpacking & --- & --- & Section~\ref{sec:exp_sparse} \\
\quad IVF, any centroids & scalar, \dotpacking & --- & --- & Section~\ref{sec:ivf_sparse} \\
\quad Flat, \hnsw, IVF & scalar, \dotpacking & scalar & --- & --- \\
\quad Flat, \hnsw, IVF & top-$a\%$ of $\ell_1$ mass, scalar or blocked & scalar or blocked residuals & --- & Section~\ref{sec:exp_sparse} \\
\midrule
\multicolumn{5}{@{}l}{\emph{Multi-vector}} \\
\quad Flat, \hnsw & scalar, \pq, two-level \pq & --- & --- & --- \\
\quad Flat, \hnsw & scalar, \pq, two-level \pq & scalar, \pq, two-level \pq & --- & --- \\
\midrule
\multicolumn{5}{@{}l}{\emph{Hybrid}} \\
\quad Flat, \hnsw, IVF & any sparse & any dense & --- & --- \\
\quad Flat, \hnsw, IVF & any sparse & any multi-vector & --- & \cite{martinico2026multivector} \\
\quad Flat, \hnsw, IVF & any dense & any sparse, any multi-vector & --- & --- \\
\midrule
\multicolumn{5}{@{}l}{\emph{Cascade}} \\
\quad \rerankindex & any \rerankindex & any & --- & --- \\
\bottomrule
\end{tabular*}
\end{table*}

\subsection{Graph Index Framework}
\label{sec:graph_framework}

\kannolo{} abstracts the storage layer of any proximity graph~\cite{fu2017fast,vamana,survey_graph_2025,wang2021comprehensive} behind a \texttt{GraphTrait} of six primitives: a node's neighbor list, the node count, the edge count, the degree bound, memory accounting, and a translation from a local graph node ID to the corresponding identifier in the underlying dataset.

The identifier translation decouples the graph's internal node ordering from the external ID space of the application. The graph can therefore be permuted without invalidating any external reference, e.g. to improve cache locality or edge compressibility, or to map the local IDs of \hnsw's upper levels onto the ground level. When no mapping is configured, the translation is a no-op. A mapping holds one entry per node, bit-packed at $\lceil \log_2 N \rceil$ bits.

The two search kernels, a greedy walk and a beam search, are implemented once on \texttt{GraphTrait}. The greedy walk is the single-best-neighbor descent used for $1$-NN search and for traversing the upper levels of \hnsw. The beam search is used by every graph index and by the ground level of \hnsw. It maintains a max-heap of the current top-$k$ candidates and a min-heap of the traversal frontier, and processes nodes in best-first order until the termination criterion fires.

A graph is built in a growable fixed-degree layout, where each node owns a fixed number of contiguous slots. While the graph is still growing, a neighbor list is located, read and extended with no offsets indirection. This layout needs no locking (Section~\ref{sec:hnsw_construction}). Once built, the graph is frozen into one of three search-time backends. The plain backend is a CSR-like layout: neighbor IDs plus an offsets array delimiting each node's list. The fixed-degree backend pads every list to a common stride and drops the offsets array, which gives direct random access to any node's list. The third is the compressed layout of Section~\ref{sec:graph_compression}. Its delta coding needs each list sorted by ID, and the conversion sorts the lists. The plain layout is the default for search, and the rest of this section writes the graph as \texttt{Graph}.

\subsection{Graph Optimizations}
\label{sec:graph_search_opts}

The beam-search kernel is implemented on \texttt{GraphTrait}, so the optimizations below apply to every graph backend and to any graph index added later. We measure all of them on the \hnsw{} index. They are presented in the order in which they act during a query.

\subsubsection{Visited Set Design and Optimization}
\label{sec:visited_set}

A graph index does not partition the search space, so the search has to track which nodes it has already evaluated. It keeps them in a ``visited set'', which prevents redundant distance computations and infinite loops. The set is queried and updated in the innermost loop of the search, so its cost is on the critical path of every expansion. We evaluate two structurally different alternatives:
\begin{itemize}
    \item \texttt{BitVec}, a dense bitmap over the whole graph. Access is $O(1)$ by direct indexing, but each query pays an $O(n)$ initialization.
    \item \texttt{FxHashSet}, a fast non-cryptographic hash table sized from the expected number of visited nodes. Access is $O(1)$ amortized and initialization is bounded by the table's capacity, independent of $n$.
\end{itemize}
A third common alternative is a global generation array. It stores, for each node, the generation of the last query that visited it. A new query increments the generation and does not clear the array, so there is no per-query initialization. We exclude it because it requires $O(n)$ \emph{permanent} memory, $200$\,MB at $n=100$M with $16$-bit generations, and introduces synchronization between threads querying the same index.

\paragraph{Sizing and choice.}
Sizing the hash table requires an estimate of the number of nodes a query visits, $n_{\text{visited}}$. Rehashing is expensive, while an oversized table wastes memory and cache. Only the ground level of \hnsw{} keeps a visited set, since the greedy walk on the upper levels cannot revisit a node. On the ground level, a query expands about $ef_{\text{search}}$ nodes and scores their unvisited neighbors. Distance-adaptive termination (Section~\ref{sec:early_termination}) multiplies the expansions by a factor that grows faster than linearly in $\lambda$. The table is therefore sized as
\[
  \text{capacity} = \bar{d} \cdot ef_{\text{search}} \cdot \left(1 + \mathds{1}_{\lambda>0}\,(\lambda + 0.05)\right)^{8} + 200,
\]
where $\bar{d} = \lfloor |E|/n \rfloor$ is the average ground-level degree and $\mathds{1}_{\lambda>0}$ is $1$ when early termination is on. Appendix~\ref{app:visited_set} derives this form and checks it per query. Without early termination, at most $7.2\%$ of the queries rehash, and almost none rehash twice.

At $n=100$M, clearing a \texttt{BitVec} costs $407$--$455\,\mu$s per query, more than the whole query budget. \texttt{BitVec} is $2.2\times$ to $4.4\times$ slower than \texttt{FxHashSet} in every configuration at this size. At $n \le 1$M it is ahead by at most $32\,\mu$s (Appendix~\ref{app:visited_set}). \kannolo{} therefore uses \texttt{FxHashSet} unconditionally.

\subsubsection{Neighbor Processing}
\label{sec:batch_prefetch}

The innermost loop of the graph beam search reads a candidate's neighbor list, fetches the bodies of the neighbors not yet visited, and scores them against the query. It executes hundreds to thousands of times per query, so its efficiency governs the throughput of the whole index. Two optimizations act on it, each on a different bottleneck:
\begin{itemize}
    \item \textbf{Batched distance computation.} The expansion buffers up to six unvisited neighbors and scores them in a single pass over six independent accumulators, through the batched kernel of Section~\ref{sec:distance_computation}.
    \item \textbf{Prefetching.} Before scoring, the loop issues explicit \texttt{PREFETCHT2} hints, targeting the L3 cache, for the batch it is about to score. The prefetch window is capped at the first $256$ bytes ($4$ cache lines) of each vector. Any remaining cache lines are left to the hardware prefetcher.
\end{itemize}
Both are built from the \texttt{Dataset} access primitives of Section~\ref{sec:datasets} and neither adds a primitive of its own.

We measure them on four collections whose vectors span $2$ to $48$ cache lines (Appendix~\ref{app:batch_prefetch}). Enabling both is the fastest or tied-fastest configuration on all four. Prefetching alone gives up to $+35\%$ QPS where a whole vector fits in the prefetch window. Batching gives $+34\%$ on the largest vectors, where it cuts retired instructions by about a quarter at constant IPC. Hardware counters show that the $4$-cache-line cap is the point where two costs balance: the first-access stall that the hints remove, and the issue slots that the hints occupy.

\subsubsection{Early Termination Strategies}
\label{sec:early_termination}

\kannolo{} implements a relaxation-based termination criterion (Section~\ref{sec:related_ann_indexes}) that generalizes Adaptive Beam Search~\cite{ee_distance_adaptive}. Adaptive Beam Search keeps only the $k$ best candidates found so far. It stops when the best unexpanded candidate is farther from the query than $1+\gamma$ times the $k$-th of them. \kannolo{} keeps the beam of $ef_{\text{search}}$ candidates and relaxes the comparison against its worst element. A candidate is expanded while its distance is within a factor $1+\lambda$ of the $ef_{\text{search}}$-th best, through the relaxation predicate of Section~\ref{sec:distance_computation}. The authors of Adaptive Beam Search leave this hybrid of beam width and distance threshold open. It contains both rules: $\lambda=0$ is the standard \hnsw{} criterion, and $ef_{\text{search}}=k$ is Adaptive Beam Search, with $1+\lambda=(1+\gamma)^2$ on squared distances. Tuning both parameters can therefore only match or improve on either rule. The approximation guarantee of Adaptive Beam Search holds for $ef_{\text{search}}=k$ and Euclidean distance on navigable graphs. \hnsw{} graphs are not provably navigable, so neither form carries the guarantee here. We compare this rule against the saturation-based family, represented by Patience-in-Proximity (PiP)~\cite{pip}. For this comparison we implemented PiP in the search loop of \kannolo; it is not part of the released library. The search stops after $\Delta$ consecutive \emph{stable} expansions, each of which admits at most $\lfloor(1-p)\,k\rfloor$ new candidates into the result set.

We evaluate the two strategies on \siftom{} and \cocond, which differ in representation and in metric (Table~\ref{tab:collections}). We measure QPS and distance computations per query at Accuracy@$10$ targets from $90\%$ to $99\%$. For each target we report the fastest configuration that reaches it and, separately, the one that reaches it with the fewest distance computations. Standard \hnsw{} sweeps $ef_{\text{search}}$ from $10$ to $400$. The relaxation grid crosses $ef_{\text{search}}$ from $k=10$ to $20$ with $\lambda$ from $0.005$ to $0.3$, so Adaptive Beam Search is among its configurations. The PiP grid crosses $ef_{\text{search}}$ from $10$ to $400$ with $\Delta$ from $3$ to $100$ and $p \in \{0.95, 0.85, 0.75\}$. All strategies run on the same indexes.

\begin{figure*}[t]
    \centering
    \includegraphics[width=\textwidth]{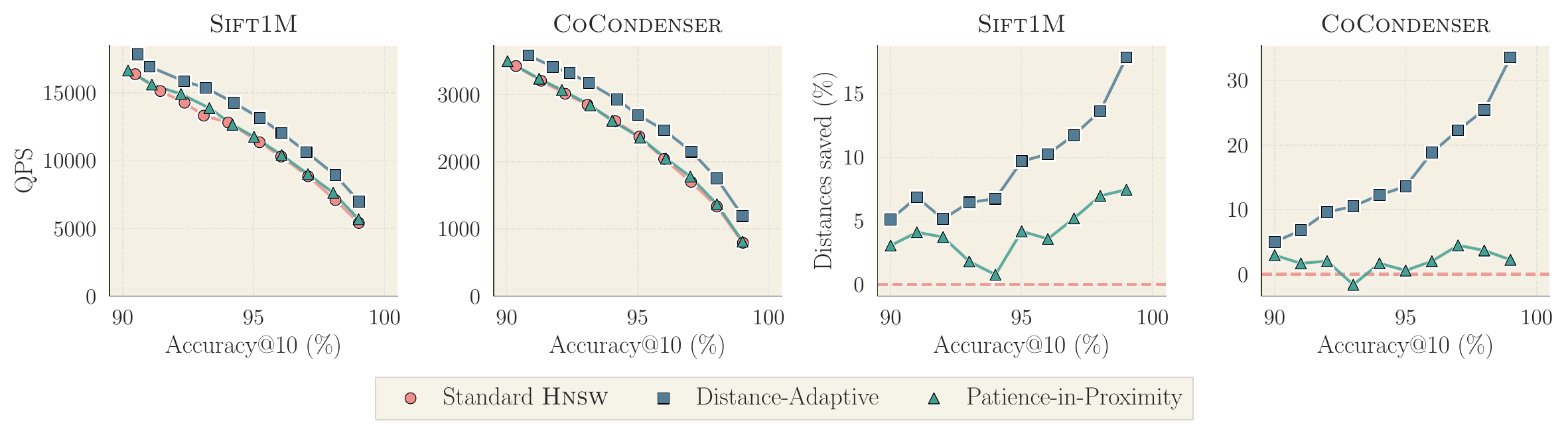}
    \caption{Early termination strategies on \siftom{} and \cocond, against standard \hnsw{} termination and Patience-in-Proximity.}
    \label{fig:et_conclusion}
\end{figure*}

Figure~\ref{fig:et_conclusion} reports QPS and the reduction in distance computations against standard \hnsw{} termination. Distance-adaptive termination is ahead on both representations and at every accuracy target. On \siftom{} it reduces distance computations by $5\%$ at $90\%$ Accuracy@$10$ and by $18\%$ at $99\%$. On the sparse \cocond{} the reduction goes from $5\%$ at $90\%$ to $34\%$ at $99\%$, where QPS rises by $50\%$. The advantage \emph{grows} with the accuracy target, and retrieval systems operate at high targets. The winning configurations keep $ef_{\text{search}}$ between $10$ and $17$, close to $k$, and the beam adds little. Restricting the grid to $ef_{\text{search}}=k$, that is, to Adaptive Beam Search, costs at most $5.4\%$ of the throughput on \siftom{} and $7.4\%$ on \cocond{} at any target. The single-parameter rule is therefore a sound default when one parameter is preferred. PiP is close to neutral. It saves at most $7.5\%$ of the distance computations on \siftom{} and $4.5\%$ on \cocond{}, and under $5\%$ at most targets. Its QPS stays within $8\%$ of standard \hnsw{} everywhere. At one target PiP computes more distances, $1.6\%$ more at $93\%$ on \cocond. This reflects the grid resolution. An unbounded patience reduces PiP to the standard criterion at the same $ef_{\text{search}}$, and the best configurations use a patience between $5$ and $30$, well inside the grid.

The difference follows from what each criterion can change. A fixed beam stops once $ef_{\text{search}}$ candidates closer than the frontier have been found, however many near-ties remain around the query. A distance threshold adapts to the query~\cite{ee_distance_adaptive}. It spends distance computations on hard queries with many near-ties and stops early on easy ones. A small beam with a relaxation thus reaches a given accuracy with fewer computations than the larger beam that the standard criterion needs. PiP can only stop a search \emph{earlier} than the standard criterion would at the same $ef_{\text{search}}$; it can never extend it. The result set stops changing at about the time the standard criterion fires. PiP therefore saves only the tail of each search and cannot lower the $ef_{\text{search}}$ that a target requires. At high accuracy, the last true neighbors are found in that tail. The patience has to grow to keep them, and PiP converges to the standard criterion.

\subsubsection{Graph Compression}
\label{sec:graph_compression}

The adjacency lists are the other half of the memory cost of a graph index (Section~\ref{sec:related_int_compression}): with the plain backend, every edge costs $4$ bytes and every node an $8$-byte offset. The compressed backend implements the scheme of~\cite{benocci2026lossless}, which was originally developed in \kannolo. After the build, the EGB permutation~\cite{egb} of the ground level relabels every level and the dataset, so graph-adjacent nodes receive close identifiers. Each sorted list is then stored as small gaps with \svb~\cite{streamvbyte}, and the per-node offsets with Elias--Fano~\cite{elias1974,CSUR21}. The permutation can also be applied to an uncompressed graph. It improves the locality of the neighbor accesses, so a compressed index can be as fast as an uncompressed, unpermuted one, or faster. Only the final top-$k$ results are mapped back to the original identifiers. A \svb{} block holds at most $256$ values, so $M$ is capped at $128$.

\subsection{\hnsw}
\label{sec:hnsw_impl}

\kannolo's \hnsw{} index~\cite{hnsw} is generic over the dataset and the graph storage. Every level is built in the mutable fixed-degree representation and frozen into the target graph type once complete.

The layered structure is stored as an array of graphs, from the topmost level down to the ground level with all $N$ nodes. A query starts from the single node of the top level, descends the upper levels with the greedy walk, and runs the beam search on the ground level. The upper levels work in local IDs, since the nodes of each are a prefix of the level below. A mapping array translates the last of them to the ground level, whose order may be permuted (Section~\ref{sec:graph_compression}).

\kannolo{} carries \hnsw's three standard parameters of Section~\ref{sec:related_ann_indexes}, with defaults $M=16$, $ef_\text{construction}=150$ and $ef_\text{search}=100$. The value of $ef_\text{search}$ is clamped to $\max(ef_\text{search},\,k)$ so that at least $k$ candidates are evaluated. A second search parameter selects the termination criterion: the standard \hnsw{} rule, which is the default, or the distance-adaptive rule of Section~\ref{sec:early_termination}, with its own $\lambda$.

\subsubsection{Index Construction}
\label{sec:hnsw_construction}

Standard implementations parallelize \hnsw{} construction by inserting nodes in parallel, assigning the next node to a free thread. They use locks on neighbor lists to make concurrent access and updates safe. The locks are a bottleneck on large collections, and they also make the resulting graph depend on thread timing. \kannolo{} follows the lock-free batched scheme of \parlayann~\cite{parlayann}, applied to each level of the hierarchy in turn. It departs from \parlayann{} in two places: the rule that sets the batch size, discussed at the end of this section, and the way the merge is made race-free. Each batch proceeds in two phases:
\begin{enumerate}
    \item \textbf{Phase 1 (parallel search).} Every node in the batch searches the current graph for its candidate neighbors. It then prunes them to the degree bound with the pruning heuristic of the original algorithm. The graph is not modified during this phase, so it needs no locks and scales across all cores.
    \item \textbf{Phase 2 (merge and commit).} The batch's edges are written. Forward edges go to slots that belong to the new nodes. Reverse edges, i.e. the links from existing nodes to the new ones, are first merged per target, as described below.
\end{enumerate}

\paragraph{Reverse-edge insertion.}
Within a batch, several new nodes often pick the same existing node $v$ as a neighbor, so several reverse edges arrive at $v$ together. The capacity of $v$'s neighbor list is fixed, so the edges cannot simply be appended. They have to be pruned against the current content of the list, and against each other. \kannolo{} therefore prunes per target: all the edges arriving at a node are considered together. The batch's reverse edges are collected as $(\textit{target},\,\textit{new node})$ pairs and sorted. Each target is then pruned once, with all of its incoming edges in the candidate set. The sort has two effects. First, it makes the merges independent, so they run in parallel over distinct targets. Second, it puts every arriving edge into the single pruning decision that concerns it, so the only edges dropped are those the heuristic rejects. Distances are computed only where the degree bound forces a choice.

With this procedure, adding reverse edges cannot cause a data race. Every write is deferred to a commit that runs afterwards over disjoint slots. The reverse targets of a batch are always already-inserted points, while its forward edges go to slots owned by the new points. The Rust borrow checker therefore verifies the absence of data races, and the build-time adjacency needs neither atomics nor interior mutability.

\paragraph{Reproducible construction.}
Determinism is a design goal of \kannolo, because prototyping and testing depend on it. Two things must not depend on thread timing. The first is the set of nodes each insertion sees, which the batch boundary already fixes. The second is the order of a target's candidate set during the merge. Sorting the reverse edges by target and then by arriving node makes this order canonical. With levels and insertion order drawn from a fixed seed, two builds of the same dataset with the same parameters produce byte-identical indexes on the same target.

\paragraph{Adaptive batch sizing.}
The batch size balances two costs. Small batches leave threads idle, because the time a vector takes to find its neighbors varies, much more on sparse collections. Large batches cost accuracy, because nodes in the same batch search a frozen graph and cannot discover each other. The damage depends on the ratio $\mathcal{B}/N$ between the batch size and the number of nodes already inserted. \parlayann{} doubles the batch until it reaches a single cap $\theta$, set empirically to $0.02n$. We keep the doubling and replace the cap with three rules:
\[
  \mathcal{B}(N) = \min\Big(\max\big(\rho N, \, B_{\text{floor}}\big), \; \rho_{\max} N\Big).
\]
The batch targets a fixed fraction $\rho$ of the graph, which bounds the part a batch cannot see. It never falls below a floor $B_\text{floor}$, stated in tasks per worker thread, so the first rounds already use every core. It never exceeds $\rho_{\max} N$, with $\rho_{\max}=0.05$, so the floor cannot make the batch too large while the graph is small.

We set $\rho$ on \siftom, whose run-to-run variation is narrow enough to resolve differences of this size. At the worst point of a sweep over $ef_\text{search}$, Accuracy@$10$ is unaffected at $\rho=0.005$, and falls by at most $0.09$ points at $\rho=0.01$ and $0.15$ at $\rho=0.02$. Build time falls as $\rho$ grows, with diminishing returns. On the $8.8$M collections the differences stay within run-to-run variation. We therefore take $\rho=0.005$, four times smaller than \parlayann's $\theta$.

\subsubsection{Validation against \hnswlib}
\label{sec:hnsw_validation}

We validate our \hnsw{} implementation against \hnswlib~\cite{hnswlib}, the reference implementation maintained by the algorithm's authors. We use the simplest configuration that both systems support: dense \texttt{f32}, uncompressed edges, no early termination, no reranking. Both sides use $M=32$ and $ef_\text{construction}=200$. \hnswlib's inner-product space returns $1-\langle q,v\rangle$, which is rank-equivalent to the dot product \kannolo{} scores with.

\hnsw's guarantees are average-case over a \emph{random} insertion order. Both the expected logarithmic search and the navigability argument assume that the inserted points arrive in random order. \hnswlib{} randomizes each point's level but inserts in the order the caller supplies, so it satisfies only half of that precondition. This makes no difference on a collection stored in random order, but it matters on one that is not, and text collections typically are not. On \dragon, which is stored by document, consecutive rows have mean cosine $0.478$ against $0.184$ for random pairs. On \siftom{}, consecutive rows are no closer than random ones.

\begin{figure}[t]
    \centering
    \includegraphics[width=\linewidth]{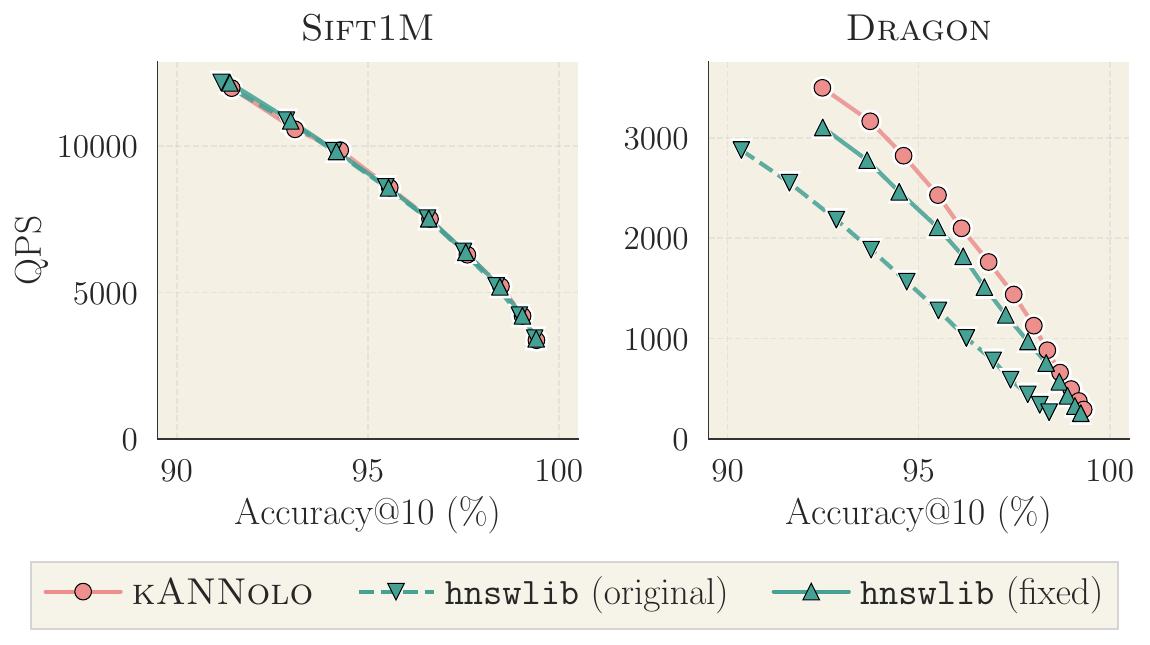}
    \caption{Accuracy@$10$ against throughput for \kannolo{} and \hnswlib, sweeping $ef_\text{search}$ at $k=10$. \emph{original} (dashed) inserts in file order, \emph{fixed} in permuted order.}
    \label{fig:v2_hnsw_frontier}
\end{figure}

We therefore report \hnswlib{} twice: in file order as shipped, and under a random permutation of the insertion sequence. \kannolo{} shuffles before assigning levels and is unaffected. On \dragon, permuting the insertion order alone adds $+1.98$ points of Accuracy@$10$ on average over the $ef_\text{search}$ ladder, from $+3.71$ at $ef_\text{search}=25$ to $+0.83$ at $520$ (Figure~\ref{fig:v2_hnsw_frontier}, dashed). On \siftom, which has no order to exploit, it adds $+0.12$. Measured in distance computations, the file order needs ${\approx}1.4\times$ as many as the permuted order for equal accuracy.

Against \hnswlib{} inserted in random order, the two implementations are indistinguishable on graph quality. Accuracy@$10$ differs by $+0.051$ points on average on \siftom{} and $+0.074$ on \dragon. Both differences are inside \hnswlib's own spread over three independent builds. At matched $ef_\text{search}$, the two perform the same number of distance computations to within $4\%$. On \siftom, \kannolo{} and \hnswlib{} in file order build in the same time, $25.56$ and $25.63$ seconds, and \hnswlib{} in permuted order takes $27.95$ (Table~\ref{tab:v2_hnsw}). On \dragon{} its build time matches that of \hnswlib{} once the insertion orders match. On \dragon, the non-permuted \hnswlib{} builds faster, in $1\,498$ seconds, but it produces a lower-quality graph. The two systems do not build the same graph, so this time is not directly comparable to \kannolo's. Throughput differs: the two are at parity at $d=128$ ($0.99\times$), and \kannolo{} is $1.15\times$ faster at $d=768$. This difference tracks the advantage of the batched kernel over sequential scoring and is not a property of the graph.

\kannolo{} stores \emph{more} edges in less memory, $289.7$M against $281.1$M on \dragon. \hnswlib{} reserves all $2M$ ground-level slots for every node, half of them unused at the measured $48\%$ occupancy, and keeps four auxiliary structures per element (${\approx}86$ bytes per vector). \kannolo's adjacency is variable-length, and its auxiliary arrays are sized to the upper levels. On resident memory, \kannolo{} is ahead by $1.37\times$ at $d=128$ and $1.07\times$ at $d=768$.

\begin{table}[t]
\centering
\small
\caption{Validation of \kannolo's \hnsw{} against \hnswlib~\cite{hnswlib}; conventions of Section~\ref{sec:methodology}. \emph{original} inserts in file order, \emph{fixed} in permuted order. Footprint is peak resident memory.}
\label{tab:v2_hnsw}
\begin{tabular}{@{}llrr@{}}
\toprule
\textbf{Collection} & \textbf{System} & \textbf{Build (s)} & \textbf{Bytes/vec} \\
\midrule
\siftom              & \kannolo                     & \textbf{25.6} & \textbf{653.8} \\
($d{=}128$, $\ell_2$) & \hnswlib{} (original)       & \textbf{25.6} & 895.1 \\
                     & \hnswlib{} (fixed)           & 28.0 & 894.9 \\
\midrule
\dragon              & \kannolo                     & 1\,710 & \textbf{3\,214.6} \\
($d{=}768$, IP)      & \hnswlib{} (original)        & \textbf{1\,498} & 3\,444.7 \\
                     & \hnswlib{} (fixed)           & 1\,726 & 3\,444.8 \\
\bottomrule
\end{tabular}
\end{table}

\subsection{Inverted Vector File Index}
\label{sec:ivf_impl}
\label{sec:ivf_sparse}

An IVF index (Section~\ref{sec:related_ivf}) is at its core a pair: a \emph{centroid index} that routes a query to a handful of clusters, and a \emph{quantizer} that encodes the vectors inside them. Around that pair the design space is large. The centroid index can be flat or graph-based. The cluster vectors can be stored at full precision or compressed. The cluster contents can be either the original vectors or centroid-relative residuals, under either metric. \faiss~\cite{douze2024faiss} exposes this space through a large family of classes (Section~\ref{sec:related_libraries}), one per combination. \kannolo{} represents every combination as one generic structure, \texttt{IVF<DQ,\,D,\,CI>}. The structure factors the space into three orthogonal axes, so a single search method and a single build method cover all of it, for every collection k-means can cluster. Like \hnsw, the index names no representation.

The first parameter, \texttt{DQ}, is the centroid and query dataset. Its encoder fixes the metric the index \emph{reports}. The second, \texttt{D}, is the cluster dataset, which holds every vector in the collection. The vectors are reordered so that each cluster occupies a contiguous range, delimited by a cluster-offsets array. This CSR layout turns a cluster scan into sequential memory access. The third, \texttt{CI}, is the centroid index. It can be a flat index, an \hnsw{} or even an \texttt{IVF} index. A second \texttt{IVF} gives a hierarchical, multi-level coarse quantizer through type composition alone.

Both reported metrics decompose in the same form, whatever the centroid index. Let $c$ be the centroid of the cluster being scanned and $\hat r$ the residual stored for a candidate, so that the candidate reconstructs as $\hat v = c + \hat r$. The inner-product decomposition is straightforward: $\langle q, \hat v\rangle = \langle q, c\rangle + \langle q, \hat r\rangle$. The squared-Euclidean one is slightly more involved:
\begin{equation}
\label{eq:ivf_decomposition}
\begin{aligned}
  \lVert q - \hat v \rVert^2 \ &= \ \lVert (q - c) - \hat r \rVert^2
  \\ &= \ \lVert q - c \rVert^2 + \lVert \hat r \rVert^2 - 2\langle q-c, \hat r\rangle
  \\ &= \; \underbrace{\lVert q - c \rVert^2}_{\text{centroid}} + \ \underbrace{\lVert \hat r \rVert^2 + 2\langle c, \hat r\rangle}_{\text{offline correction}} \ - \underbrace{2\langle q, \hat r\rangle}_{\text{candidate}}.
\end{aligned}
\end{equation}
The two decompositions have the same structure if inner product is given a zero correction term. The centroid contribution is the query's score against the centroid under the reported metric, and the centroid search has already produced it. The offline correction depends on the stored vector alone and is precomputed once per vector at build time. Under either metric, the only term that depends on both the query and the candidate is the dot product $\langle q, \hat r\rangle$. The cluster dataset therefore never evaluates the reported metric, and \kannolo{} stores it in dot-product form throughout. Plain storage (no residuals) uses the same formula with $c = 0$ and $\hat r = \hat v$. The leading term is then the query-constant $\lVert q\rVert^2$, and \kannolo{} does not compute it. It shifts every candidate equally and cannot change a ranking. Adding it once per query recovers the exact distance where needed. The metric the index reports fixes, at compile time, the sign of the dot product ($-2$ under squared-Euclidean distance and $+1$ under inner product) and whether a correction is present.

The cluster encoder itself computes the offline correction, as $\lVert \hat r + c\rVert^2 - \lVert c\rVert^2$. For scalar and PQ storage this equals the correction of Equation~\eqref{eq:ivf_decomposition}, with both terms acting on the decoded vector. For \rabitq{} and \rabitqext{} it is the encoder's own estimate. On dense data, clusters can thus be stored as \texttt{f32}, any scalar type, PQ, \rabitq{} or \rabitqext{} codes, plain or as residuals, under either metric. Residuals pay off on dense data because the centroid already carries part of each vector. The residual can therefore be compressed more strongly at a smaller cost in accuracy. On sparse data the effect is reversed. A residual $v - c$ has the nonzeros of both $v$ and $c$, so it takes more memory than the vector it replaces. Sparse clusters therefore store the vectors themselves.

\paragraph{The single-table residual decomposition.}
Under residual encoding, the candidate contribution $\langle q, \hat r\rangle$ does not depend on the cluster. For PQ storage, one $m \times 256$ table per query therefore serves every probed cluster. For any \emph{stored} vector, the offline correction is a single number. \kannolo{} and \faiss{} differ in when they materialize it.

\faiss{} adopts the same decomposition. It enables the \texttt{use\_\allowbreak precomputed\_\allowbreak table} mode whenever the table fits under its default $2$\,GB cap, as in our benchmarks. In this mode it builds the query table once and precomputes the correction during training, before any vector has been assigned to a cluster. Its inverted lists have no room for a per-vector scalar, so the correction must cover every (cluster, code) pair, as an $\text{nlist} \times m \times 256$ array. Reading that array costs an $O(m \cdot 256)$ combine on every probed cluster, linear in \texttt{nprobe}.

\kannolo{} assigns every vector's cluster and code in one batch, before the index is assembled, so the correction does not have to stay generic. At build time it becomes a single stored \texttt{f32} scalar per dataset vector. The encoder computes it from that vector's own code, so it stays consistent with the $\langle q, \hat r\rangle$ that the search evaluates. Mixing the exact $\lVert r\rVert^2$ with the approximate $\langle q, \hat r\rangle$ would introduce a $2\langle q, r - \hat r\rangle$ ranking error. The query-side table is built once per query and reused across every probed cluster. Since the correction is already stored per vector, nothing is left to combine per cluster.

At \texttt{nprobe}\,=\,512 with $m=64$, \faiss's combine alone performs ${\approx}8.4$M operations, against ${\approx}2$M multiply--adds for scoring the ${\approx}31$K candidates retrieved, which explains the growing advantage in Section~\ref{sec:ivf_validation}.

\paragraph{Construction.}
Building an \texttt{IVF} index runs \vectorium's k-means, optionally trained on a random subsample of the collection. It then assigns every vector to its nearest centroid and reorders the dataset into the CSR layout above. The centroid encoder fixes the clustering metric. On sparse data, an optional refinement prunes the centroids during training. Centroid \emph{assignment}, both during the k-means iterations and in the final pass, can switch from an exhaustive scan to an approximate search through \hnsw{}. This pays off when the cluster count makes scanning tens of thousands of centroids per point the dominant cost. Which of the two is cheaper depends on the cluster count, the dimensionality and the collection size together, so the choice is a build parameter. The convenience entry point uses the \hnsw{} path by default at $32\,768$ clusters and above. On sparse data, IVF is far slower than \kannolo's \hnsw: on the first $1$M passages of \cocond, the best sparse IVF we built reaches $92.1\%$ Accuracy@$10$ at $75$ QPS, against $93.4\%$ at $2\,053$ QPS for \hnsw.

\subsubsection{Validation against \faiss}
\label{sec:ivf_validation}

\begin{figure*}[t]
    \centering
    \includegraphics[width=\textwidth]{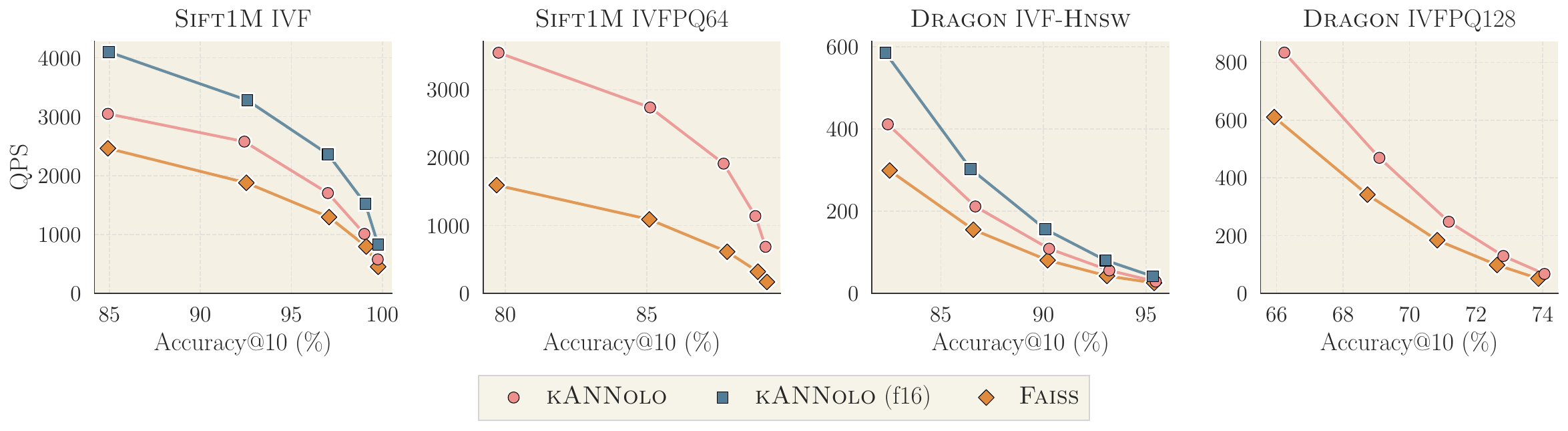}
    \caption{Accuracy@$10$ against QPS for \kannolo{} and \faiss{} under matched parameters, single-query search, on the four configurations of \siftom{} and \dragon.}
    \label{fig:ivf_recall_qps_grid}
\end{figure*}

We validated \kannolo's IVF against \faiss~\cite{douze2024faiss} to check that the generic, single-code-path design has no correctness or performance penalty. Conditions were matched: identical hardware, k-means and index parameters, accuracy targets, and single-query, single-threaded search timing on both systems. Four configurations span both metrics, both cluster-storage regimes and both kinds of centroid index. On \siftom{} (squared-Euclidean, flat centroid index) we test IVFFlat, with \texttt{f32} and \texttt{f16} cluster storage, and IVFPQ with residuals. On \dragon{} (inner product, \hnsw{} centroid index at $65\,536$ clusters) we test IVF and IVFPQ with residuals. Figure~\ref{fig:ivf_recall_qps_grid} reports accuracy against QPS for all four.

Accuracy matches \faiss{} within $0.3$ percentage points at every operating point across all four configurations. Under the same conditions, \kannolo's single-query latency is at or below \faiss's. On \siftom, \texttt{f32} cluster storage is $1.24$--$1.37\times$ faster and \texttt{f16} storage $1.6$--$1.9\times$ faster. The margin grows with \texttt{nprobe} as the scan becomes memory-bandwidth-bound. In the residual-PQ configuration the gap widens further, from $2.0\times$ to $4.1\times$ as \texttt{nprobe} grows. Here the single-table decomposition above removes \faiss's per-cluster combine. On the inner-product \dragon{} configurations, \faiss's own residual formulation is already single-table, and the gap is a flatter $1.3$--$1.5\times$. Our decomposition offers no structural advantage on that metric, as the algebra above predicts.

\subsubsection{Validation against the \rabitq{} reference}
\label{sec:ivf_rabitq_validation}

The authors of \rabitq{} provide a reference implementation of IVF over \rabitq{} codes: the IVF index of their library. We compare it with \kannolo's IVF on one collection per metric, at $4$ bits per component. We use \siftom{} under squared-Euclidean distance with $4\,096$ clusters, and \snowf{} under inner product with $16\,384$ clusters. Both sides store centroid-relative codes. The reference takes its partition from outside the index. We therefore give it a \kannolo{} k-means partition with the settings of \kannolo's own build: $10$ iterations, and on \snowf{} spherical k-means trained on a $4.2$M-vector sample. Search runs one query at a time on one core, with $k=10$, following the conventions of Section~\ref{sec:methodology}. The reference library routes with exhaustively scanned centroids below $20\,000$ clusters, and with an \hnsw{} over the centroids above. Both cluster counts here therefore use its flat routing. We compare it first with \kannolo's flat centroid index, and also report \kannolo{} with an \hnsw{} centroid index.

Table~\ref{tab:ivf_rabitq} reports the results. \emph{Accuracy matches}. At every \texttt{nprobe}, the two flat-centroid indexes differ by at most $0.35$ points on \siftom{} and $0.18$ on \snowf, and by at most $0.14$ points on \siftom{} at $8$ bits. \emph{Throughput} is split. The split depends on where each design spends the time of a query: routing it to the clusters, or scanning their candidates.

Before the reference reads a candidate's full code, it scores the candidate with a $1$-bit fast-scan estimate. It discards the candidate if the error bound of this estimate rules it out of the top $k$. Only the remaining candidates need a multi-bit estimate. \kannolo{} has no such first pass and scores every candidate with the full code. Section~\ref{sec:rabitq} notes the same difference for the flat scan. The cost per candidate of \kannolo{} is therefore higher, and the reference is faster as soon as the candidates dominate the query time. On \siftom{} this happens from \texttt{nprobe}\,=\,$16$, with a lead of $1.9\times$ at $90\%$ Accuracy@$10$. On \snowf{} it happens from \texttt{nprobe}\,=\,$128$, with a lead of up to $1.9\times$ at the highest \texttt{nprobe}.

The reference scans its centroids in \texttt{f32}, while \kannolo{} stores them in \texttt{f16}, which halves the bytes read. On \siftom{} the centroids are small, and this shows only at \texttt{nprobe}\,=\,$8$, where \kannolo{} is $1.12\times$ faster. On \snowf, $16\,384$ centroids of dimension $1\,024$ give the reference a minimum cost of about $4.4$\,ms per query, and \kannolo{} is $1.6\times$ faster at \texttt{nprobe}\,=\,$16$. Replacing the flat centroid index with an \hnsw{} changes one type parameter and leaves the rest of the index unchanged. It removes most of that minimum cost. \kannolo{} is then $5\times$ faster than the reference at \texttt{nprobe}\,=\,$16$ and $1.6\times$ at $64$, at a loss of $1.1$ and $0.75$ points of accuracy. At \texttt{nprobe}\,=\,$128$ the scan dominates again. \kannolo's index files are $12\%$ smaller on \siftom{} and $4\%$ on \snowf.

\begin{table}[t]
\centering
\small
\setlength{\tabcolsep}{3.5pt}
\caption{IVF over $4$-bit \rabitqext{} residual codes: \kannolo{} against the IVF of the \rabitq{} authors' library (\emph{ref.}), single-query search, Accuracy@$10$ (\%) and QPS. \emph{flat} and \emph{\hnsw} name \kannolo's centroid index; the reference scans its centroids exhaustively at these cluster counts.}
\label{tab:ivf_rabitq}
\begin{tabular}{@{}r rrr rrr@{}}
\toprule
& \multicolumn{3}{c}{\textbf{Accuracy@10 (\%)}} & \multicolumn{3}{c}{\textbf{QPS}} \\
\cmidrule(lr){2-4}\cmidrule(lr){5-7}
\texttt{nprobe} & ref. & flat & \hnsw & ref. & flat & \hnsw \\
\midrule
\multicolumn{7}{@{}l}{\siftom{} ($\ell_2$, $4\,096$ clusters)} \\
$8$   & 69.8 & 70.2 & --- & 10\,008 & \textbf{11\,236} & --- \\
$16$  & 80.3 & 80.6 & --- & \textbf{8\,732} & 7\,874 & --- \\
$32$  & 86.8 & 87.0 & --- & \textbf{7\,119} & 4\,926 & --- \\
$64$  & 90.0 & 90.0 & --- & \textbf{5\,343} & 2\,841 & --- \\
$128$ & 91.1 & 91.2 & --- & \textbf{3\,708} & 1\,565 & --- \\
\midrule
\multicolumn{7}{@{}l}{\snowf{} (IP, $16\,384$ clusters)} \\
$16$  & 76.7 & 76.7 & 75.6 & 222 & 362 & \textbf{1\,104} \\
$32$  & 81.5 & 81.4 & 80.1 & 214 & 288 & \textbf{633} \\
$64$  & 85.4 & 85.3 & 84.7 & 201 & 207 & \textbf{326} \\
$128$ & 88.6 & 88.5 & 88.1 & \textbf{179} & 135 & 167 \\
$256$ & 91.2 & 91.1 & 90.9 & \textbf{148} & 77 & 89 \\
\bottomrule
\end{tabular}
\end{table}

\section{Experiments}
\label{sec:exp}

This section compares \kannolo{} end to end against state-of-the-art systems on dense and sparse collections.

\subsection{Experimental Setting}
\label{sec:exp_setting}

For the dense domain, the comparison runs inside \vibe~\cite{vibe}, a benchmark for vector indexes. \vibe{} fixes the collections, the ground truth, the parameter grids of every system it ships, and the timing loop. We run \vibe{} at commit \texttt{1b60239} with four patches, each applied to all systems in the same way.

The first patch concerns build parallelism. \vibe{} pins construction to one core, while our builds use $64$ threads. Parallel builds make this volume of benchmarks feasible. They also measure construction under realistic conditions: indexes are built in parallel in practice, and making construction scale with the number of available threads is part of designing an index. The other three patches concern measurement conventions and are described in Appendix~\ref{app:harness}.

For the sparse domain there is no equivalent of \vibe, so we use our own harness with the same conventions: construction on $64$ threads under a per-collection budget, and search on a single core. Two sparse competitors, \lsp{} and \sindi{}, can only be searched through their command-line tools. We therefore time every sparse system in the same way, as one search process that runs over the whole query set (Appendix~\ref{app:sparse_timing}). Thus, dense and sparse systems cannot be compared with each other.

\subsubsection{Collections}
\label{sec:exp_collections}

Table~\ref{tab:exp_collections} lists the four dense and three sparse collections. None of them was used in the design studies of Sections~\ref{sec:data_management} and~\ref{sec:indexing}. Three of them use the \msmarcovo{} corpus of the development collections, but with a different encoder.

On the dense side, \agnews~\cite{agnews} is a collection of news articles, embedded with mxbai-embed-large~\cite{mxbai}. \msmarcoqwen{} is the passage corpus of \msmarcovo~\cite{msmarco}, drawn from Bing search results and embedded with Qwen3-Embedding-0.6B~\cite{qwen3emb}. \dpr{} is the Wikipedia passage collection of Dense Passage Retrieval~\cite{dpr}, embedded with jina-embeddings-v5-text-nano~\cite{jinav5}. On these three, \vibe{} holds out $1\,000$ random documents of the corpus as queries. \laion{} contains the CLIP~\cite{clip} embeddings of one million images from LAION-400M~\cite{laion400m}. It is searched with $1\,000$ embeddings of image captions, so its queries are out-of-distribution. The set covers four encoder families and $0.77$M to $21$M points. \msmarcoqwen{} and \dpr{} are the two largest collections that \vibe{} ships. They are closer to the scale of modern applications. They are also a challenge for some competitors: two of the four published degree settings of \symphonyqg{} cannot be built on this machine at $21$M points, and a third exceeds the build budget (Section~\ref{sec:exp_symphonyqg}).

On the sparse side, the vectors are learned term weights over a vocabulary, and queries are real user queries. \msmarcosplade{} encodes the \msmarcovo{} passages with \spladeT~\cite{lassance2024splade}. \msmarcoesplade{} encodes them with \esplade~\cite{lassance2022efficiency}, an efficiency-oriented \splade{} model whose query encoder is trained to produce queries with few nonzero entries.\footnote{\raggedright The V-large checkpoints, \url{https://huggingface.co/naver/efficient-splade-V-large-doc} and \url{https://huggingface.co/naver/efficient-splade-V-large-query}.} \nq{} is the BEIR~\cite{beir} version of Natural Questions~\cite{nq}: Wikipedia passages searched with Google queries, encoded with \splade\,\cocondo~\cite{splade,cocondenser}. \msmarcoesplade{} is the hardest case for a graph index because its queries are much shorter than its documents, making the search similar to an out-of-distribution task.

Every collection is searched with all of its queries, against a ground truth we verified with an exhaustive scan.

\begin{table*}[t]
\centering
\footnotesize
\setlength{\tabcolsep}{4pt}
\caption{Evaluation collections. nnz is the average number of nonzero entries. Size is the collection in raw format: \texttt{f32} vectors for dense data, a $32$-bit component and an \texttt{f32} value per nonzero entry for sparse data. $B$ is the per-collection construction budget. $^{\dagger}$ ships a learn set drawn from the query distribution.}
\label{tab:exp_collections}
\begin{tabular}{@{}lllrcrrrrrr@{}}
\toprule
\textbf{Collection} & \textbf{Encoder} & \textbf{Type} & $d$ & \textbf{Metric} & $n$ & \textbf{Queries}
  & \multicolumn{2}{c}{\textbf{nnz}} & \textbf{Size} (GiB) & $B$ (s) \\
\cmidrule(lr){8-9}
 & & & & & & & doc & query & & \\
\midrule
\agnews            & mxbai    & dense  & $1\,024$ & $\ell_2$ & $769\,382$     & $1\,000$ & --- & --- & $2.9$  & $300$ \\
\msmarcovo         & Qwen3    & dense  & $1\,024$ & IP       & $8\,840\,823$  & $1\,000$ & --- & --- & $33.7$ & $2\,400$ \\
\dpr               & Jina     & dense  & $768$  & IP       & $20\,969\,760$ & $1\,000$ & --- & --- & $60.0$ & $5\,400$ \\
\laion$^{\dagger}$ & CLIP     & dense  & $512$  & IP       & $1\,000\,448$  & $1\,000$ & --- & --- & $1.9$  & $300$ \\
\midrule
\msmarcovo         & \spladeT & sparse & ---    & IP       & $8\,841\,823$  & $6\,980$ & $169.8$ & $24.5$ & $11.2$ & $5\,400$ \\
\nq                & \cocondo & sparse & ---    & IP       & $2\,680\,893$  & $3\,452$ & $153.6$ & $50.7$ & $3.1$  & $1\,800$ \\
\msmarcovo         & \esplade & sparse & ---    & IP       & $8\,841\,823$  & $6\,980$ & $182.8$ & $5.9$  & $12.0$ & $6\,000$ \\
\bottomrule
\end{tabular}
\end{table*}

\subsubsection{Competitors}
\label{sec:exp_systems}

The dense competitors cover the main designs of modern ANN indexes. Each design is represented by the strongest implementation we know of. \hnswlib~\cite{hnsw,hnswlib} is the reference implementation of \hnsw, over uncompressed \texttt{f32} vectors. \glass~\cite{pyglass} has the same design as our \kannolorerank: a graph searched over compressed vectors, with a rescoring stage on top. It is also the broadest library here, with four graph algorithms and fourteen quantizers, any of which can serve either stage. \svsleanvec~\cite{leanvec} combines \vamana~\cite{vamana} with dimensionality reduction. It uses LVQ~\cite{lvq} codes in the first stage and a wider secondary dataset for rescoring. \rabitqhnsw{} is the reference implementation of \rabitq~\cite{rabitq,rabitq_ext}, which \kannolo{} also uses in its first stage. Unlike our implementation, it stores additional metadata with each code to drive early exit. \symphonyqg~\cite{symphonyqg} fuses the \rabitq{} codes into the adjacency lists of the graph. \lorann~\cite{lorann} builds an IVF~\cite{ivf_gpu} index with a reduced-rank regression per cluster. It is the only system that can be trained on the query distribution, from a learn set, and only \laion{} ships one. \faiss{} is not among the competitors. Its IVF indexes are compared with ours in Section~\ref{sec:ivf_validation}, and its \hnsw{} index was already slower than the first version of \kannolo~\cite{kannolo}.

We benchmark two configurations of \kannolo{}. \kannolohnsw{} is our \hnsw index built on \texttt{f16} representations. \kannolorerank{} is a two-stage index: a graph over quantized vectors retrieves a set of candidates, which are then reranked with a higher-precision representation. We use \kannolo{}~0.15.1 with \vectorium{}~0.11.1, \hnswlib~0.8.0, \glass~2.1.0, \symphonyqg~0.1.0, \svsleanvec~0.4.0, \rabitqhnsw~0.1.1 and \lorann~0.4.6. \glass, \symphonyqg{} and \rabitqhnsw{} are built from source at the commits pinned by \vibe. 

On sparse data, we compare against four inverted indexes: \seismic~\cite{seismic} and \seismicw~\cite{seismic_knn} (both at $0.6.0$), \sindi~\cite{li2025sindi} (commit \texttt{4d22e55}) and \lsp~\cite{dsbp,carlson2026lsp} (\texttt{fa0ede5}). We also compare against one graph index, \pyann~\cite{pyann} (\texttt{cbb5c81}). Our group develops \seismic{} and \seismicw{} on top of \vectorium, so both also exercise the performance of \vectorium{} under different indexing strategies. All the systems built on \vectorium{} store sparse collections with the blocked encoder (Section~\ref{sec:blocked_sparse}).

Our two sparse configurations mirror the dense ones. \kannolohnsw{} is an \hnsw{} graph over the collection. The collection is stored with \texttt{f16} values and \texttt{u16} components. \kannolorerank{} splits each vector into two parts: the first part holds its largest entries, up to $a\%$ of its $\ell_1$ norm, and the residual holds the remaining ones. The graph is built over the first part. The residual rescores the candidates through the residual mode of \rerankindex{} (Section~\ref{sec:system_overview}), and the two partial scores sum to the exact inner product.

\subsubsection{Measures}
\label{sec:exp_measures}

Accuracy@$k$, throughput and build time are as defined in Section~\ref{sec:measures}, with $k = 10$ throughout. A query with no results has zero accuracy. The following conventions differ from those of the design studies.

\paragraph{Footprint.} Footprint is the increase in the unique set size (USS) of the search process across index construction: the private memory the process holds after the build minus the memory it held before. This measure charges an index for what it allocates. For every system here but one, this is the system's own representation of the collection. The exception is \lorann{}, as it does not copy the collection from \vibe, and it re-ranks its candidates against the array of the harness. The difference therefore does not include the memory its queries read. We add the raw \texttt{f32} collection to its footprint, since any deployment would have to hold it. On sparse data, indexes are built and searched in separate processes, so for sparse data we report the resident memory of the search process. Both measures differ from those of the design studies (Section~\ref{sec:measures}), so footprints are compared only within a table or figure.

\paragraph{Construction time.} On dense data, construction time is the time \vibe{} measures to build an index from the in-memory collection. On sparse data, it excludes the conversion of the collection into the input format of each library. In both cases it includes any processing of the data that an index relies on, such as encoding, pruning or document reordering.

\paragraph{Throughput per unit of memory.} This measure is QPS per GiB of index. A system can gain throughput by using more memory, and Section~\ref{sec:exp_results} shows that the memory use of the systems here differs by up to an order of magnitude.

\paragraph{Repetition.} Every search configuration is timed over two passes of the query set, and we report the higher QPS of the two. Where a warm-up pass was run, it was discarded. The median spread between the two passes is $0.11$--$0.19\%$ per collection, except on \msmarcoesplade{} ($0.97\%$). Our graph builds are deterministic (Section~\ref{sec:hnsw_construction}), and rebuilds reproduce throughput within $1.0\%$. We therefore treat smaller differences as parity.

\subsection{Methodology}
\label{sec:exp_methodology}

For each system and accuracy target, we report one configuration, chosen from a parameter grid within a construction budget.

\subsubsection{Parameter Grids}
\label{sec:exp_grids}

For dense benchmarks, grids come from \vibe{} verbatim. Where the shipped grid omits the documented default of a library, we add that default, which can only improve the results of the competitors. Grids do not change across collections. The grid of \kannolohnsw{} is $M \in \{8, 16, 32\}$, $ef_\text{construction} \in \{200, 500\}$, and graph representation in $\{ \text{\texttt{standard}}, \text{\texttt{permuted}} \} $. The grid of \kannolorerank{} has $M \in \{16, 32, 64\}$, the same values of $ef_\text{construction}$ and the \texttt{permuted} representation (Section~\ref{sec:exp_graph_repr}), and sweeps the first-stage encoder over $1$-bit \rabitq{} and \rabitqext{} at $2$ and $4$ bits. This gives $18$ builds, against $36$ for \rabitqhnsw, the largest competitor grid. It is searched over the number of candidates, the beam width, $\alpha$ and $\lambda$, and with one-bit \rabitq{} also over the query width.

The search ladders of \vibe{} are designed for $k = 100$, and we run them at $k = 10$. They do not limit the competitors. No selected configuration of a competitor is at an end of its search ladder, with two exceptions, both of \lorann{}: at $90\%$ on \agnews{} it uses its smallest reranking depth, and at $99\%$ on \laion{} its largest number of probed clusters.

On sparse data there is no shared benchmark, and only \seismic{} comes with configurations released by its authors for each collection, which we use unchanged. We define the grids of the other competitors. To avoid weakening a system because we know it less well than its authors, we start from the settings they recommend and extend an axis on a collection wherever a selected configuration is at its end. Axes defined by the authors of a system are reported at their limits. The sparse grids therefore differ across collections. For \sindi{} we sweep its two pruning parameters and its search parameter around the configuration recommended in its paper. For \lsp{}, which provides no grid, we sweep around the settings of its documentation, with $8$-bit weights as its authors recommend. For \seismicw{} we reuse the construction settings of \seismic{} and sweep the number of $\kappa$-NN neighbors expanded at search time. \pyann{} ships no construction grid, and its published search points reach at most $50\%$ accuracy on \nq. We build it with our \hnsw{} construction parameters and search it over a wider beam, with which it reaches $99\%$ on \nq, $96.52\%$ on \msmarcosplade{} and $90.92\%$ on \msmarcoesplade. \kannolohnsw{} is built with the \texttt{standard} representation, with $M \in \{16, 32, 64\}$ and $ef_\text{construction} \in \{200, 500, 1\,000, 2\,000\}$. Not every combination is built on every collection. It is searched over the beam width and $\lambda$. \kannolorerank{} is built with $M \in \{16, 32\}$, $ef_\text{construction} \in \{1\,000, 2\,000\}$ and pruning fraction $a \in \{60, 70, 80\}$. It is searched over the number of candidates, the beam width and $\lambda$. The same extension rule applies to the search axes of our two configurations.

The full grid of every system on every collection is listed in the supplementary material, and code, results and sparse collections are publicly available.\footnote{\url{https://github.com/TusKANNy/papers_reproducibility/tree/main/information-systems-vectorium-kannolo}}

\subsubsection{Build Budget}
\label{sec:exp_budget}

\vibe{} bounds every configuration with the same time limit of $24$ hours, and builds every index on a single thread~\cite{vibe}. Our builds use $64$ threads, and our collections range from $0.77$M to $21$M points. We therefore replace this limit with a construction budget $B$ per collection, scaled with its size (Table~\ref{tab:exp_collections}). Every system has the same budget, and a configuration whose build exceeds $B$ is excluded from the comparison. The budgets reflect the build time of graph indexes. On sparse data we derive $B$ from the dense budgets, since a graph builds in similar time at the same construction parameters, and scale it with the number of nonzero entries per vector, which sets the cost of a distance computation. Inverted indexes build much faster by design (Section~\ref{sec:exp_sparse}). To measure the effect of the budget, Appendix~\ref{app:budget} re-selects every reported dense configuration at $B/2$, $2B$ and $4B$.

For each system and each accuracy target $t \in \{90, 95, 99\}$, Tables~\ref{tab:exp_results_dense} and~\ref{tab:exp_results_sparse} in Appendix~\ref{app:results} report the configuration maximizing QPS subject to Accuracy@$k \ge t$ over that system's whole grid. The configuration behind every cell is listed in the supplementary material.

\subsection{Experimental Results}
\label{sec:exp_results}

We compare \kannolo{} with the competitors on dense data, then on sparse data, and conclude with ablations.

\subsubsection{Dense Comparisons}
\label{sec:exp_comparisons}

\paragraph{Search.}

\begin{figure*}[t]
\centering
\includegraphics[width=\textwidth]{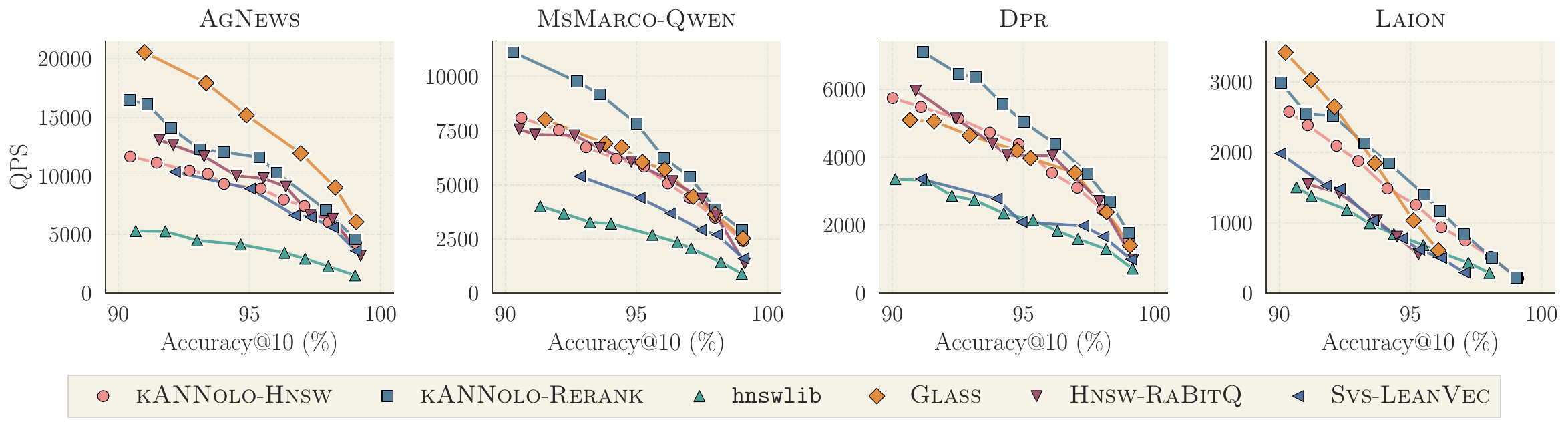}
\caption{Accuracy--throughput frontiers of the graph systems on the four dense collections. Each
point is a system's fastest configuration at the accuracy on its axis.}
\label{fig:exp_frontier_dense}
\end{figure*}

Figure~\ref{fig:exp_frontier_dense} draws the frontiers of the graph systems on the four collections. It leaves out \lorann, the only system that does not build a graph, and \symphonyqg, the only one whose index is several times larger than the collection. Table~\ref{tab:exp_results_dense} reports every system at the three accuracy targets.

Which system leads depends on the size of the collection. Below $1.1$M points the field is tight and we do not always lead. On \agnews{}, \glass{} is faster than \kannolorerank{} at all three targets. At $99\%$ it reaches $6\,085$ QPS against our $4\,585$ ($1.33\times$ more), with $2.0\times$ the footprint. \laion{} is the harder of the two small collections because its queries are out-of-distribution. Only our two configurations and \lorann{} reach $99\%$. The other five systems stop between $95.84$ and $98.74\%$ over their whole grids. We lead at $95$ and $99\%$, with $1\,398$ and $220$ QPS. \glass{} leads at $90\%$, with $3\,417$ QPS against our $2\,987$.

At $8.8$M and $21.0$M points, the systems separate clearly and we lead. Excluding \symphonyqg, \kannolorerank{} is the fastest system at every target on both collections. On \msmarcoqwen{} it reaches $11\,114$, $7\,837$ and $2\,896$ QPS, ahead of the next system by $1.39$, $1.29$ and $1.15\times$. On \dpr{} it reaches $7\,095$, $5\,039$ and $1\,776$ QPS, ahead by $1.19$, $1.25$ and $1.27\times$.

Against \hnswlib{} at $99\%$, \kannolohnsw{} reaches $2.84\times$ its throughput on \agnews, $2.70\times$ on \msmarcoqwen{} and $2.17\times$ on \dpr. \kannolorerank{} reaches $3.04$, $3.26$ and $2.47\times$. In every case our indexes are $1.9$ to $2.2\times$ smaller. On \laion, \hnswlib{} never reaches $99\%$ and tops out at $98.74\%$. The gap combines several factors. On \texttt{f32} data, our \hnsw{} is on par with \hnswlib{} on \siftom{} and faster on \dragon{} (Section~\ref{sec:hnsw_validation}). Here it also stores the vectors in \texttt{f16}, and it uses distance-adaptive termination and the graph permutation. Table~\ref{tab:exp_hnswlib_gap} adds these ingredients one at a time on \agnews{} and \msmarcoqwen, timing one query at a time as in \vibe. On \agnews, the fastest measured configuration of the last row that reaches $99\%$ runs at $4\,405$ QPS, within $3\%$ of the $4\,277$ QPS that \vibe{} measures for the same configuration. Inserting the points in random order makes no measurable difference for \hnswlib{} on these collections. At equal $ef_\text{search}$, the two builds have the same throughput within $1\%$. Their accuracy differs by at most $0.41$ points on \agnews{} and $1.06$ on \msmarcoqwen, with no consistent sign, so the $0.89\times$ on \msmarcoqwen{} is variation between builds. In total, the ingredients give $2.96$ and $3.53\times$ at these build parameters, against $2.84$ and $2.70\times$ with each system at its fastest configuration.

\begin{table}[t]
\centering
\small
\setlength{\tabcolsep}{4pt}
\caption{Throughput at Accuracy@$10 = 99\%$ as the ingredients of \kannolohnsw{} are added one at a time, every index built with $M=16$ and $ef_\text{construction}=200$. Each value is interpolated between the two configurations of its row that bracket $99\%$ on the accuracy--throughput frontier. \emph{Step} is the ratio to the row above, and for the random-order row the ratio to file order.}
\label{tab:exp_hnswlib_gap}
\begin{tabular}{@{}lrrrr@{}}
\toprule
 & \multicolumn{2}{c}{\agnews} & \multicolumn{2}{c}{\msmarcoqwen} \\
\cmidrule(lr){2-3}\cmidrule(l){4-5}
 & \textbf{QPS} & \textbf{Step} & \textbf{QPS} & \textbf{Step} \\
\midrule
\hnswlib, file order & $1\,501$ & & $716$ & \\
\kannolo, \texttt{f32} & $2\,050$ & $1.37\times$ & $1\,016$ & $1.42\times$ \\
\quad + \texttt{f16} & $2\,803$ & $1.37\times$ & $1\,401$ & $1.38\times$ \\
\quad + early termination & $4\,201$ & $1.50\times$ & $2\,253$ & $1.61\times$ \\
\quad + permutation & $4\,449$ & $1.06\times$ & $2\,529$ & $1.12\times$ \\
\midrule
\hnswlib, random order & $1\,470$ & $0.98\times$ & $634$ & $0.89\times$ \\
\bottomrule
\end{tabular}
\end{table}

We also compare against \glass{} at the same \texttt{f16} rescoring precision, which isolates the index from the choice of rescoring representation. In this setting \kannolorerank{} is $1.39$, $1.27$ and $1.27\times$ faster on \dpr, with $1.17$ to $1.36\times$ less memory. On \msmarcoqwen{} it is $1.33\times$ faster at $99\%$, with $1.27\times$ less memory. The $95\%$ point of \glass{} on \dpr{} uses an \texttt{f32} refine stage ($73.44$ GiB), and it is within the pass-to-pass spread of its \texttt{f16} point.

\paragraph{Memory.}

\begin{figure*}[t]
\centering
\includegraphics[width=\textwidth]{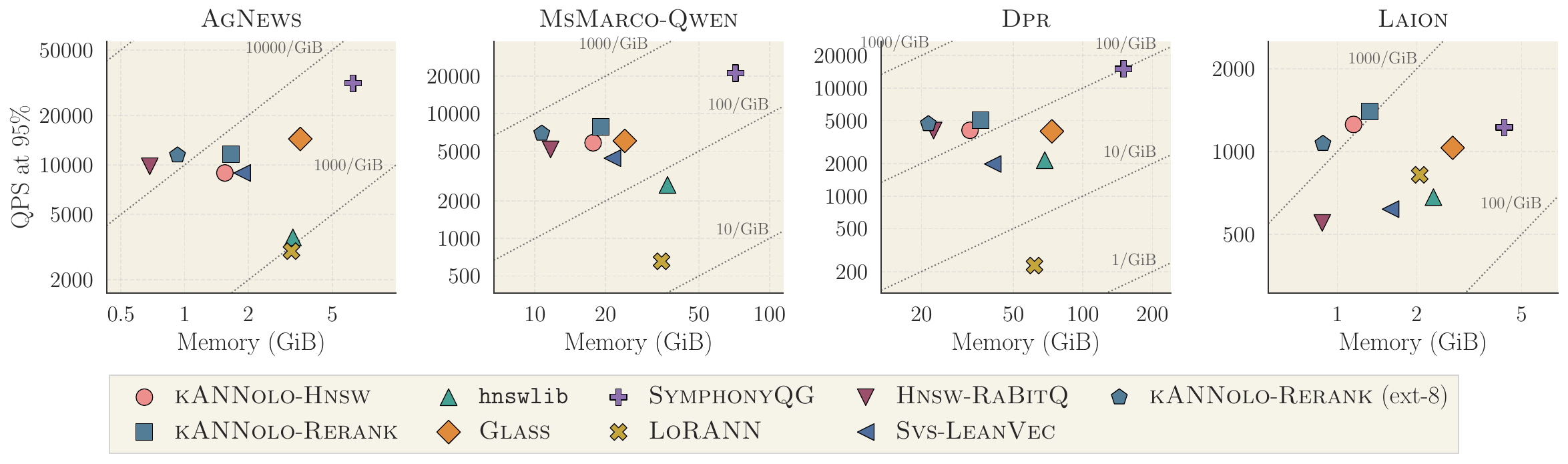}
\caption{Footprint (increase in USS, Section~\ref{sec:exp_measures}) against throughput at Accuracy@$k \ge 95\%$ on the four dense collections, one
point per system. Dotted lines are constant QPS per GiB of index. \emph{ext-8} marks \kannolorerank{} with \rabitqext{} at $8$ bits as the rescoring dataset (Section~\ref{sec:exp_rerank_precision}).}
\label{fig:exp_tradeoff_dense}
\end{figure*}

Figure~\ref{fig:exp_tradeoff_dense} places every system by footprint and throughput at $95\%$, where all eight reach the target on every collection. Among the competitors, only \rabitqhnsw{} builds a smaller index than \kannolorerank{} with its \texttt{f16} rescoring stage, on every collection and at every target it reaches. At $99\%$, \kannolorerank{} holds $1.75$ GiB on \agnews{} and $36.96$ GiB on \dpr, $0.62\times$ the size of that collection. On \dpr, \hnswlib{} holds $74.64$ GiB, \glass{} $50.40$, \svsleanvec{} $43.23$ and \rabitqhnsw{} $22.58$.

\symphonyqg{} needs much more memory than the other systems. In throughput per GiB, both our configurations are ahead of it at every target on every collection. \rabitqhnsw{} has the best trade-off among the competitors. It is ahead of our \texttt{f16} configurations in throughput per GiB at $90$ and $95\%$ on \agnews, \msmarcoqwen{} and \dpr, and at $99\%$ on \agnews. It is not ahead on \laion, where it stops at $95.84\%$. In \kannolo{} the rescoring dataset is chosen independently of the index, so changing it requires no new index. If the \texttt{f16} rescoring dataset is replaced with \rabitqext{} at $8$ bits (Section~\ref{sec:exp_rerank_precision}), our configurations lead every system in throughput per GiB in $9$ of the $12$ collection--target pairs. \rabitqext{} rescoring leads in eight of them. In the ninth, \laion{} at $99\%$, \rabitqext{} rescoring does not reach the target and \kannolohnsw{} leads. \rabitqhnsw{} still leads on \agnews{} at $90$ and $95\%$ and on \dpr{} at $90\%$. At $99\%$ these configurations are $25$--$38\%$ slower than with \texttt{f16}.

\paragraph{Construction.}

We do not lead on construction time (Table~\ref{tab:exp_results_dense}). At $99\%$, \svsleanvec{} builds fastest wherever it reaches the target, and \hnswlib{} slowest on the two large collections. Our builds are in the middle: across collections and targets they rank between second and sixth of the eight systems, most often fourth or fifth. At $99\%$ on \dpr, ours take $4\,814$ and $4\,891$ s, against $2\,982$ s for \svsleanvec{} and $5\,238$ s for \hnswlib. \svsleanvec{} builds its \vamana{} graph on vectors reduced to $384$ dimensions and compressed with LVQ to one byte per dimension. The collections have $768$ or $1\,024$ dimensions at four bytes each.
\paragraph{Summary.}
On the two small collections no system dominates. \glass{} is the fastest system after \symphonyqg{} on \agnews{} at every target, and the fastest on \laion{} at $90\%$, by $1.14$ to $1.33\times$ over us. We lead on \laion{} at $95$ and $99\%$. On the two large collections, \kannolorerank{} is faster than every system except \symphonyqg{} at every target, by $1.15$ to $1.39\times$. Its index is $0.57$ to $0.62\times$ the size of the collection, and its build time ranks third to sixth of the eight systems.

\symphonyqg{} is the fastest system at every target on \agnews, \msmarcoqwen{} and \dpr. This speed comes at a cost in memory, and the extra memory is proportionally larger than the extra throughput\label{sec:exp_symphonyqg}. On \msmarcoqwen{} it has $3.82$, $3.62$ and $2.39\times$ the throughput of \kannolohnsw{} at the three targets, for $4.00\times$ its footprint. On \dpr{} it has $3.36$ down to $1.81\times$ the throughput of \kannolorerank, for $4.06\times$ its footprint. At $99\%$ it is therefore behind both our configurations in throughput per GiB: $80.1$ against $134.1$ and $151.9$ on \msmarcoqwen, and $21.4$ against $46.9$ and $48.0$ on \dpr. Each node stores the \rabitq{} code of every neighbor next to its own vector. Each code is therefore replicated about as many times as the average degree. On \dpr{} the index takes $149.99$ GiB, $2.50\times$ the size of the collection. At degree $64$ the index grows to $240$ GiB and the build to $6\,485$ s, over $B$. The two largest settings would need about $490$ and $600$ GiB, close to or above the memory of the machine, and were not run. The smallest setting was measured with its memory interleaved across both nodes, because it could not be placed on one. This makes its \dpr{} row a lower bound. On \laion{} it is behind \glass{} at $90\%$ and behind us at $95\%$, and stops at $97.60\%$.

\subsubsection{Sparse Comparisons}
\label{sec:exp_sparse}

\paragraph{Search.}

\begin{figure*}[t]
\centering
\includegraphics[width=\textwidth]{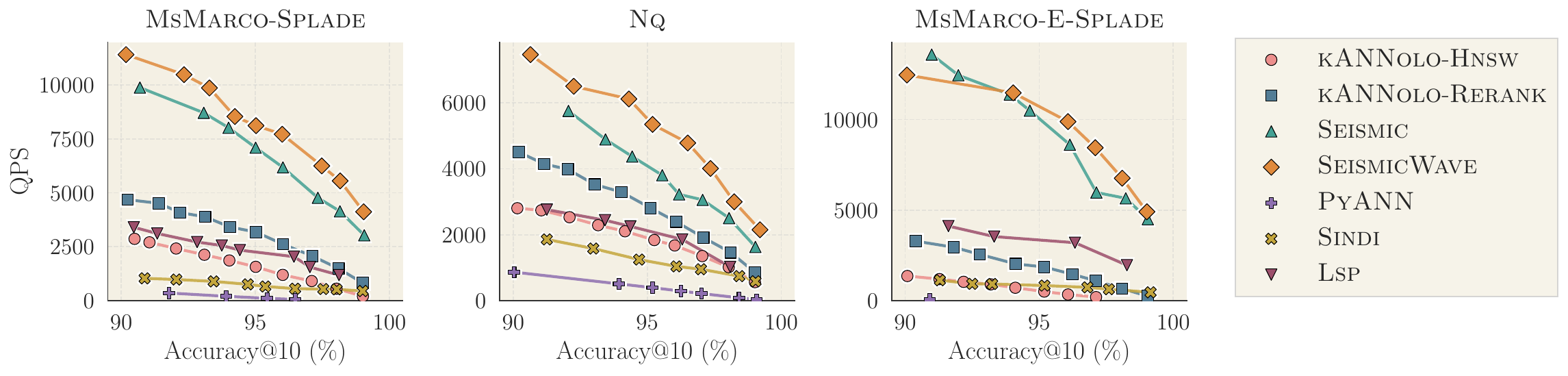}
\caption{Accuracy--throughput frontiers on the three sparse collections. Each point is a system's
fastest configuration at the accuracy on its axis.}
\label{fig:exp_frontier_sparse}
\end{figure*}

Figure~\ref{fig:exp_frontier_sparse} draws the frontiers on the three collections, and Table~\ref{tab:exp_results_sparse} reports every system at the three accuracy targets. The ranges below are taken over the whole frontier from $90$ to $99\%$, so they include targets between those of the table. On \msmarcosplade{} and \nq{} the ranking is the same at every target. The two \seismic{} variants lead. \seismicw{} is $2.32$--$5.04\times$ faster than \kannolorerank{} on \msmarcosplade{} and $1.57$--$2.50\times$ on \nq. \seismic{} is $1.93$--$3.71\times$ and $1.28$--$1.89\times$ faster. The gap grows with the target. \kannolorerank{} is faster than every other system. It is $1.26$--$1.87\times$ faster than \lsp{} wherever \lsp{} reaches the target (it never reaches $99\%$). It is $1.46$--$4.82\times$ faster than \sindi, and $5.2$--$44.5\times$ faster than \pyann, the other graph index. \kannolohnsw{} is compared with \kannolorerank{} in Section~\ref{sec:exp_ablations}.

On \msmarcoesplade{} the inverted indexes pull further ahead. \seismic{} and \seismicw{} are within $9\%$ of each other up to $95\%$, and \seismicw{} leads from $96$ to $99\%$. They are $4.13$--$17.6\times$ and $3.79$--$19.2\times$ faster than \kannolorerank, and again the gap grows with the target. \lsp{} is third up to $98\%$, $1.25$--$2.95\times$ faster than \kannolorerank, and never reaches $99\%$. \kannolorerank{} is $2.90\times$ faster than \sindi{} at $90\%$ and stays ahead of it up to $98\%$. At $99\%$, \sindi{} is faster, with $463$ against $255$ QPS. \kannolohnsw{} does not reach $98\%$. \pyann{} reaches $90\%$ at $68$ QPS, $48.4\times$ slower than \kannolorerank. Only the largest build of \pyann, of degree $64$, fails on the two \msmarcovo{} collections, where it returns no correct result: its graph loader computes the size of the graph in bytes with a signed $32$-bit integer, which overflows at $8.8$M documents. The \pyann{} results above come from its other builds. \lsp{} is not robust here. Every configuration that keeps at most half of the query terms aborts ($241$ of $504$), because the shortest queries are pruned to no term at all. Of the other configurations, $8$ leave two queries unanswered. The configurations in Table~\ref{tab:exp_results_sparse} are not affected.

\paragraph{Memory.}

\begin{figure*}[t]
\centering
\includegraphics[width=\textwidth]{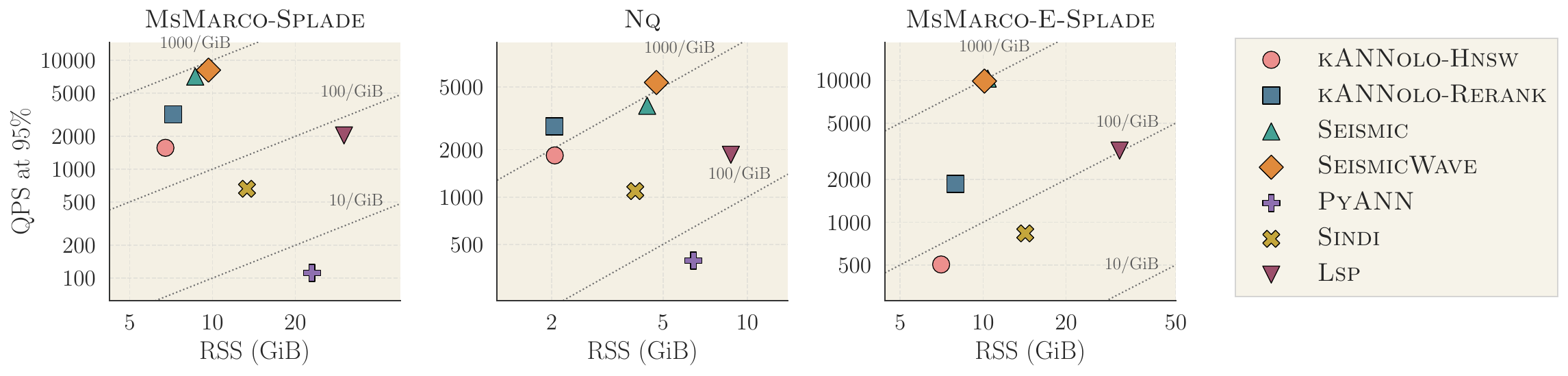}
\caption{Resident memory (RSS) of the search process against throughput at Accuracy@$k \ge 95\%$ on the three sparse
collections, one point per system. Dotted lines are constant QPS per GiB.}
\label{fig:exp_tradeoff_sparse}
\end{figure*}

Our indexes use the least memory on every collection and at every target (Figure~\ref{fig:exp_tradeoff_sparse}). On \msmarcosplade{} they take $6.8$--$7.6$ GiB, against $7.9$--$10.6$ GiB for \seismic, $9.2$--$11.8$ for \seismicw, $13.4$ for \sindi, $23.0$ for \pyann{} and $30.1$ for \lsp. On \nq{} they take $2.0$--$2.1$ GiB, against $3.4$--$8.7$ GiB for every other system. This is less than half the memory of \seismic, and enough to give \kannolorerank{} the highest throughput per GiB at all three targets. On \msmarcoesplade{} they take $7.0$--$8.0$ GiB, against $8.4$--$11.5$ GiB for \seismic, $10.1$--$11.6$ for \seismicw, $13.2$--$14.2$ for \sindi, $25.9$ for \pyann{} and $31.3$ for \lsp.

\paragraph{Construction.}

Graph indexes take far longer to build than inverted indexes, which build in seconds to minutes. The fastest builds are those of \sindi{} ($26$--$104$ s) and \seismic{} ($63$--$156$ s). \seismicw{} and \lsp, which add a $\kappa$-NN graph and a document reordering, take $292$--$1\,844$ s and $259$--$1\,118$ s. Our builds take $481$ to $5\,428$ s. The two-stage index builds its graph over the pruned vectors, and is therefore cheaper to build than the single-stage one. On \msmarcosplade, at the same $M = 16$ and $ef_\text{construction} = 1\,000$, it takes $2\,268$ s against $4\,835$ s. Within $B = 5\,400$ s it can therefore afford $M = 32$ or $ef_\text{construction} = 2\,000$, which push the single-stage build beyond $B$. On \msmarcosplade, the selected builds of \pyann{} are cheaper than those of our single-stage index, $2\,054$ against $4\,835$ s, because they use smaller construction parameters. They also search $7.6$--$20.3\times$ more slowly.

\paragraph{Summary.} On sparse data the inverted indexes are faster. \seismic{} and \seismicw{} beat \kannolorerank{} at every target on every collection they run on, and on \msmarcoesplade{} \lsp{} does too, up to $98\%$. \kannolorerank{} is the fastest graph index we measured. Together with \kannolohnsw, it is also the most compact index on every collection, but it takes much longer to build.

\subsubsection{Ablations}
\label{sec:exp_ablations}

\paragraph{Two-stage over single-stage search.}
At $99\%$, \kannolorerank{} is $1.20\times$ faster than \kannolohnsw{} on \msmarcoqwen, $1.14\times$ on \dpr{} and $1.07\times$ on \agnews{} and \laion. It gains more at the two lower targets: $+41\%$ and $+30\%$ on \agnews, $+37\%$ and $+34\%$ on \msmarcoqwen, $+24\%$ and $+24\%$ on \dpr, $+16\%$ and $+11\%$ on \laion.

On sparse data the gap is larger. \kannolorerank{} is $1.63$ to $3.97\times$ faster than \kannolohnsw{} on \msmarcosplade{} and $1.42$ to $1.61\times$ on \nq, and at equal construction parameters it is also faster to build, because its graph is built over the first part only. Its selected builds can take longer (Table~\ref{tab:exp_results_sparse}), because within $B$ it can afford larger construction parameters. Memory is roughly the same, since the two parts together hold all the nonzero entries. On \msmarcoesplade{} it is $2.41$ and $3.69\times$ faster at $90$ and $95\%$. On this collection \kannolohnsw{} reaches neither $98$ nor $99\%$ within $B$.

\paragraph{Pruning fraction.}
The fastest configurations use $a = 60$ or $70$ on \msmarcosplade, $a = 70$ or $80$ on \nq{} and $a = 60$ on \msmarcoesplade. On \msmarcoesplade, at $M = 16$ and $ef_\text{construction} = 1\,000$, $a = 60$ is the fastest at $90$, $95$ and $98\%$, by $1$--$10\%$ over $a = 70$ and $20$--$36\%$ over $a = 80$. Pruning further hurts: on \msmarcosplade, at the same parameters, $a = 60$ is $30$--$71\%$ faster than $a = 40$. With $a = 100$ the residual is empty and the index reduces to \kannolohnsw.

\paragraph{Graph representation.}
\label{sec:exp_graph_repr}
We compare \texttt{permuted} with \texttt{standard} on \agnews{} at $M = 16$ and $ef_\text{construction} = 200$, over $117$ matched pairs of search parameters. \texttt{permuted} gains $+11.81\%$ median throughput, and never less than $+9.04\%$. It costs $+20.9\%$ build time and $+4.2\%$ USS. The permutation is a separate pass after graph construction, and it scales worse than the graph: at $21.0$M points it takes $1\,391$--$2\,386$ s for $M = 8$ to $32$, which roughly doubles the build time. \texttt{permuted} is still faster at every accuracy target at $21.0$M, and our \dpr{} rows in Table~\ref{tab:exp_results_dense} use it. We leave the compressed layout of Section~\ref{sec:graph_compression} out of the grid. At $d \in \{768, 1\,024\}$ the adjacency is $4.5\%$ of the index, so compressing it by $51.6\%$ saves $2.3\%$ overall. On sparse data we use only \texttt{standard}, because build times there are already high and the permutation would increase them further.

\paragraph{Early termination.}
For each collection and target we compare the fastest configuration with the fastest one with $\lambda$ disabled. For \kannolohnsw{} on the dense collections, $\lambda$ gives $4$--$9\%$ more throughput at $90\%$ and $25$--$28\%$ at $95\%$. On \msmarcosplade{} and \nq{} it gives $20$--$67\%$ at every target, and on \msmarcoesplade{} $93$ and $161\%$ at $90$ and $95\%$. At the other dense targets, our grid reaches the target only with $\lambda$ on, because its beam widths are sized for search with early termination. On \laion{} this also holds beyond our grid. With $\lambda$ off and $ef_\text{search}$ swept up to $1\,000$, \kannolohnsw{} plateaus at $98.36\%$ and \hnswlib{} at $98.74\%$. With early termination, \kannolohnsw{} reaches $99.11\%$ at $ef_\text{search} = 12$. \kannolorerank{} depends less on $\lambda$ on dense data: $\lambda$ improves its fastest configuration by more than $1\%$ in $4$ of the $10$ pairs measured both ways, by $3.5$--$9\%$, and on \laion{} it reaches $95$ and $99\%$ only with $\lambda$ on. It depends more on $\lambda$ on sparse data. There $\lambda$ improves its fastest configuration in $5$ of the $6$ pairs measured both ways, by $7$--$34\%$, and \kannolorerank{} reaches $99\%$ only with $\lambda$ on, on all three collections.

\paragraph{Tail latency.}
Early termination widens the distribution of per-query latency, as the spread of visited nodes in Section~\ref{sec:visited_set} predicts. We read the per-query latencies that \vibe{} records for the best pass of each reported configuration, over the $1\,000$ queries. A 99th percentile therefore rests on the ten slowest queries. On \agnews{} at $95\%$, $\lambda$ lowers the median latency of \kannolohnsw{} from $141$ to $100\,\mu$s but raises its 99th percentile from $226$ to $308\,\mu$s. At $99\%$, which our grid reaches only with $\lambda$ on, the 99th percentile of \kannolohnsw{} is $4.2$--$5.4\times$ its median on \agnews, \msmarcoqwen{} and \dpr, against $1.7$--$2.0\times$ for \hnswlib. It still stays below that of \hnswlib: $957$ against $1\,211\,\mu$s, $1\,402$ against $2\,144\,\mu$s and $2\,180$ against $2\,310\,\mu$s. \kannolorerank{} keeps the ratio between $1.5$ and $2.4$ on the same three collections. On \laion{} at $99\%$, a target no other graph system reaches, the ratio is $5.3$ for \kannolorerank{} and $8.0$ for \kannolohnsw. Sparse data shows the same pattern. Here we time each query inside the search tool, in a re-run of the reported configuration. The re-run reproduces the accuracy of Table~\ref{tab:exp_results_sparse} exactly and its throughput within $3.2\%$. At $99\%$, the 99th percentile of \kannolohnsw{} is $8.7\times$ its median on \msmarcosplade{} and $5.6\times$ on \nq. \kannolorerank{} keeps the ratio at $1.7$ and $2.4\times$, below \seismic{} ($3.9$ and $3.8\times$) and \seismicw{} ($3.2$ and $3.1\times$). On \msmarcoesplade, where \kannolohnsw{} does not reach $99\%$, the ratio is $4.4\times$ for \kannolorerank, against $5.0\times$ for \seismic{} and $4.1\times$ for \seismicw. \pyann{} reaches $99\%$ only on \nq. There it keeps the ratio at $1.5\times$, but its 99th percentile is $47$\,ms, $18\times$ that of \kannolorerank. \lsp{} and \sindi{} can only be timed per pass and are not included.

\paragraph{Rescoring precision.}
\label{sec:exp_rerank_precision}
The rescoring dataset dominates the footprint of \kannolorerank, so we test whether it needs to be near-exact. We replace \texttt{f16} with \rabitqext{} at $8$ bits per component on all four collections, keeping the selected builds and the search grid, so the candidate lists are unchanged. At the same search parameters the quantized stage is as fast as \texttt{f16}, within $2\%$. It loses, however, an almost constant amount of accuracy: a median of $0.25$--$0.49$ points on \agnews, \msmarcoqwen{} and \dpr, and $1.1$ points on \laion. At $90\%$ it keeps $0.95$--$0.99\times$ the throughput of \texttt{f16} on the first three collections, with $1.69$--$1.79\times$ less memory. At $99\%$ it costs $25$, $25$ and $38\%$ of the throughput. Throughput per GiB is $1.42\times$ better on \agnews{} ($3\,721$ against $2\,617$), $1.21\times$ better on \msmarcoqwen{} ($184.6$ against $151.9$) and $1.04\times$ better on \dpr{} ($49.9$ against $48.0$). On \laion, whose queries are out-of-distribution, it costs $24\%$ of the throughput already at $95\%$, for a footprint $1.51\times$ smaller, and stops at $97.73\%$, against $99.45\%$ with \texttt{f16}. We report \texttt{f16} because it keeps $1.3$--$1.6\times$ the throughput at $99\%$ and is the only one of the two that reaches $99\%$ on \laion. Rescoring with encoded residuals, through the residual mode of \rerankindex, gives mixed results: $58.9$ QPS/GiB at $99\%$ on \dpr, but $99\%$ is never reached on \msmarcoqwen.

\paragraph{The first-stage encoder does not transfer.}
\label{sec:exp_encoder_transfer}
The fastest first-stage encoder differs across collections. One-bit \rabitq{} is selected on \agnews{} and \msmarcoqwen{} at almost every target, \rabitqext{} at two bits on \dpr{} and at four bits on \laion. The share of the exact top-$k$ that the top $100$ candidates of each encoder cover explains this choice. One-bit \rabitq{} covers at least $99.7\%$ on \agnews{} and \msmarcoqwen. On \dpr{} it saturates at $99.11\%$, against $99.93\%$ for \rabitqext{} at two bits. On \laion{} it collapses to $35$--$65\%$ at every query width. The failure on \dpr{} is not due to modality. Like \msmarcoqwen, \dpr{} is text retrieval under inner product, but the encoder that is sufficient on \msmarcoqwen{} is not sufficient on \dpr. We do not sweep \pq. At the code width of one-bit \rabitq{} ($128$ subspaces at $d = 1\,024$), it was $1.4$--$2.3\times$ slower than \rabitq{} on \agnews. At $64$ subspaces it covers at most $96.59\%$ on every collection.

\section{Conclusions and Future Work}
\label{sec:conclusions}

We presented \vectorium{} and \kannolo, two Rust libraries that factorize approximate nearest neighbor search into composable abstractions. Users compose them into indexes for dense, sparse and multi-vector representations, covering many trade-offs between construction time, memory, accuracy and speed. Because the compiler specializes every combination, the factoring adds no run-time overhead and still reaches state-of-the-art performance. Our Product Quantizer matches the accuracy of \faiss{} at competitive speed, our \rabitq{} and \hnsw{} match or outperform their reference implementations, and our IVF outperforms \faiss.

End to end (Section~\ref{sec:exp}), our two-stage \hnsw{} is faster than every dense system except \symphonyqg{} on the two large collections, with an index of $0.57$--$0.62\times$ the size of the collection. \symphonyqg{} needs about four times our memory. On sparse data, the systems built on \vectorium{} lead. \seismic{} and \seismicw{} are the fastest systems on every collection, our indexes are the most compact on every collection, and our two-stage index is the fastest graph index we measured. Our indexes are not the fastest to build. On dense data they rank in the middle of the field, and on sparse data they take far longer to build than inverted indexes.

In \kannolo, the same components build both the dense and the sparse two-stage indexes, with no index code specific to either representation. This is useful in practice. The first-stage encoder that is sufficient differs from one collection to the next, and the rescoring dataset trades accuracy for memory. In both cases, a change requires only a new configuration and no new index.

Future work goes in three directions. First, \hnsw{} is so far the only algorithm built on the shared graph layer. We plan to add further graph indexes on it and to support dynamic collections, with insertions and deletions after construction. Second, we will keep both libraries current with new quantization and indexing techniques. We will also evaluate the capabilities that are already implemented but not measured here. Third, and most important, we plan to refine and extend the abstractions themselves, so that they cover designs that currently fall outside them. Specialized indexes such as \symphonyqg{}, \seismic{} and \tachiom{} cannot yet be expressed with our framework's abstractions, even though \seismic{} and \tachiom{} are already built on \vectorium. Bringing them inside the framework would let us study and develop these techniques further, and combine their components in ways not explored so far.

\appendix
\counterwithin*{table}{section}\renewcommand{\thetable}{\thesection.\arabic{table}}
\counterwithin*{figure}{section}\renewcommand{\thefigure}{\thesection.\arabic{figure}}
\counterwithin*{equation}{section}\renewcommand{\theequation}{\thesection.\arabic{equation}}

\section{Harness and Measurement Details}
\label{app:harness}

Three of the four \vibe{} patches of Section~\ref{sec:exp_setting} are measurement conventions, applied to every system alike.

\paragraph{Page placement.}
\label{sec:exp_placement}
Our machine has two NUMA nodes, and the $64$ build threads run on both. If an index were allocated on one node only, the threads of the other node would read remote memory at every access. Graph builds consist mostly of such reads and would take roughly $50\%$ longer, while the build of \lorann{} would be almost unchanged. This would distort the comparison of build costs across index families. Builds therefore spread their memory over both nodes, and the finished index is then migrated to the node of the core that runs the search. We discard any run whose migration cannot be verified from the process's page map, and every reported configuration is at least $99\%$ local to its node. With migrated pages, graph indexes lose $1.5\%$ (ours) to $2.5\%$ (\hnswlib) of the throughput they reach with pages allocated locally from the start, so our reported ratio against \hnswlib{} is inflated by about $1\%$. Only the index of \symphonyqg{} on \dpr{} cannot be migrated (Section~\ref{sec:exp_symphonyqg}).

\paragraph{Memory measurement and allocator settings.}
\label{app:allocator}
Build and search are recorded in two separate passes, so that the memory we report belongs to the index and not to the benchmark harness. Two settings of the C memory allocator, applied to every system alike, make this measurement reliable. The first caps the number of allocation arenas (\texttt{MALLOC\_ARENA\_MAX}); without it, a multi-threaded build leaves memory reserved but unused. The second (\texttt{MALLOC\_MMAP\_THRESHOLD\_}) serves allocations of up to $256$~MiB from the allocator's own pools. Without it, the $64$ build threads queue on the lock that guards the process's memory map, which slowed the construction of \glass{} on \dpr{} by a factor of $14$. Rebuilt with both settings, the cheapest configuration of every system measured before the change moved by $-4.6$ to $+1.6\%$ in build time, so those measurements stand.

\subsection{Budget Sensitivity}
\label{app:budget}
\label{app:pruning}

We also measure configurations over the budget, and this makes the analysis possible. On dense data, each library's grid is built cheapest first, in ascending order of its cost-monotone parameters. A build that exceeds $B$ is still completed and measured, up to a time limit of about $2B$, but it is excluded from the selection. All configurations at least as large as it in every one of those parameters are then skipped without being built, since they cannot be cheaper. A configuration that is larger in some parameters and smaller in others is still built. Because of this order, the set of skipped configurations does not depend on chance.

We re-select every reported configuration on \dpr{} at $B/2$, $2B$ and $4B$ from the measured configurations, taking the maximum QPS at each accuracy target. On \agnews{} our whole grid already builds within $B$. On dense data, skipped configurations cannot be recovered, so the analysis bounds the sensitivity of the selection but not that of the full grid. Above $B$ the ranking changes only once: on \dpr{} at $99\%$, \rabitqhnsw{} overtakes \svsleanvec{} from $2B$ on ($1\,109$ against $993$ QPS). The largest gains leave every position unchanged. \symphonyqg{}, already the fastest on \dpr, goes from $3\,213$ to $6\,030$ QPS at $99\%$, and at $4B$ our configurations gain at most $0.1\%$. At $B/2$, \kannolohnsw{} loses $13$--$17\%$, again without changing its position, while \kannolorerank, \glass, \symphonyqg, \lorann{} and \rabitqhnsw{} have no configuration that reaches $90\%$ on \dpr.

\subsection{Sparse Timing}
\label{app:sparse_timing}

Every sparse system is timed in the same way: one search process runs over the whole query set on a single core, with its memory local to that core's node. The command-line tools of \lsp{} and \sindi{} pass over the query set several times in one invocation. We take the end of each pass from their output, discard the first pass and report the best of the next two. We time every other system with the same rule, through a command-line driver of our own. Systems that can be called in-process (ours, \seismic, \seismicw{} and \pyann) are tuned in-process, and only their fastest configurations at each target are timed from the command line. The systems built on \vectorium{} use the blocked encoder. \pyann{} is called one query at a time. Its batch interface prunes each query against the largest weight in the whole batch, which changes its results. Calling it per query costs it $5.9\%$ of its throughput on \msmarcosplade.

\section{Full Results}
\label{app:results}

Tables~\ref{tab:exp_results_dense} and~\ref{tab:exp_results_sparse} report, for every system and collection of Section~\ref{sec:exp}, the fastest configuration at each accuracy target, and Table~\ref{tab:exp_tail} the tail latency of the same configurations (Section~\ref{sec:exp_ablations}).

\begin{table*}[t]
\centering
\scriptsize
\setlength{\tabcolsep}{2pt}
\caption{Throughput, footprint and construction time of each system's fastest configuration at
each accuracy target on the dense collections, two collections per cell; footprint is the increase
in unique set size (USS), for \lorann{} plus the raw collection (Section~\ref{sec:exp_measures}); -- where a system does
not reach the target, $\cdot$ where it was not run. On \laion{} the \lorann{} row is its learned
variant. \emph{+ ext-8} is \kannolorerank{} rescoring with \rabitqext{} at $8$ bits instead of \texttt{f16}
(Section~\ref{sec:exp_rerank_precision}). $^{\dagger}$: control configuration, not drawn in
Figure~\ref{fig:exp_frontier_dense}: on \laion{} \lorann{} without the learned variant.}
\label{tab:exp_results_dense}
\begin{tabular}{@{}lr@{}rr@{}rr@{}rr@{}rr@{}rr@{}rr@{}rr@{}rr@{}r@{}}
\toprule
& \multicolumn{6}{c}{QPS} & \multicolumn{6}{c}{Memory (GiB)} & \multicolumn{6}{c}{Build (s)} \\
\cmidrule(lr){2-7}\cmidrule(lr){8-13}\cmidrule(lr){14-19}
System & \multicolumn{2}{c}{$90\%$} & \multicolumn{2}{c}{$95\%$} & \multicolumn{2}{c}{$99\%$} & \multicolumn{2}{c}{$90\%$} & \multicolumn{2}{c}{$95\%$} & \multicolumn{2}{c}{$99\%$} & \multicolumn{2}{c}{$90\%$} & \multicolumn{2}{c}{$95\%$} & \multicolumn{2}{c}{$99\%$} \\
\midrule
\multicolumn{19}{@{}l}{\agnews{} / \laion} \\
\quad \kannolohnsw & 11\,665 & /2\,578 & 8\,930 & /1\,254 & 4\,277 & /206 & 1.55 & /1.15 & 1.55 & /1.15 & 1.67 & /1.15 & 60 & /127 & 60 & /127 & 78 & /127 \\
\quad \kannolorerank & 16\,464 & /2\,987 & 11\,595 & /1\,398 & 4\,585 & /220 & 1.65 & /1.33 & 1.65 & /1.33 & 1.75 & /1.34 & 79 & /129 & 79 & /129 & 97 & /144 \\
\quad \quad + ext-8 & 15\,949 & /2\,835 & 11\,450 & /1\,069 & 3\,442 & /-- & 0.92 & /0.87 & 0.92 & /0.88 & 0.92 & /-- & 80 & /130 & 80 & /146 & 80 & /-- \\
\quad \hnswlib & 5\,307 & /1\,504 & 3\,624 & /682 & 1\,507 & /-- & 3.16 & /2.26 & 3.25 & /2.31 & 3.39 & /-- & 58 & /142 & 124 & /176 & 157 & /-- \\
\quad \glass & 20\,547 & /3\,417 & 14\,374 & /1\,032 & 6\,085 & /-- & 2.45 & /1.64 & 3.52 & /2.74 & 3.48 & /-- & 39 & /57 & 80 & /63 & 63 & /-- \\
\quad \symphonyqg & 38\,898 & /3\,116 & 31\,356 & /1\,223 & 13\,685 & /-- & 6.25 & /4.30 & 6.25 & /4.30 & 9.55 & /-- & 136 & /93 & 136 & /93 & 197 & /-- \\
\quad \lorann & 6\,218 & /1\,696 & 2\,989 & /823 & 912 & /211 & 3.28 & /2.05 & 3.20 & /2.05 & 3.23 & /2.07 & 45 & /153 & 85 & /153 & 148 & /192 \\
\quad \rabitqhnsw & 13\,090 & /1\,547 & 9\,802 & /550 & 3\,188 & /-- & 0.59 & /0.87 & 0.68 & /0.88 & 0.96 & /-- & 214 & /103 & 215 & /239 & 217 & /-- \\
\quad \svsleanvec & 10\,357 & /1\,984 & 8\,924 & /617 & 3\,608 & /-- & 1.83 & /1.59 & 1.87 & /1.59 & 1.83 & /-- & 35 & /88 & 40 & /88 & 35 & /-- \\
\quad \lorann{}$^{\dagger}$ & $\cdot$ & /1\,593 & $\cdot$ & /761 & $\cdot$ & /208 & $\cdot$ & /2.07 & $\cdot$ & /2.05 & $\cdot$ & /2.05 & $\cdot$ & /198 & $\cdot$ & /140 & $\cdot$ & /140 \\
\midrule
\multicolumn{19}{@{}l}{\msmarcoqwen{} / \dpr} \\
\quad \kannolohnsw & 8\,092 & /5\,737 & 5\,845 & /4\,065 & 2\,405 & /1\,562 & 17.71 & /31.66 & 17.71 & /32.46 & 17.92 & /33.33 & 1\,738 & /4\,058 & 1\,738 & /3\,983 & 1\,335 & /4\,814 \\
\quad \kannolorerank & 11\,114 & /7\,095 & 7\,837 & /5\,039 & 2\,896 & /1\,776 & 19.06 & /36.15 & 19.06 & /36.15 & 19.06 & /36.96 & 1\,622 & /3\,938 & 1\,622 & /3\,938 & 1\,622 & /4\,891 \\
\quad \quad + ext-8 & 10\,510 & /7\,046 & 6\,953 & /4\,659 & 2\,165 & /1\,106 & 10.70 & /21.42 & 10.70 & /21.42 & 11.73 & /22.16 & 1\,629 & /3\,970 & 1\,629 & /3\,970 & 1\,631 & /5\,034 \\
\quad \hnswlib & 4\,015 & /3\,355 & 2\,686 & /2\,149 & 889 & /719 & 36.76 & /66.55 & 36.76 & /68.40 & 38.86 & /74.64 & 1\,292 & /2\,226 & 1\,292 & /3\,663 & 2\,107 & /5\,238 \\
\quad \glass & 8\,020 & /5\,098 & 6\,058 & /3\,977 & 2\,527 & /1\,400 & 24.17 & /42.40 & 24.17 & /73.44 & 42.18 & /50.40 & 964 & /3\,949 & 964 & /4\,311 & 1\,384 & /4\,358 \\
\quad \symphonyqg & 30\,937 & /23\,858 & 21\,148 & /15\,061 & 5\,742 & /3\,213 & 71.67 & /149.99 & 71.67 & /149.99 & 71.67 & /149.99 & 1\,944 & /4\,331 & 1\,944 & /4\,331 & 1\,944 & /4\,331 \\
\quad \lorann & 1\,841 & /550 & 656 & /227 & 135 & /-- & 34.68 & /61.73 & 34.68 & /61.73 & 34.68 & /-- & 1\,990 & /4\,996 & 1\,990 & /4\,996 & 1\,990 & /-- \\
\quad \rabitqhnsw & 7\,568 & /5\,953 & 5\,168 & /4\,047 & 1\,365 & /986 & 10.61 & /16.95 & 11.66 & /22.58 & 10.61 & /22.58 & 1\,413 & /4\,841 & 2\,030 & /4\,868 & 1\,413 & /4\,868 \\
\quad \svsleanvec & 5\,395 & /3\,361 & 4\,400 & /1\,981 & 1\,601 & /993 & 22.45 & /40.72 & 21.39 & /40.72 & 22.45 & /43.23 & 1\,102 & /1\,377 & 454 & /1\,377 & 1\,102 & /2\,982 \\
\bottomrule
\end{tabular}

\vspace{1.5ex}
\caption{Throughput, resident memory of the search process and construction time of each
system's fastest configuration at each accuracy target on the sparse collections, two collections
per cell in the first block; -- where a system does not reach the target.}
\label{tab:exp_results_sparse}
\begin{tabular}{@{}lr@{}rr@{}rr@{}rr@{}rr@{}rr@{}rr@{}rr@{}rr@{}r@{}}
\toprule
& \multicolumn{6}{c}{QPS} & \multicolumn{6}{c}{RSS (GiB)} & \multicolumn{6}{c}{Build (s)} \\
\cmidrule(lr){2-7}\cmidrule(lr){8-13}\cmidrule(lr){14-19}
System & \multicolumn{2}{c}{$90\%$} & \multicolumn{2}{c}{$95\%$} & \multicolumn{2}{c}{$99\%$} & \multicolumn{2}{c}{$90\%$} & \multicolumn{2}{c}{$95\%$} & \multicolumn{2}{c}{$99\%$} & \multicolumn{2}{c}{$90\%$} & \multicolumn{2}{c}{$95\%$} & \multicolumn{2}{c}{$99\%$} \\
\midrule
\multicolumn{19}{@{}l}{\msmarcosplade{} / \nq} \\
\quad \kannolohnsw & 2\,867 & /2\,804 & 1\,570 & /1\,839 & 206 & /563 & 6.75 & /1.96 & 6.75 & /2.05 & 6.75 & /2.05 & 4\,835 & /481 & 4\,835 & /1\,640 & 4\,835 & /1\,640 \\
\quad \kannolorerank & 4\,682 & /4\,506 & 3\,183 & /2\,811 & 818 & /861 & 7.15 & /2.04 & 7.21 & /2.04 & 7.60 & /2.05 & 2\,268 & /994 & 4\,160 & /994 & 3\,218 & /1\,574 \\
\quad \seismic & 9\,877 & /5\,755 & 7\,091 & /3\,804 & 3\,032 & /1\,631 & 7.94 & /4.39 & 8.66 & /4.39 & 10.59 & /6.40 & 98 & /63 & 110 & /63 & 139 & /92 \\
\quad \seismicw & 11\,407 & /7\,462 & 8\,113 & /5\,345 & 4\,121 & /2\,151 & 9.18 & /6.37 & 9.68 & /4.74 & 11.82 & /6.37 & 858 & /364 & 1\,157 & /292 & 1\,742 & /364 \\
\quad \pyann & 357 & /866 & 112 & /395 & -- & /33 & 23.00 & /6.42 & 23.00 & /6.42 & -- & /6.42 & 2\,054 & /1\,248 & 2\,054 & /1\,248 & -- & /1\,248 \\
\quad \sindi & 1\,036 & /1\,858 & 660 & /1\,091 & 449 & /588 & 13.35 & /3.63 & 13.35 & /3.98 & 13.35 & /3.98 & 92 & /26 & 92 & /28 & 92 & /28 \\
\quad \lsp & 3\,402 & /2\,761 & 2\,040 & /1\,857 & -- & /-- & 30.08 & /3.42 & 30.08 & /8.74 & -- & /-- & 1\,045 & /259 & 1\,045 & /270 & -- & /-- \\
\midrule
\multicolumn{19}{@{}l}{\msmarcoesplade} \\
\quad \kannolohnsw & \multicolumn{2}{r}{1\,363} & \multicolumn{2}{r}{504} & \multicolumn{2}{r}{--} & \multicolumn{2}{r}{7.04} & \multicolumn{2}{r}{7.04} & \multicolumn{2}{r}{--} & \multicolumn{2}{r}{4\,150} & \multicolumn{2}{r}{4\,150} & \multicolumn{2}{r}{--} \\
\quad \kannolorerank & \multicolumn{2}{r}{3\,286} & \multicolumn{2}{r}{1\,862} & \multicolumn{2}{r}{255} & \multicolumn{2}{r}{7.92} & \multicolumn{2}{r}{7.92} & \multicolumn{2}{r}{7.95} & \multicolumn{2}{r}{5\,428} & \multicolumn{2}{r}{5\,428} & \multicolumn{2}{r}{5\,428} \\
\quad \seismic & \multicolumn{2}{r}{13\,585} & \multicolumn{2}{r}{10\,302} & \multicolumn{2}{r}{4\,501} & \multicolumn{2}{r}{8.37} & \multicolumn{2}{r}{10.39} & \multicolumn{2}{r}{11.49} & \multicolumn{2}{r}{103} & \multicolumn{2}{r}{135} & \multicolumn{2}{r}{156} \\
\quad \seismicw & \multicolumn{2}{r}{12\,457} & \multicolumn{2}{r}{9\,885} & \multicolumn{2}{r}{4\,908} & \multicolumn{2}{r}{10.11} & \multicolumn{2}{r}{10.11} & \multicolumn{2}{r}{11.63} & \multicolumn{2}{r}{1\,562} & \multicolumn{2}{r}{1\,562} & \multicolumn{2}{r}{1\,844} \\
\quad \pyann & \multicolumn{2}{r}{68} & \multicolumn{2}{r}{--} & \multicolumn{2}{r}{--} & \multicolumn{2}{r}{25.86} & \multicolumn{2}{r}{--} & \multicolumn{2}{r}{--} & \multicolumn{2}{r}{1\,474} & \multicolumn{2}{r}{--} & \multicolumn{2}{r}{--} \\
\quad \sindi & \multicolumn{2}{r}{1\,132} & \multicolumn{2}{r}{836} & \multicolumn{2}{r}{463} & \multicolumn{2}{r}{13.23} & \multicolumn{2}{r}{14.23} & \multicolumn{2}{r}{14.23} & \multicolumn{2}{r}{99} & \multicolumn{2}{r}{104} & \multicolumn{2}{r}{104} \\
\quad \lsp & \multicolumn{2}{r}{4\,118} & \multicolumn{2}{r}{3\,201} & \multicolumn{2}{r}{--} & \multicolumn{2}{r}{31.28} & \multicolumn{2}{r}{31.28} & \multicolumn{2}{r}{--} & \multicolumn{2}{r}{1\,118} & \multicolumn{2}{r}{1\,118} & \multicolumn{2}{r}{--} \\
\bottomrule
\end{tabular}

\end{table*}

\begin{table*}[t]
\centering
\scriptsize
\setlength{\tabcolsep}{2pt}
\caption{Tail latency of the configurations of Tables~\ref{tab:exp_results_dense}
and~\ref{tab:exp_results_sparse} at $95$ and $99\%$ accuracy: 99th-percentile latency in \textmu s,
and its ratio to the median in parentheses. Dense: per-query times recorded by \vibe{} for the best
pass, $1\,000$ queries. Sparse: each query timed inside the search tool over the timed passes of
the reported configuration (in-process for \pyann); \lsp{} and
\sindi{} can only be timed per pass and are not included. -- where a system
does not reach the target.}
\label{tab:exp_tail}
\begin{tabular}{@{}lr@{\,}l@{\hspace{7pt}}r@{\,}l@{\hspace{7pt}}r@{\,}l@{\hspace{7pt}}r@{\,}l@{\hspace{7pt}}r@{\,}l@{\hspace{7pt}}r@{\,}l@{\hspace{7pt}}r@{\,}l@{\hspace{7pt}}r@{\,}l@{}}
\toprule
 & \multicolumn{4}{c}{\agnews} & \multicolumn{4}{c}{\msmarcoqwen} & \multicolumn{4}{c}{\dpr} & \multicolumn{4}{c}{\laion} \\
\cmidrule(lr){2-5}\cmidrule(lr){6-9}\cmidrule(lr){10-13}\cmidrule(lr){14-17}
 & \multicolumn{2}{c}{$95\%$} & \multicolumn{2}{c}{$99\%$} & \multicolumn{2}{c}{$95\%$} & \multicolumn{2}{c}{$99\%$} & \multicolumn{2}{c}{$95\%$} & \multicolumn{2}{c}{$99\%$} & \multicolumn{2}{c}{$95\%$} & \multicolumn{2}{c}{$99\%$} \\
\midrule
\kannolohnsw & $308$ & {\scriptsize $(3.1)$} & $957$ & {\scriptsize $(5.4)$} & $392$ & {\scriptsize $(2.5)$} & $1\,402$ & {\scriptsize $(4.3)$} & $486$ & {\scriptsize $(2.1)$} & $2\,180$ & {\scriptsize $(4.2)$} & $2\,342$ & {\scriptsize $(3.4)$} & $26\,235$ & {\scriptsize $(8.0)$} \\
\kannolorerank & $121$ & {\scriptsize $(1.4)$} & $488$ & {\scriptsize $(2.4)$} & $211$ & {\scriptsize $(1.7)$} & $519$ & {\scriptsize $(1.6)$} & $301$ & {\scriptsize $(1.5)$} & $850$ & {\scriptsize $(1.6)$} & $1\,445$ & {\scriptsize $(2.1)$} & $18\,187$ & {\scriptsize $(5.3)$} \\
\quad + ext-8 & $119$ & {\scriptsize $(1.4)$} & $450$ & {\scriptsize $(1.6)$} & $226$ & {\scriptsize $(1.6)$} & $1\,439$ & {\scriptsize $(3.7)$} & $326$ & {\scriptsize $(1.5)$} & $1\,777$ & {\scriptsize $(2.1)$} & $1\,894$ & {\scriptsize $(2.1)$} & \multicolumn{2}{c}{--} \\
\hnswlib & $446$ & {\scriptsize $(1.7)$} & $1\,211$ & {\scriptsize $(1.9)$} & $641$ & {\scriptsize $(1.8)$} & $2\,144$ & {\scriptsize $(2.0)$} & $691$ & {\scriptsize $(1.5)$} & $2\,310$ & {\scriptsize $(1.7)$} & $1\,996$ & {\scriptsize $(1.3)$} & \multicolumn{2}{c}{--} \\
\glass & $102$ & {\scriptsize $(1.5)$} & $228$ & {\scriptsize $(1.4)$} & $237$ & {\scriptsize $(1.5)$} & $756$ & {\scriptsize $(1.9)$} & $333$ & {\scriptsize $(1.3)$} & $1\,024$ & {\scriptsize $(1.4)$} & $1\,341$ & {\scriptsize $(1.4)$} & \multicolumn{2}{c}{--} \\
\symphonyqg & $47$ & {\scriptsize $(1.5)$} & $121$ & {\scriptsize $(1.7)$} & $77$ & {\scriptsize $(1.7)$} & $290$ & {\scriptsize $(1.8)$} & $103$ & {\scriptsize $(1.6)$} & $454$ & {\scriptsize $(1.4)$} & $1\,096$ & {\scriptsize $(1.3)$} & \multicolumn{2}{c}{--} \\
\lorann & $362$ & {\scriptsize $(1.1)$} & $1\,193$ & {\scriptsize $(1.1)$} & $1\,805$ & {\scriptsize $(1.2)$} & $8\,265$ & {\scriptsize $(1.1)$} & $5\,132$ & {\scriptsize $(1.2)$} & \multicolumn{2}{c}{--} & $1\,385$ & {\scriptsize $(1.1)$} & $5\,247$ & {\scriptsize $(1.1)$} \\
\rabitqhnsw & $170$ & {\scriptsize $(1.7)$} & $442$ & {\scriptsize $(1.4)$} & $349$ & {\scriptsize $(1.9)$} & $1\,195$ & {\scriptsize $(1.7)$} & $426$ & {\scriptsize $(1.8)$} & $1\,621$ & {\scriptsize $(1.6)$} & $3\,283$ & {\scriptsize $(1.9)$} & \multicolumn{2}{c}{--} \\
\svsleanvec & $156$ & {\scriptsize $(1.4)$} & $346$ & {\scriptsize $(1.3)$} & $336$ & {\scriptsize $(1.5)$} & $894$ & {\scriptsize $(1.5)$} & $710$ & {\scriptsize $(1.4)$} & $1\,367$ & {\scriptsize $(1.4)$} & $2\,020$ & {\scriptsize $(1.2)$} & \multicolumn{2}{c}{--} \\
\midrule
 & \multicolumn{4}{c}{\msmarcosplade} & \multicolumn{4}{c}{\nq} & \multicolumn{4}{c}{\msmarcoesplade} & \multicolumn{4}{c}{} \\
\cmidrule(lr){2-5}\cmidrule(lr){6-9}\cmidrule(lr){10-13}
 & \multicolumn{2}{c}{$95\%$} & \multicolumn{2}{c}{$99\%$} & \multicolumn{2}{c}{$95\%$} & \multicolumn{2}{c}{$99\%$} & \multicolumn{2}{c}{$95\%$} & \multicolumn{2}{c}{$99\%$} & \multicolumn{4}{c}{} \\
\midrule
\kannolohnsw & $2\,082$ & {\scriptsize $(3.6)$} & $28\,446$ & {\scriptsize $(8.7)$} & $1\,106$ & {\scriptsize $(2.1)$} & $7\,778$ & {\scriptsize $(5.6)$} & $11\,287$ & {\scriptsize $(8.4)$} & \multicolumn{2}{c}{--} & \multicolumn{4}{c}{} \\
\kannolorerank & $574$ & {\scriptsize $(1.9)$} & $2\,047$ & {\scriptsize $(1.7)$} & $623$ & {\scriptsize $(1.8)$} & $2\,635$ & {\scriptsize $(2.4)$} & $1\,302$ & {\scriptsize $(2.6)$} & $14\,992$ & {\scriptsize $(4.4)$} & \multicolumn{4}{c}{} \\
\seismic & $425$ & {\scriptsize $(3.6)$} & $1\,073$ & {\scriptsize $(3.9)$} & $760$ & {\scriptsize $(3.3)$} & $1\,944$ & {\scriptsize $(3.8)$} & $301$ & {\scriptsize $(3.6)$} & $891$ & {\scriptsize $(5.0)$} & \multicolumn{4}{c}{} \\
\seismicw & $327$ & {\scriptsize $(3.0)$} & $683$ & {\scriptsize $(3.2)$} & $492$ & {\scriptsize $(3.0)$} & $1\,259$ & {\scriptsize $(3.1)$} & $287$ & {\scriptsize $(3.2)$} & $698$ & {\scriptsize $(4.1)$} & \multicolumn{4}{c}{} \\
\pyann & $14\,864$ & {\scriptsize $(1.7)$} & \multicolumn{2}{c}{--} & $4\,040$ & {\scriptsize $(1.6)$} & $47\,220$ & {\scriptsize $(1.5)$} & \multicolumn{2}{c}{--} & \multicolumn{2}{c}{--} & \multicolumn{4}{c}{} \\
\bottomrule
\end{tabular}

\end{table*}

\section{Design Studies}
\label{app:design}

This appendix reports the measurements behind design decisions stated in Sections~\ref{sec:data_management} and~\ref{sec:indexing}: the distance kernels, the Product-Quantization table build, dataset addressing, the visited set and the neighbor-processing loop of the graph search.

\subsection{Distance Kernels}
\label{app:kernels}

We compile each arithmetic kernel in two variants that differ \emph{only} in their reduction operators. We run both on a handful of \emph{cache-resident} vectors, which isolates pure compute. This measurement and the batched one below sweep synthetic vectors from $64$ to $2\,048$ dimensions, and we confirm them on real \sift, \cocond{} and higher-dimensional embeddings (micro-benchmark conventions of Section~\ref{sec:methodology}). Since the loop shape is the same, the reassociation barrier alone decides whether the loop vectorizes; Section~\ref{sec:distance_computation} reports the resulting speedups. The strict kernel retires several times more instructions per distance \emph{and} stalls on its dependency chain: its IPC falls from $1.6$ to below $0.9$ as the reduction lengthens. On the exhaustive top-$10$ results, whose rankings are identical, the mean relative score deviation is ${\approx}3\times10^{-7}$ on \dragon{} and exactly $0$ on the integer-valued \sift{} vectors, whose sums stay exactly representable.

\paragraph{Batched evaluation.} The batch-of-six kernel of Section~\ref{sec:distance_computation} cuts the instructions per \texttt{f32} distance from $392$ to $207$ at $d{=}1\,024$, while IPC stays flat or drops slightly. Its gain therefore comes from query reuse and not from instruction-level parallelism. Table~\ref{tab:batch6} reports the speedup over six single-candidate calls in three access regimes. In the \emph{cache-resident} regime, a few vectors are rescored without leaving L1/L2. In the \emph{sequential} regime, a scan runs over a buffer far larger than the last-level cache, as in exhaustive search. In the \emph{random} regime, access is scattered over the same buffer, as among graph neighbors. With cache-resident vectors, the \texttt{f32} gain grows with dimension, from $1.22\times$ at $d{=}64$ to $1.73\times$ at $d{=}1\,024$, because there is more compute per query load to amortize. \texttt{f16} gains less ($1.14$--$1.45\times$ up to $d{=}2\,048$). Batching frees the single memory operand of the fused multiply--add for the vector, but an \texttt{f16} value cannot use it without a separate widening instruction. The sequential scan is bandwidth-bound and the prefetcher already hides latency, so executing fewer instructions gains little. Under random access every vector is a cold miss, and the batched kernel wins by issuing six independent loads that overlap. The densified-query sparse kernel (\cocond, nominal dimension ${\approx}30$k) has no query reuse: at each step, its six documents index six scattered positions of the query buffer. Batching therefore retires \emph{more} instructions, and its only benefit comes from overlapping loads, so it helps under scattered access alone. This is the setting where the system uses it: the neighbors of a graph node, whose learned-sparse embeddings also share much of their support.

\begin{table}[t]
\centering
\small
\caption{Speedup of the batch-of-six dot-product kernel over six single-candidate calls, by access regime; micro-benchmark conventions of Section~\ref{sec:methodology}. \emph{Sparse}: densified-query kernel on \cocond.}
\label{tab:batch6}
\setlength{\tabcolsep}{3pt}
\begin{tabular}{@{}lrrrrrrr@{}}
\toprule
& \multicolumn{3}{c}{\texttt{f32}} & \multicolumn{3}{c}{\texttt{f16}} & \\
\cmidrule(lr){2-4}\cmidrule(lr){5-7}
$d$ & 128 & 768 & 1\,024 & 128 & 768 & 1\,024 & \textbf{Sparse} \\
\midrule
cache & 1.32 & 1.55 & 1.73 & 1.14 & 1.19 & 1.15 & 1.02 \\
sequential & 1.03 & 1.16 & 1.25 & 0.99 & 0.97 & 1.16 & 0.95 \\
random & 1.24 & 1.54 & 1.43 & 1.41 & 1.59 & 1.58 & 1.22 \\
\bottomrule
\end{tabular}
\end{table}

\paragraph{Other instruction sets.} The same source vectorizes for other instruction sets without changes. We rebuilt the binaries for \texttt{AVX2} alone, with no \texttt{AVX-512} instruction. The algebraic \texttt{f32} kernels are then $13\times$ faster than the strict ones at $d{=}128$ and $18$--$20\times$ at higher dimensions, as in the native build. The batch-of-six kernel gains $1.20$--$1.84\times$ on cache-resident vectors, against $1.12$--$1.76\times$ for the native build in the same run. The PQ table build of Section~\ref{sec:product_quantization} is $4.5$--$5.2\times$ faster than the natural AoS build, against $4.4$--$5.2\times$ for the native build. Absolute times differ by at most $7\%$, because on this CPU the compiler emits $256$-bit vectors for these kernels even in the native build. On a Raspberry Pi~5 (Armv8, $128$-bit NEON), the algebraic \texttt{f32} kernels are $4.4$--$6.3\times$ faster than the strict ones, the batch-of-six kernel gains $1.65$--$1.70\times$, and the PQ table build is $2.6$--$4.2\times$ faster than the natural AoS build. The \texttt{f16} kernels widen each value with Rust's built-in \texttt{f16} conversion, which the compiler lowers to NEON's widening instruction. Their algebraic version is $3.5$--$3.8\times$ faster than the strict one.

\subsection{Product-Quantization Table Build}
\label{app:pq_table}

We time the table build alone. We sweep $m$ over the full divisor range at $d \in \{128, 768, 1\,024\}$ for both metrics. The $256d$ multiply--adds do not depend on $m$, so the sweep changes only the reduction length $\text{dsub}$ and the loop nest around it. The natural AoS build has a single IEEE-strict accumulator, which keeps the reduction serial. Besides this build, we compile an AoS variant with the algebraic operators of Section~\ref{sec:distance_computation}, which separates the effect of the layout from that of the reassociation. All variants produce the same table up to \texttt{f32} round-off, ${\approx}5\times10^{-7}$ (micro-benchmark conventions of Section~\ref{sec:methodology}). Loops this short are sensitive to code placement. The alignment of branch targets alone changes the time of a kernel by up to $1.65\times$, through instruction-fetch stalls and not through its arithmetic. Each figure is therefore the minimum over three builds that differ only in code alignment. Table~\ref{tab:bpq1} reports the three deployed shapes, all at $\text{dsub}=8$, the least favorable end of the sweep (Section~\ref{sec:product_quantization} gives their speedups). Across the $36$ shapes, the gain over natural AoS ranges from $4.5\times$ to $12.5\times$. Against the fast-math variant it never falls below $2.5\times$ at any $\text{dsub} \in \{4,8,16\}$. Transposing alone, with neither build tiled, gives $3.4$--$5.1\times$ on the deployed shapes, $3.4$--$13.8\times$ across the sweep and $2.9$--$4.1\times$ against the fast-math variant. Tiling with $T{=}32$ adds a further $1.02$--$1.33\times$, while $T{=}8$ is slower than no tiling. Fast-math itself speeds up AoS by $1.26\times$ on $\ell_2$ but only by $1.12\times$ on IP. At $\text{dsub}=8$ both AoS builds are bound by instruction issue, and the two fast-math kernels run in nearly the same time. The gain therefore measures the extra work of the strict kernel. The strict $\ell_2$ kernel also pays a subtraction per element, which the vectorized form amortizes across lanes.
AoS wins only at $\text{dsub} \gtrsim 96$, where a long reduction finally amortizes the horizontal sum. This regime means $m \le 8$, a code of at most eight bytes for a $768$-dimensional vector, and is of no practical interest.

\begin{table}[t]
\centering
\small
\setlength{\tabcolsep}{4pt}
\caption{Product-Quantization distance-table build, \textmu s per table; micro-benchmark conventions of Section~\ref{sec:methodology}, minimum over three builds. \emph{fm}: fast-math; $T$: SoA tile width, \emph{none} for the untiled SoA build.}
\label{tab:bpq1}
\begin{tabular}{@{}llrrrrr@{}}
\toprule
& & \multicolumn{2}{c}{\textbf{AoS}} & \multicolumn{3}{c}{\textbf{SoA}} \\
\cmidrule(lr){3-4}\cmidrule(lr){5-7}
\textbf{Shape} & \textbf{Metric} & plain & fm & none & $T{=}8$ & $T{=}32$ \\
\midrule
\siftom                 & $\ell_2$ & 12.4 &  9.8 &  2.6 &  3.8 &  \textbf{2.4} \\
$d{=}128$, $m{=}16$     & IP       & 11.0 &  9.8 &  2.5 &  3.4 &  \textbf{2.2} \\
\midrule
\dragon                  & $\ell_2$ & 74.2 & 58.9 & 14.5 & 22.9 & \textbf{14.2} \\
$d{=}768$, $m{=}96$     & IP       & 66.1 & 58.8 & 14.5 & 20.4 & \textbf{13.3} \\
\midrule
\snowf                  & $\ell_2$ & 99.3 & 78.6 & 26.7 & 33.5 & \textbf{20.4} \\
$d{=}1\,024$, $m{=}128$   & IP       & 88.7 & 78.5 & 26.4 & 30.3 & \textbf{19.8} \\
\bottomrule
\end{tabular}
\end{table}

\subsection{Identifier and Range Addressing}
\label{app:range_access}

We isolate the effect on the full \cocond{} collection stored at \texttt{f16}. We perform five million accesses in random order, which is the scattered graph-neighbor pattern, and in storage order as a control. Each access is scored by the same \texttt{QueryEvaluator}. The three strategies fetch by identifier, by a range precomputed once, and by range with a prefetch issued $32$ vectors ahead of use. We follow the micro-benchmark conventions of Section~\ref{sec:methodology}.

Under random access, an access costs $331$\,ns by identifier, $226$\,ns by range and $160$\,ns by range with prefetch. Range addressing removes the ${\approx}1.8$ last-level-cache misses per access caused by the offsets table (${\approx}70$\,MB), which does not fit in cache. The prefetch removes ${\approx}4$ more misses. It executes slightly more instructions in fewer cycles, and IPC rises from $0.18$ to $0.43$. The gain therefore comes from hiding latency, and the amount of work does not decrease. In storage order the three strategies tie at $142$\,ns, so the benefit is specific to scattered access. Section~\ref{sec:batch_prefetch} measures the same prefetch inside the graph beam search.

\subsection{Visited Set}
\label{app:visited_set}

\paragraph{Sizing the visited set.}
Only the ground level of \hnsw{} keeps a visited set. On the upper levels, the greedy walk moves strictly closer to the query and cannot revisit a node. On the ground level without early termination ($\lambda=0$), the search expands about $ef_{\text{search}}$ nodes and scores the unvisited neighbors of each. We therefore expect $n_{\text{visited}}$ to follow $ef_{\text{search}}$ and the degree of the graph, reduced by the overlap between neighbor lists.

We instrumented \kannolo{} to record, per query, $n_{\text{visited}}$, the number of expanded nodes and the length of the neighbor lists they scan. We vary $ef_{\text{search}} \in \{10,\allowbreak 20,\allowbreak 40,\allowbreak 80,\allowbreak 160,\allowbreak 320\}$ and $\lambda \in \{0, 0.05, 0.1, 0.15, 0.2, 0.3\}$ on a $10$M sample of \siftob, on \dragon{} and on \cocond{}, all indexed with $M=32$ and $ef_{\text{construction}}=200$. At $\lambda=0$ a query expands between $1.0$ and $1.5$ times $ef_{\text{search}}$ nodes, at most $7$ beyond it. Between $62\%$ and $90\%$ of the neighbors it scans are new, and this share decreases as $ef_{\text{search}}$ grows. $n_{\text{visited}}$ therefore follows $ef_{\text{search}}$ times the average ground-level degree. The estimate reads this degree from the index, because the pruning heuristic leaves most nodes below the $2M$ bound.

Adaptive termination (Section~\ref{sec:early_termination}) keeps expanding while a candidate is within a factor $1+\lambda$ of the $ef_{\text{search}}$-th best distance. Its extra work therefore also scales with $ef_{\text{search}}$. At a fixed $\lambda$, the number of expansions is multiplied by a factor that barely depends on $ef_{\text{search}}$: between $3.8$ and $4.7$ at $\lambda=0.15$ on \sift{} at $10$M. The factor grows faster than linearly in $\lambda$. At $ef_{\text{search}}=40$ it is $1.6$, $2.7$, $4.6$, $7.7$ and $19.7$ for $\lambda$ from $0.05$ to $0.3$. Its growth rate depends on the collection: at $ef_{\text{search}}=20$ and $\lambda=0.3$ the factor is $19$ on \sift{} at $10$M, $30$ on \cocond{} and $108$ on \dragon{} at $8.8$M.

The capacity estimate of Section~\ref{sec:visited_set} follows from these observations. The term $\bar{d} \cdot ef_{\text{search}}$ counts the neighbors of the expanded nodes, ignoring their overlap. The constant $200$ covers the few expansions beyond $ef_{\text{search}}$. The $\lambda$ term models the growth of the expansions. Suppose that the number of candidates within a threshold $t$ grows as $t^{D}$, with $D$ the local intrinsic dimensionality of the data around the query~\cite{intrinsic_dim_levina,AUMULLER2021101807}. Then a factor $1+\lambda$ on a squared-Euclidean score multiplies them by $(1+\lambda)^{D/2}$. Under inner product, scores are measured from zero and not from the query, so the effective exponent also grows with the ratio of the scores to their spread. Measured on the grid as $\ln(\text{factor})/\ln(1+\lambda)$, the exponent ranges from $8$ to $11$ on \sift, $9$ to $14$ on \cocond{} and $12$ to $18$ on \dragon. We take the conservative value $8$. It never over-predicts the growth, but it causes more rehashes where the exponent is larger. The offset $0.05$, like the constant $200$, gives small relaxations more room. We chose the constants by hand on these development collections and did not fit them. The estimate balances rehashing against oversized tables. Small sizing errors have little effect on search time, which is dominated by distance computations.

We judge the estimate by two errors. A query that outgrows its table pays a rehash, which is a full copy of everything inserted so far. A table much larger than needed costs more to clear and spreads the inserts over more cache lines. For each query we record whether it rehashes. If it does not, we record its overshoot: the slots the table holds divided by the query's $n_{\text{visited}}$. The allocator rounds the capacity up to a power of two, so a table holds one to about two times the estimate. A good estimate keeps rehashes rare and the overshoot between $2\times$ and $4\times$, with $10\times$ as the exception.

\begin{table}[t]
\centering
\small
\caption{Sizing quality of the capacity estimate, $ef_{\text{search}}$ from $10$ to $320$; the $\lambda>0$ rows pool $\lambda \in \{0.05, 0.1, 0.15\}$. Rehash: share of queries rehashing at least once and at least twice. Overshoot: over the queries that do not rehash.}
\label{tab:visited_sizing}
\setlength{\tabcolsep}{3.5pt}
\begin{tabular}{@{}llrr*{3}{>{\raggedleft\arraybackslash}p{2.7em}}@{}}
\toprule
 & & \multicolumn{2}{c}{\textbf{Rehash (\%)}} & \multicolumn{3}{c}{\textbf{Overshoot} ($\times\,n_{\text{visited}}$)} \\
\cmidrule(lr){3-4}\cmidrule(l){5-7}
& $\boldsymbol{\lambda}$ & \textbf{any} & $\boldsymbol{\ge 2}$ & \textbf{med.} & \textbf{p99} & \textbf{max} \\
\midrule
\sift   & $=0$ & 5.1  & 0.0 & 1.7 & 4.8  & 11.6 \\
        & $>0$ & 10.9 & 0.7 & 2.1 & 10.3 & 41.0 \\
\dragon & $=0$ & 5.5  & 0.0 & 1.8 & 4.3  & 8.4 \\
        & $>0$ & 21.0 & 4.0 & 1.9 & 6.1  & 16.3 \\
\cocond & $=0$ & 7.2  & 0.2 & 2.1 & 5.5  & 9.4 \\
        & $>0$ & 7.4  & 0.6 & 2.5 & 8.4  & 20.1 \\
\bottomrule
\end{tabular}
\end{table}

\begin{figure*}[t]
    \centering
    \includegraphics[width=\textwidth]{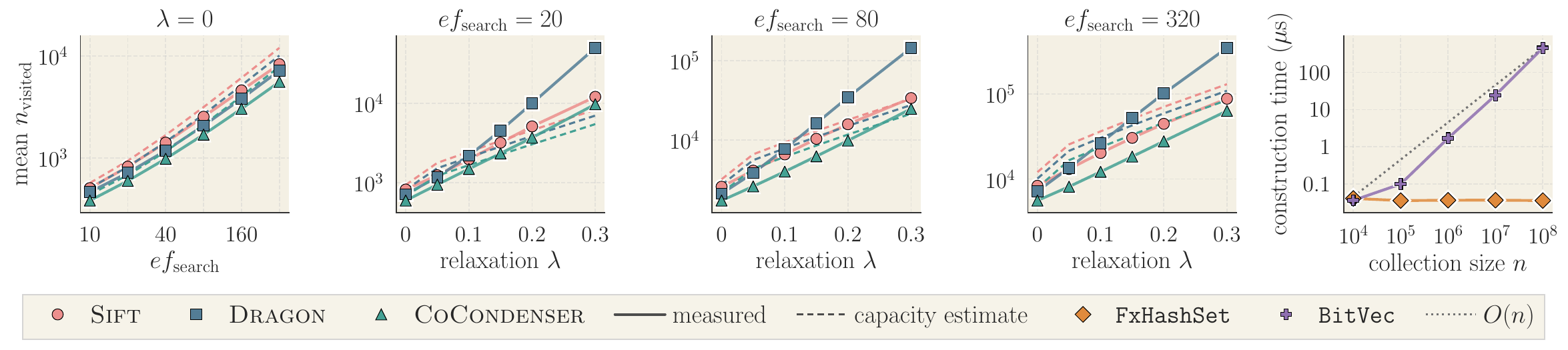}
    \caption{Mean $n_{\text{visited}}$ (solid) against the capacity estimate (dashed) on \sift{} at $10$M and \dragon{} and \cocond{} at $8.8$M: against $ef_{\text{search}}$ at $\lambda=0$, then against $\lambda$ at three fixed $ef_{\text{search}}$. Right: per-query construction cost of the two visited sets against $n$.}
    \label{fig:n_visited_scaling}
\end{figure*}

Figure~\ref{fig:n_visited_scaling} compares the estimate with the measured mean, and Table~\ref{tab:visited_sizing} reports both errors. Without early termination the estimate meets this target on all three collections: few queries rehash, almost none twice, and no table is oversized beyond $12\times$. Adaptive termination makes both errors more frequent, for two reasons. First, it widens the distribution of $n_{\text{visited}}$ within a configuration: its $99$th percentile is a median $2.6$ times the mean, against $1.9$ at $\lambda=0$. A table sized once per configuration therefore fits fewer queries well: more of them rehash, and more of them are oversized. Second, the growth with $\lambda$ differs between collections. On \sift{}, $1.0\%$ of queries are oversized beyond $10\times$, against $0.03\%$ on \dragon{} and $0.35\%$ on \cocond. On \dragon{}, a fifth of the queries rehash. At $\lambda=0.2$ and $0.3$ the difference grows: $4.2$ and $7.6\%$ of \sift{} queries are oversized beyond $10\times$, and $48$ and $92\%$ of \dragon{} queries rehash. A constant that served one collection better would serve the other worse. Even a perfect estimate of the mean would leave the easiest queries oversized and the hardest ones rehashing, since within one configuration the $99$th percentile of $n_{\text{visited}}$ is $1.6$ to $3.8$ times its mean.

\paragraph{\texttt{BitVec} against \texttt{FxHashSet}.}
With the capacity settled, we can compare the two structures end to end. At $n=10^8$ and $ef_{\text{search}}=20$, \texttt{BitVec} has the cheaper insert and hit ($4.2$ against $4.9$\,ns and $1.3$ against $2.3$\,ns), because every hash-set access computes a hash. It has the more expensive miss ($2.8$ against $2.1$\,ns). It also needs an $O(n)$ initialization on every query. Figure~\ref{fig:n_visited_scaling} (right) isolates this initialization. The initialization of \texttt{FxHashSet} is bounded by the capacity: $36$ to $62$\,ns for tables of $896$ to $3\,584$ slots, at every collection size. \texttt{BitVec} clears $n/8$ bytes per query. This takes $1.7\,\mu$s at $n=1$M, $23$ to $24\,\mu$s at $10$M, and $407$ to $455\,\mu$s at $100$M, where it is a $12.5$\,MB \texttt{memset}.
The end-to-end comparison crosses over at the same point as these two curves. At $n \le 1$M, \texttt{BitVec} is ahead by at most $10\,\mu$s for $ef_{\text{search}} \le 80$ and by up to $32\,\mu$s at $ef_{\text{search}}=160$. At $n=8.8$M, \texttt{FxHashSet} is faster in eight of nine configurations, by $2$ to $69\%$; \texttt{BitVec} wins the ninth by $13\%$. At $n=100$M, \texttt{BitVec} is slower in every configuration by $414$ to $424\,\mu$s per query (Section~\ref{sec:visited_set}).

\subsection{Neighbor Processing}
\label{app:batch_prefetch}

Four collections span the memory footprints the loop can encounter, from a vector that fits inside the prefetch window to one far larger: \snowf{} compressed with PQ128 ($2$ cache lines per vector), \siftom{} uncompressed at \texttt{f16} ($4$), the sparse \cocond{} (${\approx}7.5$ on average), and \dragon{} uncompressed at \texttt{f32} ($48$).

\begin{figure*}[t]
    \centering
    \includegraphics[width=\textwidth]{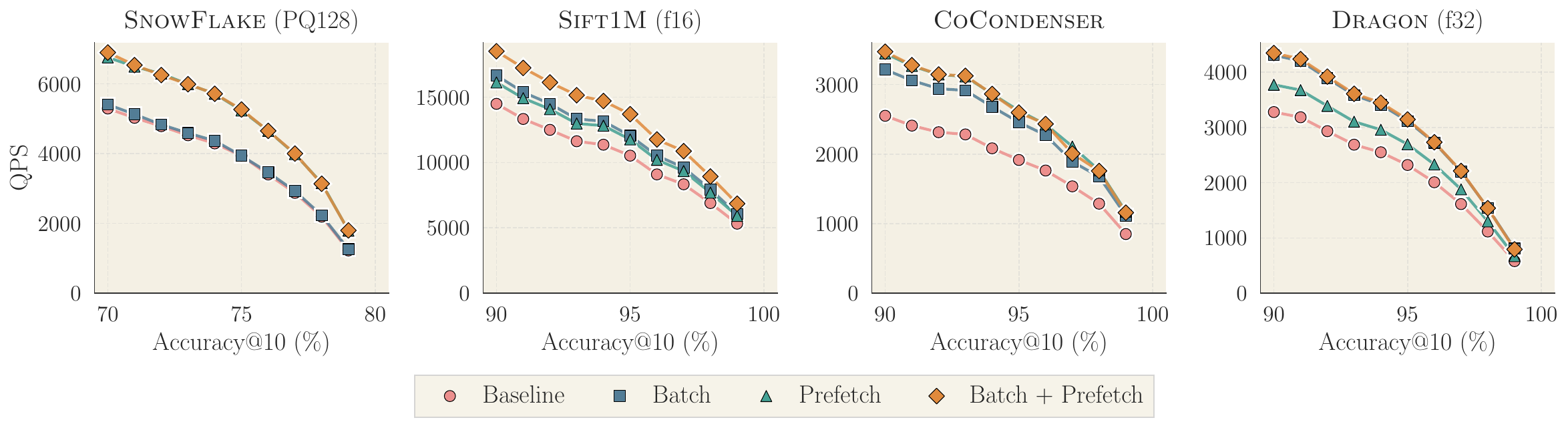}
    \caption{Batching and prefetching against QPS on four collections spanning the range of per-vector memory footprints.}
    \label{fig:batch_prefetch_comparison}
\end{figure*}

Figure~\ref{fig:batch_prefetch_comparison} reports QPS with each feature enabled alone and with both together. Which single feature wins depends on the collection's physical size and vector type.

The four collections separate the effects of the two mechanisms. On the compact \snowf{} codes, prefetching alone gives $+35\%$ QPS. The PQ codes span only $128$ bytes, so one hint covers the whole vector. Batching adds nothing, since the PQ encoder has no single-pass batched kernel. The vectors of \siftom{} also fit inside the window, and as plain dense vectors they also use the batched kernel; the two contributions add up to $29\%$. On \cocond{} both features help, prefetching more than batching. A batch of sparse vectors has six different sets of active components, so the gain of batching is limited to memory-level parallelism (Appendix~\ref{app:kernels}). Combining the two still costs nothing compared with the better of them. On \dragon, whose vectors are much larger than the window, batching gives the largest gain at $+34\%$ QPS, and prefetching contributes $+16\%$.

Hardware performance counters (IPC, last-level-cache load misses and retired instructions) show the mechanism behind each gain. As for the isolated kernel of Appendix~\ref{app:kernels}, batching improves instruction efficiency. On \dragon{} the IPC stays at ${\approx}3.3$ whether batching is on or off, while retired instructions fall by ${\approx}26\%$. Prefetching helps only where the CPU stalls on memory latency. On \cocond{} the irregular sparse accesses hold the baseline IPC down to $2.1$, and the $4$-cache-line prefetch cuts LLC misses by $45\%$ and raises QPS by $37\%$. On \dragon{} we can test the cap itself: prefetching all $48$ cache lines of a vector cuts LLC misses by $46\%$ but gains only $0.5\%$. The out-of-order engine already hid the misses that this prefetch removes. The $1\,536$ \texttt{PREFETCHT2} instructions it issues per expansion also lower the IPC enough to cancel the rest of the gain.

\printcredits

\section*{Declaration of generative AI and AI-assisted technologies in the manuscript preparation process}
During the preparation of this work the authors used Claude (Anthropic) in order to assist with parts of the software implementation, the setup of the experiments and the collection of their results, the preparation of the figures, and the preparation of the manuscript. After using this tool, the authors reviewed and edited the content as needed and take full responsibility for the content of the published article.

\section*{Acknowledgements}
We thank Leonardo Delfino and Domenico Erriquez for their work on the preliminary version of \kannolo~\cite{kannolo}.
The authors were partially supported by the European Union's Horizon Europe Research and Innovation program, under the RenAI project (GA No.~101291378) and by the ENRESFOOD project, funded under the FutureFoodS partnership. The project is co-funded by the Austrian Science Fund (FWF; 10.55776/KIN4661525), the Dutch Ministry of Agriculture, Fisheries, Food Security and Nature (BO-43-219-026), the Italian Ministry of University and Research (MUR), the German Federal Ministry of Research, Technology and Space (031B1717), and the European Union's Horizon Europe research and innovation programme via FutureFoodS, the European Partnership for a Sustainable Future of Food Systems (GA No.~101136361). The funders had no role in the study design, in the collection, analysis and interpretation of the data, in the writing of the article, or in the decision to submit it for publication.

\bibliographystyle{unsrtnat}
\bibliography{references}

\end{document}